\pdfoutput=1

\documentclass[12pt,a4paper]{article}

\usepackage{ifthen} 
\newboolean{pdflatex}
\setboolean{pdflatex}{true} 

\newboolean{articletitles}
\setboolean{articletitles}{true} 

\newboolean{uprightparticles}
\setboolean{uprightparticles}{false} 

\newboolean{inbibliography}
\setboolean{inbibliography}{false} 

\def\paperauthors{LHCb collaboration} 
\def\paperasciititle{Measurement of $C\!P$ asymmetries in two-body $B_{(s)}^(0)$-meson decays to charged pions and kaons} 
\def\papertitle{Measurement of \CP asymmetries in two-body \Bds-meson decays to charged pions and kaons} 
\def\paperkeywords{{High Energy Physics}, {LHCb}, {CP violation}} 
\def\papercopyright{\the\year\ CERN for the benefit of the LHCb collaboration} 
\def\paperlicence{CC-BY-4.0 licence}
\def\paperlicenceurl{https://creativecommons.org/licenses/by/4.0/}

\usepackage[top=1in, bottom=1.25in, left=1in, right=1in]{geometry}

\usepackage{microtype}
\usepackage{lineno}  
\usepackage{xspace} 
\usepackage{caption} 

\usepackage{graphicx}  
\usepackage{color}
\usepackage{colortbl}
\graphicspath{{./figs/}} 
\DeclareGraphicsExtensions{.eps}

\usepackage{amsmath} 
\usepackage{amssymb}
\usepackage{amsfonts}
\usepackage{upgreek} 

\newcommand*\patchAmsMathEnvironmentForLineno[1]{%
\expandafter\let\csname old#1\expandafter\endcsname\csname #1\endcsname
\expandafter\let\csname oldend#1\expandafter\endcsname\csname
end#1\endcsname
 \renewenvironment{#1}%
   {\linenomath\csname old#1\endcsname}%
   {\csname oldend#1\endcsname\endlinenomath}%
}
\newcommand*\patchBothAmsMathEnvironmentsForLineno[1]{%
  \patchAmsMathEnvironmentForLineno{#1}%
  \patchAmsMathEnvironmentForLineno{#1*}%
}
\AtBeginDocument{%
\patchBothAmsMathEnvironmentsForLineno{equation}%
\patchBothAmsMathEnvironmentsForLineno{align}%
\patchBothAmsMathEnvironmentsForLineno{flalign}%
\patchBothAmsMathEnvironmentsForLineno{alignat}%
\patchBothAmsMathEnvironmentsForLineno{gather}%
\patchBothAmsMathEnvironmentsForLineno{multline}%
\patchBothAmsMathEnvironmentsForLineno{eqnarray}%
}

\usepackage{hyperxmp}

\usepackage[pdftex,
            pdfauthor={\paperauthors},
            pdftitle={\paperasciititle},
            pdfkeywords={\paperkeywords},
            pdfcopyright={Copyright (C) \papercopyright},
            pdflicenseurl={\paperlicenceurl}]{hyperref}

\usepackage[all]{hypcap} 

\usepackage{xspace} 
\usepackage{upgreek}

\def\lhcb {\mbox{LHCb}\xspace}

\def\babar  {\mbox{BaBar}\xspace}
\def\belle  {\mbox{Belle}\xspace}

\def\cdf    {\mbox{CDF}\xspace}

\def\lhc    {\mbox{LHC}\xspace}

\def\MagUp {\mbox{\em Mag\kern -0.05em Up}\xspace}

\ifthenelse{\boolean{uprightparticles}}%
{

 \def\Pnu         {\ensuremath{\upnu}\xspace}                 
                  
 \def\Ppi         {\ensuremath{\uppi}\xspace}

 \def\PDelta      {\ensuremath{\Delta}\xspace}                 
 \def\PXi      {\ensuremath{\Xi}\xspace}                 
 \def\PLambda      {\ensuremath{\Lambda}\xspace}                 
 \def\PSigma      {\ensuremath{\Sigma}\xspace}                 
 \def\POmega      {\ensuremath{\Omega}\xspace}                 
 \def\PUpsilon      {\ensuremath{\Upsilon}\xspace}                 
 
 \def\PB      {\ensuremath{\mathrm{B}}\xspace}                 
                  
 \def\PD      {\ensuremath{\mathrm{D}}\xspace}

 \def\PK      {\ensuremath{\mathrm{K}}\xspace}

 \def\Pb      {\ensuremath{\mathrm{b}}\xspace}                 
 \def\Pc      {\ensuremath{\mathrm{c}}\xspace}                 
 \def\Pd      {\ensuremath{\mathrm{d}}\xspace}

 \def\Pi      {\ensuremath{\mathrm{i}}\xspace}

 \def\Pp      {\ensuremath{\mathrm{p}}\xspace}

 \def\Ps      {\ensuremath{\mathrm{s}}\xspace}

}
{

 \def\Pnu         {\ensuremath{\nu}\xspace}                 
                  
 \def\Ppi         {\ensuremath{\pi}\xspace}

 \mathchardef\PDelta="7101
 \mathchardef\PXi="7104
 \mathchardef\PLambda="7103
 \mathchardef\PSigma="7106
 \mathchardef\POmega="710A
 \mathchardef\PUpsilon="7107
                  
 \def\PB      {\ensuremath{B}\xspace}                 
                  
 \def\PD      {\ensuremath{D}\xspace}

 \def\PK      {\ensuremath{K}\xspace}

 \def\Pb      {\ensuremath{b}\xspace}                 
 \def\Pc      {\ensuremath{c}\xspace}                 
 \def\Pd      {\ensuremath{d}\xspace}

 \def\Pi      {\ensuremath{i}\xspace}

 \def\Pp      {\ensuremath{p}\xspace}

 \def\Ps      {\ensuremath{s}\xspace}

}

\makeatletter
\ifcase \@ptsize \relax
  \newcommand{\miniscule}{\@setfontsize\miniscule{4}{5}}
\or
  \newcommand{\miniscule}{\@setfontsize\miniscule{5}{6}}
\or
  \newcommand{\miniscule}{\@setfontsize\miniscule{5}{6}}
\fi
\makeatother

\DeclareRobustCommand{\optbar}[1]{\shortstack{{\miniscule (\rule[.5ex]{1.25em}{.18mm})}
  \\ [-.7ex] $#1$}}

\def\ellp       {{\ensuremath{\ell^+}}\xspace}

\def\neu        {{\ensuremath{\Pnu}}\xspace}

\def\dquark    {{\ensuremath{\Pd}}\xspace}
\def\dquarkbar {{\ensuremath{\overline \dquark}}\xspace}

\def\squark    {{\ensuremath{\Ps}}\xspace}
\def\squarkbar {{\ensuremath{\overline \squark}}\xspace}

\def\cquark    {{\ensuremath{\Pc}}\xspace}

\def\bquark    {{\ensuremath{\Pb}}\xspace}
\def\bquarkbar {{\ensuremath{\overline \bquark}}\xspace}
\def\bbbar     {{\ensuremath{\bquark\bquarkbar}}\xspace}

\def\pion   {{\ensuremath{\Ppi}}\xspace}
\def\piz    {{\ensuremath{\pion^0}}\xspace}

\def\pip    {{\ensuremath{\pion^+}}\xspace}
\def\pim    {{\ensuremath{\pion^-}}\xspace}

\def\pimp   {{\ensuremath{\pion^\mp}}\xspace}

\def\kaon    {{\ensuremath{\PK}}\xspace}
  \def\Kbar    {{\kern 0.2em\overline{\kern -0.2em \PK}{}}\xspace}

\def\KorKbar    {\kern 0.18em\optbar{\kern -0.18em K}{}\xspace}
\def\Kz      {{\ensuremath{\kaon^0}}\xspace}
\def\Kzb     {{\ensuremath{\Kbar{}^0}}\xspace}
\def\Kp      {{\ensuremath{\kaon^+}}\xspace}
\def\Km      {{\ensuremath{\kaon^-}}\xspace}
\def\Kpm     {{\ensuremath{\kaon^\pm}}\xspace}

  \def\Dbar    {{\kern 0.2em\overline{\kern -0.2em \PD}{}}\xspace}
\def\D       {{\ensuremath{\PD}}\xspace}

\def\DorDbar    {\kern 0.18em\optbar{\kern -0.18em D}{}\xspace}
\def\Dz      {{\ensuremath{\D^0}}\xspace}

\def\Dp      {{\ensuremath{\D^+}}\xspace}
\def\Dm      {{\ensuremath{\D^-}}\xspace}

\def\Dstarp  {{\ensuremath{\D^{*+}}}\xspace}

\def\Dsm     {{\ensuremath{\D^-_\squark}}\xspace}

\def\B       {{\ensuremath{\PB}}\xspace}
\def\Bbar    {{\ensuremath{\kern 0.18em\overline{\kern -0.18em \PB}{}}}\xspace}
\def\Bb      {{\ensuremath{\Bbar}}\xspace}
\def\BorBbar    {\kern 0.18em\optbar{\kern -0.18em B}{}\xspace}
\def\Bz      {{\ensuremath{\B^0}}\xspace}

\def\Bu      {{\ensuremath{\B^+}}\xspace}

\def\Bp      {{\ensuremath{\Bu}}\xspace}

\def\Bd      {{\ensuremath{\B^0}}\xspace}
\def\Bs      {{\ensuremath{\B^0_\squark}}\xspace}
\def\Bsb     {{\ensuremath{\Bbar{}^0_\squark}}\xspace}
\def\Bdb     {{\ensuremath{\Bbar{}^0}}\xspace}
\def\Bds     {{\ensuremath{\B^0_{(\squark)}}}\xspace}
\def\Bdsb    {{\ensuremath{\Bbar{}^0_{(\squark)}}}\xspace}

  \def\Y#1S{\ensuremath{\PUpsilon{(#1S)}}\xspace}

\def\proton      {{\ensuremath{\Pp}}\xspace}

\def\Lz          {{\ensuremath{\PLambda}}\xspace}
\def\Lbar        {{\ensuremath{\kern 0.1em\overline{\kern -0.1em\PLambda}}}\xspace}
\def\LorLbar    {\kern 0.18em\optbar{\kern -0.18em \PLambda}{}\xspace}

\def\Lb      {{\ensuremath{\Lz^0_\bquark}}\xspace}

\newcommand{\decay}[2]{\ensuremath{#1\!\to #2}\xspace}         

\def\to                 {\ensuremath{\rightarrow}\xspace}

\def\CP                {{\ensuremath{C\!P}}\xspace}
\def\CPT               {{\ensuremath{C\!PT}}\xspace}

\newcommand{\dms}{{\ensuremath{\Delta m_{\squark}}}\xspace}
\newcommand{\dmd}{{\ensuremath{\Delta m_{\dquark}}}\xspace}
\newcommand{\DG}{{\ensuremath{\Delta\Gamma}}\xspace}
\newcommand{\DGs}{{\ensuremath{\Delta\Gamma_{\squark}}}\xspace}
\newcommand{\DGd}{{\ensuremath{\Delta\Gamma_{\dquark}}}\xspace}
\newcommand{\Gs}{{\ensuremath{\Gamma_{\squark}}}\xspace}
\newcommand{\Gd}{{\ensuremath{\Gamma_{\dquark}}}\xspace}

\def\BdTopipi     {\decay{\Bd}{\pip\pim}}
\def\BsTopipi     {\decay{\Bs}{\pip\pim}}
\def\BsToKK       {\decay{\Bs}{\Kp\!\Km}}
\def\BdToKK       {\decay{\Bd}{\Kp\!\Km}}
\def\BdToKpi      {\decay{\Bd}{\Kp\pim}}
\def\BsTopiK      {\decay{\Bs}{\pip\!\Km}}
\def\LbTopK       {\decay{\Lb}{\proton\Km}}

\def\BdToDpi      {\decay{\Bd}{\Dm\pip}}
\def\BsToDspi     {\decay{\Bs}{\Dsm\pip}}

\def\Cpipi        {\ensuremath{C_{\pip\pim}\xspace}}
\def\Spipi        {\ensuremath{S_{\pip\pim}\xspace}}
\def\CKK          {\ensuremath{C_{\Kp\!\Km}\xspace}}
\def\SKK          {\ensuremath{S_{\Kp\!\Km}\xspace}}
\def\ADGKK        {\ensuremath{A_{\Kp\!\Km}^{\DG}\xspace}}
\def\ACPBd        {\ensuremath{A_{\CP}^{\Bd}\xspace}}
\def\ACPBs        {\ensuremath{A_{\CP}^{\Bs}\xspace}}

\def\AT#1     {\ensuremath{A_{\mathrm{T}}^{#1}}\xspace}           

\def\C#1      {\ensuremath{\mathcal{C}_{#1}}\xspace}                       
\def\Cp#1     {\ensuremath{\mathcal{C}_{#1}^{'}}\xspace}                    
\def\Ceff#1   {\ensuremath{\mathcal{C}_{#1}^{\mathrm{(eff)}}}\xspace}        
\def\Cpeff#1  {\ensuremath{\mathcal{C}_{#1}^{'\mathrm{(eff)}}}\xspace}       
\def\Ope#1    {\ensuremath{\mathcal{O}_{#1}}\xspace}                       
\def\Opep#1   {\ensuremath{\mathcal{O}_{#1}^{'}}\xspace}                    

\newcommand{\tev}{\ifthenelse{\boolean{inbibliography}}{\ensuremath{~T\kern -0.05em eV}}{\ensuremath{\mathrm{\,Te\kern -0.1em V}}}\xspace}
\newcommand{\gev}{\ensuremath{\mathrm{\,Ge\kern -0.1em V}}\xspace}
\newcommand{\mev}{\ensuremath{\mathrm{\,Me\kern -0.1em V}}\xspace}
\newcommand{\kev}{\ensuremath{\mathrm{\,ke\kern -0.1em V}}\xspace}
\newcommand{\ev}{\ensuremath{\mathrm{\,e\kern -0.1em V}}\xspace}
\newcommand{\gevc}{\ensuremath{{\mathrm{\,Ge\kern -0.1em V\!/}c}}\xspace}
\newcommand{\mevc}{\ensuremath{{\mathrm{\,Me\kern -0.1em V\!/}c}}\xspace}
\newcommand{\gevcc}{\ensuremath{{\mathrm{\,Ge\kern -0.1em V\!/}c^2}}\xspace}
\newcommand{\gevgevcccc}{\ensuremath{{\mathrm{\,Ge\kern -0.1em V^2\!/}c^4}}\xspace}
\newcommand{\mevcc}{\ensuremath{{\mathrm{\,Me\kern -0.1em V\!/}c^2}}\xspace}

\def\mum  {\ensuremath{{\,\upmu\mathrm{m}}}\xspace}

\def\invfb   {\ensuremath{\mbox{\,fb}^{-1}}\xspace}

\def\ps   {\ensuremath{{\mathrm{ \,ps}}}\xspace}
\def\fs   {\ensuremath{\mathrm{ \,fs}}\xspace}

\def\invps{\ensuremath{{\mathrm{ \,ps^{-1}}}}\xspace}

\newcommand{\chisq}{\ensuremath{\chi^2}\xspace}

\newcommand{\chisqip}{\ensuremath{\chi^2_{\text{IP}}}\xspace}

\def\gsim{{~\raise.15em\hbox{$>$}\kern-.85em
          \lower.35em\hbox{$\sim$}~}\xspace}
\def\lsim{{~\raise.15em\hbox{$<$}\kern-.85em
          \lower.35em\hbox{$\sim$}~}\xspace}

\def\ptot       {\mbox{$p$}\xspace}
\def\pt         {\mbox{$p_{\mathrm{ T}}$}\xspace}

\def\evtgen     {\mbox{\textsc{EvtGen}}\xspace}

\def\geant      {\mbox{\textsc{Geant4}}\xspace}

\def\photos     {\mbox{\textsc{Photos}}\xspace}

\def\pythia     {\mbox{\textsc{Pythia}}\xspace}

\def\tell1  {TELL1\xspace}
\def\ukl1   {UKL1\xspace}

\newcommand{\ie}{\mbox{\itshape i.e.}\xspace}

\usepackage{cite} 
\usepackage{mciteplus}

\usepackage{longtable} 
\usepackage{lscape}

\mathchardef\mhyphen="2D
\def\threebody {{\rm 3\mhyphen body}}
\begin{document}

\renewcommand{\thefootnote}{\fnsymbol{footnote}}
\setcounter{footnote}{1}


\begin{titlepage}
\pagenumbering{roman}

\vspace*{-1.5cm}
\centerline{\large EUROPEAN ORGANIZATION FOR NUCLEAR RESEARCH (CERN)}
\vspace*{1.5cm}
\noindent
\begin{tabular*}{\linewidth}{lc@{\extracolsep{\fill}}r@{\extracolsep{0pt}}}
\ifthenelse{\boolean{pdflatex}}
{\vspace*{-1.5cm}\mbox{\!\!\!\includegraphics[width=.14\textwidth]{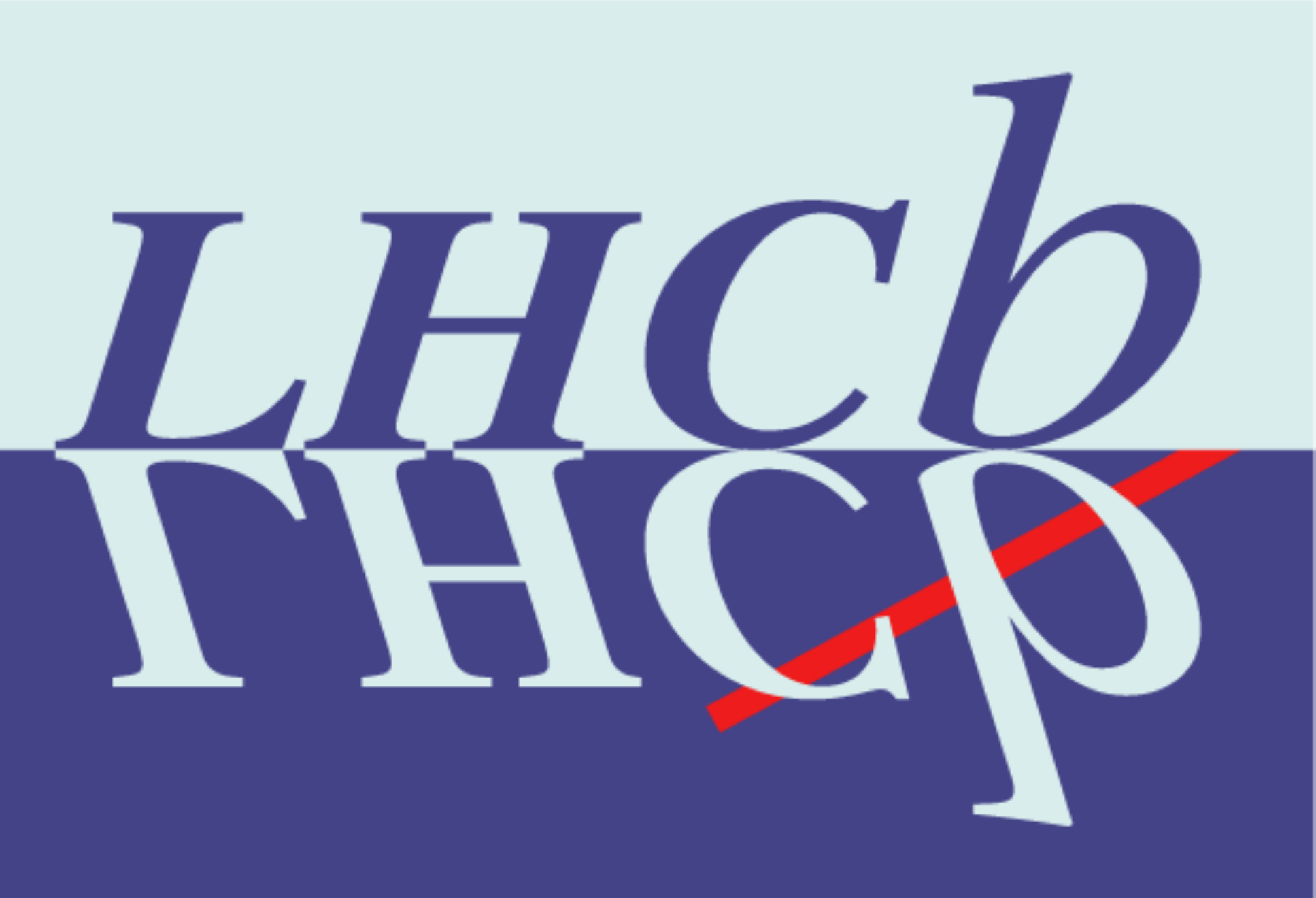}} & &}%
{\vspace*{-1.2cm}\mbox{\!\!\!\includegraphics[width=.12\textwidth]{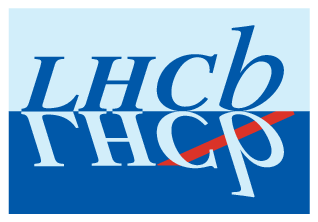}} & &}%
\\
 & & CERN-EP-2018-086 \\  
 & & LHCb-PAPER-2018-006 \\  
 & & August 21, 2018 \\ 
 & & \\
\end{tabular*}

\vspace*{4.0cm}

{\normalfont\bfseries\boldmath\huge
\begin{center}
  \papertitle 
\end{center}
}

\vspace*{2.0cm}

\begin{center}
\paperauthors\footnote{Authors are listed at the end of this paper.}
\end{center}

\vspace{\fill}

\begin{abstract}
  \noindent
  The time-dependent \CP asymmetries in \BdTopipi and \BsToKK decays are measured using a data sample of \proton\proton collisions corresponding to an integrated luminosity of 3.0\invfb, collected with the \lhcb detector at centre-of-mass energies of 7 and 8\tev. The same data sample is used to measure the time-integrated \CP asymmetries in \BdToKpi and \BsTopiK decays. The results are $\Cpipi = -0.34 \pm 0.06 \pm 0.01$, $\Spipi = -0.63 \pm 0.05 \pm 0.01$, $\CKK = 0.20 \pm 0.06 \pm 0.02$, $\SKK = 0.18 \pm 0.06 \pm 0.02$, $\ADGKK = -0.79 \pm 0.07 \pm 0.10$, $\ACPBd = -0.084 \pm 0.004 \pm 0.003$, and $\ACPBs = 0.213 \pm 0.015 \pm 0.007$, where the first uncertainties are statistical and the second systematic. Evidence for \CP violation is found in the \BsToKK decay for the first time.
\end{abstract}

\vspace*{2.0cm}

\begin{center}
  Published in Phys.~Rev.~D98 (2018) 032004 
\end{center}

\vspace{\fill}

{\footnotesize 
\centerline{\copyright~\papercopyright. \href{\paperlicenceurl}{\paperlicence}.}}
\vspace*{2mm}

\end{titlepage}


\newpage
\setcounter{page}{2}
\mbox{~}
%
%
%
%

\cleardoublepage


\renewcommand{\thefootnote}{\arabic{footnote}}
\setcounter{footnote}{0}



\pagestyle{plain} 
\setcounter{page}{1}
\pagenumbering{arabic}


%



\section{Introduction}

The study of \CP violation in charmless decays of \Bds mesons to charged two-body final states represents a powerful tool to test the Cabibbo-Kobayashi-Maskawa (CKM) picture~\cite{Cabibbo:1963yz,Kobayashi:1973fv} of the quark-flavour mixing in the Standard Model (SM) and to investigate the presence of physics lying beyond~\cite{PhysRevLett.75.1703,He1999,Fleischer:1999pa,GRONAU200071,LIPKIN2005126,Fleischer:2007hj,Fleischer:2010ib}. As discussed in Refs.~\cite{Fleischer:1999pa,Fleischer:2007hj,Fleischer:2010ib}, the hadronic parameters entering the \BdTopipi and \BsToKK decay amplitudes are related by U-spin symmetry, \ie by the exchange of \dquark and \squark quarks in the decay diagrams.\footnote{Unless stated otherwise, the inclusion of charge-conjugate decay modes is implied throughout this paper.} It has been shown that a combined analysis of the branching fractions and \CP asymmetries in two-body \B-meson decays, accounting for U-spin breaking effects, allows stringent constraints on the CKM angle $\gamma$ and on the \CP-violating phase $-2\beta_s$ to be set~\cite{Ciuchini:2012gd,LHCb-PAPER-2014-045}. 
More recently, it has been proposed to combine the \CP asymmetries of the \BdTopipi and \BsToKK decays with information provided by the semileptonic decays \decay{\Bd}{\pim\ellp\neu}
and \decay{\Bs}{\Km\ellp\neu}, in order to achieve a substantial reduction of the theoretical uncertainty on the determination of $-2\beta_s$~\cite{Fleischer:2016jbf,Fleischer:2016ofb}.
The \CP asymmetry in the \BdTopipi decay is also a relevant input to the determination of the CKM angle $\alpha$, when combined with other measurements from the isospin-related decays \decay{\Bd}{\piz\piz} and \decay{\Bu}{\pip\piz}~\cite{PhysRevLett.65.3381,PhysRevD.76.014015,Charles2017}.

In this paper, measurements of the time-dependent \CP asymmetries in \BdTopipi and \BsToKK decays and of the time-integrated \CP asymmetries in \BdToKpi and \BsTopiK decays are presented. The analysis is based on a data sample of \proton\proton collisions corresponding to an integrated luminosity of 3.0\invfb, collected with the \lhcb detector at centre-of-mass energies of 7 and 8\tev. The results supersede those from previous analyses performed with 1.0\invfb of integrated luminosity at \lhcb~\cite{LHCb-PAPER-2013-040,LHCb-PAPER-2013-018}.

Assuming \CPT invariance, the \CP asymmetry as a function of decay time for \Bds mesons decaying to a \CP eigenstate $f$ is given by
\begin{equation}\label{eq:acpTDDef}
A_{\CP}(t)=\frac{\Gamma_{\Bdsb \to f}(t)-\Gamma_{\Bds \to f}(t)}{\Gamma_{\Bdsb \to f}(t)+\Gamma_{\Bds \to f}(t)}=\frac{-C_f \cos(\Delta m_{d,s} t) + S_f \sin(\Delta m_{d,s} t)}{\cosh\left(\frac{\Delta\Gamma_{d,s}}{2} t\right) + A^{\Delta\Gamma}_f \sinh\left(\frac{\Delta\Gamma_{d,s}}{2} t\right)},
\end{equation}
where $\Delta m_{d,s}$ and $\Delta\Gamma_{d,s}$ are the mass and width differences of the mass eigenstates in the $\Bds-\Bdsb$ system. The quantities $C_f$, $S_f$ and $A^{\Delta\Gamma}_f$ are defined as
\begin{equation}
C_{f} \equiv \frac{1-|\lambda_f|^2}{1+|\lambda_f|^2},\qquad
S_{f} \equiv  \frac{2 {\rm Im} \lambda_f}{1+|\lambda_f|^2},\qquad
A^{\Delta\Gamma}_f \equiv  - \frac{2 {\rm Re} \lambda_f}{1+|\lambda_f|^2},
\label{eq:adirmix} 
\end{equation}
where $\lambda_f$ is given by
\begin{equation}
\lambda_f \equiv \frac{q}{p}\frac{\bar{A}_f}{A_f}.
\end{equation}
The two mass eigenstates of the effective Hamiltonian in the $\Bds-\Bdsb$ system are $p|\Bds\rangle \pm q|\Bdsb\rangle$, where $p$ and $q$ are complex parameters. The parameter $\lambda_f$ is thus related to $\Bds-\Bdsb$ mixing (via $q/p$) and to the decay amplitudes of the \decay{\Bds}{f} decay~($A_f$) and of the \decay{\Bdsb}{f} decay~($\bar{A}_f$). Assuming negligible \CP violation in the mixing ($\left|q/p\right| = 1$), as expected in the SM and confirmed by current experimental determinations~\cite{HFLAV16,LHCb-PAPER-2016-013,LHCb-PAPER-2014-053}, the terms $C_f$ and $S_f$ parameterise \CP violation in the decay and in the interference between mixing and decay, respectively. 
The quantities $C_f$, $S_f$ and $A_f^{\DG}$ must satisfy the condition $\left(C_f\right)^2+\left(S_f\right)^2+\left(A_f^{\DG}\right)^2 = 1$. This constraint is not imposed in this analysis, but its validity is verified {\it a posteriori} as a cross-check. In this paper a negligible value of \DGd is assumed, as supported by current experimental knowledge~\cite{HFLAV16}. Hence the expression of the time-dependent \CP asymmetry for the \BdTopipi decay simplifies to $A_{\CP}(t)=-\Cpipi\cos(\dmd t) + \Spipi\sin(\dmd t)$. The time-integrated \CP asymmetry for a \Bds decay to a flavour-specific final state $f$, such as \BdToKpi and \BsTopiK, is defined as
\begin{equation}\label{eq:acpDef}
A_{\CP} = \frac{\left|\bar{A}_{\bar{f}}\right|^2-\left|A_f\right|^2}{\left|\bar{A}_{\bar{f}}\right|^2+\left|A_f\right|^2},
\end{equation}
where $A_{f}$ ($\bar{A}_{\bar{f}}$) is the decay amplitude of the \decay{\Bds}{f} (\decay{\Bdsb}{\bar{f}}) transition. The current experimental knowledge on $C_f$ and $S_f$ for the \BdTopipi and \BsToKK decays, and on $A_{\CP}$ for the \BdToKpi~(\ACPBd) and \BsTopiK~(\ACPBs) decays, is summarised in Tables~\ref{tab:currentStatusMeasurementsTD} and~\ref{tab:currentStatusMeasurementsTI}, respectively. Only \lhcb measured \CKK~and \SKK, while no previous measurement of \ADGKK~is available to date.
\begin{table}[t]
	\begin{center}
		\caption{\small Current experimental knowledge on \Cpipi, \Spipi, \CKK~and \SKK. For the experimental measurements, the first uncertainties are statistical and the second systematic, whereas for the averages the uncertainties include both contributions. The correlation factors, denoted as $\rho$, are also reported.}\label{tab:currentStatusMeasurementsTD}
		\begin{tabular}{l|ccc}
			Reference                        & \Cpipi                    & \Spipi                    & $\rho\left(\Cpipi,\,\Spipi\right)$ \\
			\hline
			\babar~\cite{Lees:2012mma}       & $-0.25 \pm 0.08 \pm 0.02$ & $-0.68 \pm 0.10 \pm 0.03$ & $-0.06$ \\
			\belle~\cite{Adachi:2013mae}     & $-0.33 \pm 0.06 \pm 0.03$ & $-0.64 \pm 0.08 \pm 0.03$ & $-0.10$ \\
			\lhcb~\cite{LHCb-PAPER-2013-040} & $-0.38 \pm 0.15 \pm 0.02$ & $-0.71 \pm 0.13 \pm 0.02$ & $\phantom{-}0.38$ \\
			\hline  
			HFLAV average~\cite{HFLAV16}     & $-0.31 \pm 0.05$          & $-0.66 \pm 0.06$          & $\phantom{-}0.00$ \\
			\noalign{\smallskip}

			                                 & \CKK                      & \SKK                      & $\rho\left(\CKK,\,\SKK\right)$ \\
			\hline
			\lhcb~\cite{LHCb-PAPER-2013-040} & $\phantom{-}0.14 \pm 0.11 \pm 0.03$ & $\phantom{-}0.30 \pm 0.12 \pm 0.04$ & $\phantom{-}0.02$ \\
		\end{tabular}
	\end{center}
\end{table}
\begin{table}[t]
	\begin{center}
		\caption{\small Current experimental knowledge on $A_{\CP}$ for \BdToKpi and \BsTopiK decays. For the experimental measurements, the first uncertainties are statistical and the second systematic, whereas for the averages the uncertainties include both contributions.}\label{tab:currentStatusMeasurementsTI}
		\begin{tabular}{l|cc}
			Experiment                       & \ACPBd                                & \ACPBs \rule[-1.3ex]{0pt}{0pt}\\
			\hline
			\babar~\cite{Lees:2012mma}       & $-0.107 \pm 0.016^{\,\,\,+\,\,\,0.006}_{\,\,\,-\,\,\,0.004}\phantom{0}\hspace{0.5mm}$ & $-$\rule{0pt}{2.6ex}\\
			\belle~\cite{Duh:2012ie}         & $-0.069 \pm 0.014 \pm 0.007$          & $-$ \\
			\cdf~\cite{Aaltonen:2014vra}     & $-0.083 \pm 0.013 \pm 0.004$          & $0.22 \pm 0.07 \pm 0.02$ \\
			\lhcb~\cite{LHCb-PAPER-2013-018} & $-0.080 \pm 0.007 \pm 0.003$          & $0.27 \pm 0.04 \pm 0.01$\rule[-1.3ex]{0pt}{0pt}\\
			\hline  
			HFLAV average~\cite{HFLAV16}     & $-0.082 \pm 0.006$                    & $0.26 \pm 0.04$ \rule{0pt}{2.6ex}\\
		\end{tabular}
	\end{center}
\end{table}

This paper is organised as follows. After a brief introduction to the \lhcb detector, trigger
and simulation in Sec.~\ref{sec:detector}, the event selection is described in Sec.~\ref{sec:selection}. The \CP asymmetries are determined by means of a simultaneous unbinned maximum likelihood fit to the distributions of candidates reconstructed in the \pip\pim, $\Kp\!\Km$ and \Kp\pim final-state hypotheses, with the fit model described in Sec.~\ref{sec:fitModel}. The measurement of time-dependent \CP asymmetries with \Bds mesons requires that the flavour of the decaying meson at the time of production is identified (flavour tagging), as discussed in Sec.~\ref{sec:flavourTagging}. In Sec.~\ref{sec:timeResolution}, the procedure to calibrate the per-event decay-time uncertainty is presented. The determination of the detection asymmetry between the \Kp\pim and \Km\pip final states, necessary to measure $A_{\CP}$, is discussed in Sec.~\ref{sec:asymmetrydet}. The results of the fits are given in Sec.~\ref{sec:fitResult} and the assessment of systematic uncertainties in Sec.~\ref{sec:systematics}. Finally, conclusions are drawn in Sec.~\ref{sec:conclusions}.


\section{Detector, trigger and simulation}\label{sec:detector}

The \lhcb detector~\cite{Alves:2008zz,LHCb-DP-2014-002} is a single-arm forward spectrometer covering the \mbox{pseudorapidity} range $2<\eta <5$, designed for the study of particles containing \bquark or \cquark quarks. The detector includes a high-precision tracking system
consisting of a silicon-strip vertex detector surrounding the $pp$ interaction region, a large-area silicon-strip detector located upstream of a dipole magnet with a bending power of about $4{\mathrm{\,Tm}}$, and three stations of silicon-strip detectors and straw drift tubes placed downstream of the magnet. The tracking system provides a measurement of momentum, \ptot, of charged particles with a relative uncertainty that varies from 0.5\% at low momentum to 1.0\% at 200\gevc.
The minimum distance of a track to a primary vertex (PV), the impact parameter (IP), is measured with a resolution of $(15+29/\pt)\mum$,
where \pt is the component of the momentum transverse to the beam, in\,\gevc. Different types of charged hadrons are distinguished using information from two ring-imaging Cherenkov (RICH) detectors. Photons, electrons and hadrons are identified by a calorimeter system consisting of scintillating-pad and preshower detectors, an electromagnetic calorimeter and a hadronic calorimeter. Muons are identified by a system composed of alternating layers of iron and multiwire proportional chambers.
The online event selection is performed by a trigger~\cite{LHCb-DP-2012-004}, which consists of a hardware stage, based on information from the calorimeter and muon systems, followed by a software stage, which applies a full event reconstruction.

At the hardware trigger stage, events are required to have a muon with high \pt or a hadron, photon or electron with high transverse energy in the calorimeters. For hadrons, the transverse energy threshold is 3.5\gev. The software trigger requires a two-track secondary vertex with a significant displacement from the PVs. At least one charged particle must have a transverse momentum $\pt>1.7\gevc$ in the 7\tev or $\pt>1.6\gevc$ in the 8\tev data, and be inconsistent with originating from a PV. A multivariate algorithm~\cite{BBDT} is used for the identification of secondary vertices consistent with the decay of a \bquark hadron. In order to improve the efficiency on signal, a dedicated trigger selection for two-body \bquark-hadron decays is implemented, imposing requirements on the quality of the reconstructed tracks, their \pt and IP, the distance of closest approach between the decay products, and the \pt, IP and proper decay time of the \bquark-hadron candidate.

Simulation is used to study the discrimination between signal and background events, and to assess the small differences between signal and calibration decays. The $pp$ collisions are generated using \pythia~\cite{Sjostrand:2007gs,Sjostrand:2006za} with a specific \lhcb configuration~\cite{LHCb-PROC-2010-056}.  Decays of hadronic particles are described by \evtgen~\cite{Lange:2001uf}, in which final-state radiation is generated using \photos~\cite{Golonka:2005pn}. The interaction of the generated particles with the detector, and its response, are implemented using the \geant toolkit~\cite{Allison:2006ve, *Agostinelli:2002hh} as described in Ref.~\cite{LHCb-PROC-2011-006}.

\section{Event selection}\label{sec:selection}

The candidates selected online by the trigger are filtered offline to reduce the amount of combinatorial background by means of a loose preselection. In addition, the decay products of the candidates, generically called \B, are required either to be responsible for the positive decision of the hadronic hardware trigger, or to be unnecessary for an affirmative decision of any of the hardware trigger requirements. Candidates that pass the preselection are then classified into mutually exclusive samples of different final states (\pip\pim, $\Kp\!\Km$, \Kp\pim and \Km\pip) by means of the particle identification (PID) capabilities of the \lhcb detector. Finally, a boosted decision tree (BDT) algorithm~\cite{Breiman,Roe} is used to separate signal from combinatorial background.

Three types of backgrounds are considered: other two-body \bquark-hadron decays with misidentified pions, kaons or protons in the final state (cross-feed background); pairs of randomly associated, oppositely charged tracks (combinatorial background); and pairs of oppositely charged tracks from partially reconstructed three-body decays of \bquark hadrons (three-body background). Since the three-body background gives rise to candidates with invariant-mass values well separated from the signal mass peak, the event selection is customised to reject mainly the cross-feed and combinatorial backgrounds, which affect the invariant mass region around the \Bd and \Bs masses.

The main cross-feed background in the \pip\pim ($\Kp\!\Km$) spectrum is the \BdToKpi decay, where a kaon (pion) is misidentified as a pion (kaon). The PID requirements are optimised in order to reduce the amount of this cross-feed background to approximately 10\% of the \BdTopipi and \BsToKK signals, respectively. The same strategy is adopted to optimise the PID requirements for the \Kp\pim final state, reducing the amount of the \BdTopipi and \BsToKK cross-feed backgrounds to approximately 10\% of the \BsTopiK yield. The PID efficiencies and misidentification probabilities for kaons and pions are determined using samples of \decay{\Dstarp}{\Dz(\to\Km\pip)\pip} decays\cite{Anderlini:2202412}. 

The BDT exploits the following properties of the decay products: the \pt of the two tracks; the minimum and maximum \chisqip of the two tracks with respect to all primary vertices, where \chisqip is defined as the difference in vertex-fit \chisq of a given PV reconstructed with and without the considered particle; the distance of closest approach between the two tracks and the quality of their common vertex fit. The BDT also uses properties of the reconstructed \B candidate, namely the \pt, the \chisqip with respect to the associated PV,\footnote{The associated PV is that with the smallest \chisqip with respect to the \B candidate.} and the \chisq of the distance of flight with respect to the associated PV, for a total of 9 variables.
A single BDT is used to select the four signal decay modes. This is trained with \BdTopipi simulated events to model the signal, and data in the high-mass sideband (from $5.6$ to $5.8\gevcc$) of the \pip\pim sample to model the combinatorial background. The possibility to use a different BDT selection for each signal has been investigated, finding no sizeable differences in the sensitivities on the \CP-violating quantities under study.
The optimal threshold on the BDT response is chosen to maximise $S/\sqrt{S+B}$, where $S$ and $B$ represent the estimated numbers of \BdTopipi signal and combinatorial background events, respectively, within $\pm 60\mevcc$ (corresponding to about $\pm 3$ times the invariant mass resolution) around the \Bd mass. Multiple candidates are present in less than 0.05\% of the events in the final sample. Only one candidate is accepted for each event on the basis of a reproducible pseudorandom sequence.

\section{Fit model}\label{sec:fitModel}

For each signal and relevant background component, the distributions of invariant
mass, decay time, flavour-tagging assignment with the associated mistag probability, and per-event decay-time uncertainty are modelled. The flavour-tagging assignment and its associated mistag probability are provided by two classes of algorithms, so-called opposite-side (OS) and same-side (SS) tagging, as discussed in Sec.~\ref{sec:flavourTagging}. Hence for each component it is necessary to model two flavour-tagging decisions and the associated mistag probabilities.

Signals are the \BdToKpi and \BsTopiK decays in the \Kp\pim sample, the \BdTopipi decay in the \pip\pim sample, and the \BsToKK decay in the $\Kp\!\Km$ sample. 
In the \pip\pim and $\Kp\!\Km$ samples, small but non-negligible components of \BsTopipi and \BdToKK decays, respectively, are present and must be taken into account. 
Apart from the cross-feed backgrounds from \B-meson decays considered in the optimisation of the event selection, the only other relevant source of cross-feed background is the \LbTopK decay with the proton misidentified as a kaon in the $\Kp\!\Km$ sample. Considering the PID efficiencies, the branching fractions and the relative hadronisation probabilities~\cite{HFLAV16}, this background is expected to give a contribution of about 2.5\% relative to the \BsToKK decay.
This component is also modelled in the fit. Two components of three-body backgrounds need to be modelled in the \Kp\pim sample: one due to \Bz and \Bp decays, and one due to \Bs decays. The only relevant contributions of three-body backgrounds to the \pip\pim and $\Kp\!\Km$ samples are found to be \Bz and \Bp decays, and \Bs decays, respectively. Components describing the combinatorial background are necessary in all of the three final states. 


\subsection{Mass model}\label{sec:massModel}

The signal component for each two-body decay is modelled by the probability density function (PDF) for the candidate mass $m$
\begin{equation}\label{eq:signalMassModel}
\mathcal{P}_{\rm sig}(m) = (1-f_{\rm tail})G(m;\mu,\sigma_1,\sigma_2,f_{\rm g})+f_{\rm tail}J(m;\mu,\sigma_1,\alpha_1,\alpha_2),
\end{equation}
where $G(m;\mu,\sigma_1,\sigma_2,f_{\rm g})$ is the sum of two Gaussian functions with common mean $\mu$ and widths $\sigma_1$ and $\sigma_2$, respectively; $f_{\rm g}$ is the relative fraction between the two Gaussian functions; $f_{\rm tail}$ is the relative fraction of the Johnson function $J(m;\mu,\sigma_1,\alpha_1,\alpha_2)$, defined as~\cite{JohnsonFunc}
\begin{equation}\label{eq:johnsonFunction}
 J(m;\mu,\sigma_1,\alpha_1,\alpha_2) = \frac{\alpha_2}{\sigma_1\sqrt{2\pi\left(1+z^2\right)}}\exp{\left[-\frac{1}{2}\left(\alpha_1+\alpha_2\sinh^{-1}{z}\right)^2\right]},
\end{equation}
where $z \equiv \left[\frac{m-\mu}{\sigma_1}\right]$, $\mu$ and $\sigma_1$ are in common with the dominant Gaussian function in Eq.~\eqref{eq:signalMassModel}, and $\alpha_1$ and $\alpha_2$ are two parameters governing the left- and right-hand side tails. In the fit to data, the parameters $\alpha_1$, $\alpha_2$ and $f_{\rm tail}$ are fixed to the values determined by fitting the model to samples of simulated decays, whereas the other parameters are left free to be adjusted by the fit.

The invariant-mass model of the cross-feed backgrounds is based on a kernel estimation method~\cite{Cranmer:2001aa} applied to simulated decays. The amount of each cross-feed background component is determined by rescaling the yields of the decay in the correct spectrum by the ratio of PID efficiencies for the correct and wrong mass hypotheses. For example, the yields of the \BdToKpi decay in the \pip\pim spectrum are determined through the equation
\begin{equation}
N_{\pip\pim}(\BdToKpi) = N(\BdToKpi)\,\,\frac{\varepsilon_{\pip\pim}(\BdToKpi)}{\varepsilon_{\Kp\pim}(\BdToKpi)},
\end{equation}
where $N_{\pip\pim}(\BdToKpi)$ is the number of \BdToKpi decays present in the \pip\pim sample, $N(\BdToKpi)$ is the number of \BdToKpi decays identified in the \Kp\pim sample, $\varepsilon_{\pip\pim}(\BdToKpi)$ is the probability to assign the \pip\pim hypothesis to a \BdToKpi decay, and $\varepsilon_{\Kp\pim}(\BdToKpi)$ is the probability to assign the correct hypothesis to a \BdToKpi decay. 

The components due to three-body \B decays are described by convolving a sum of two Gaussian functions, defined using the same parameters as those used in the signal model, with ARGUS functions~\cite{Albrecht:1994aa}. For the \Kp\pim sample two three-body background components are used: one describing three-body \Bd and \Bp decays and one describing three-body \Bs decays. For the \pip\pim and $\Kp\!\Km$ samples a single ARGUS component is found to be sufficient to describe the invariant-mass shape in the low-mass region. The combinatorial background is modelled by exponential functions with an independent slope for each final-state hypothesis.


\subsection{Decay-time model}\label{sec:timeModel}

The time-dependent decay rate of a flavour-specific \decay{\B}{f} decay and of its \CP conjugate \decay{\Bb}{\bar{f}}, as for the cases of \BdToKpi and \BsTopiK decays, is given by the PDF
\begin{equation}\label{eq:decayTimeB2KPI}
  \begin{split}
    f_{\rm FS}\left(t,\,\delta_t,\,\psi,\,\vec{\xi},\,\vec{\eta}\right) = & K_{\rm FS}\left(1-\psi A_{\CP}\right)\left(1-\psi A_{\rm F}\right) \times \\
    & \left\{\left[\left(1\!-\!A_{\rm P}\right)\!\Omega_{\rm sig}(\vec{\xi},\vec{\eta})\!+\!\left(1\!+\!A_{\rm P}\right)\!\bar{\Omega}_{\rm sig}(\vec{\xi},\vec{\eta})\right]\!H_{+}\left(t,\,\delta_t\right)\!+\!\right. \\
    & \left. \psi \!\left[\left(1\!-\!A_{\rm P}\right)\!\Omega_{\rm sig}(\vec{\xi},\vec{\eta})\!-\!\left(1\!+\!A_{\rm P}\right)\!\bar{\Omega}_{\rm sig}(\vec{\xi},\vec{\eta})\right]\!H_{-}\left(t,\,\delta_t\right) \!\right \}, 
  \end{split}
\end{equation}
where $K_{\rm FS}$ is a normalisation factor and the discrete variable $\psi$ assumes the value $+1$ for the final state $f$ and $-1$ for the final state $\bar{f}$. The direct \CP asymmetry, $A_{\CP}$, is defined in Eq.~\eqref{eq:acpDef}, while the final-state detection asymmetry, $A_{\rm F}$, and the \Bds-meson production asymmetry, $A_{\rm P}$, are defined as
\begin{equation}
  A_{\rm F} = \frac{\varepsilon_{\rm tot}\left(\bar{f}\right)-\varepsilon_{\rm tot}\left(f\right)}{\varepsilon_{\rm tot}\left(\bar{f}\right)+\varepsilon_{\rm tot}\left(f\right)},\qquad
  A_{\rm P} = \frac{\sigma_{\Bds}-\sigma_{\Bdsb}}{\sigma_{\Bds}+\sigma_{\Bdsb}},\label{eq:nuisanceAsymDef}
\end{equation}
where $\varepsilon_{\rm tot}$ is the time-integrated efficiency in reconstructing and selecting the final state $f$ or $\bar{f}$, and $\sigma_{\Bds}$ ($\sigma_{\Bdsb}$) is the production cross-section of the given \Bds (\Bdsb) meson. The asymmetry $A_{\rm P}$ arises because production rates of \Bds and \Bdsb mesons are not expected to be identical in proton-proton collisions. It is measured to be order of percent at \lhc energies~\cite{LHCb-PAPER-2016-062}. 
Although $A_{\CP}$ can be determined from a time-integrated analysis, its value needs to be disentangled from the contribution of the production asymmetry. By studying the more general time-dependent decay rate, the production asymmetry can be determined simultaneously. 

The variable $\vec{\xi}=\left(\xi_{\rm OS},\,\xi_{\rm SS}\right)$ is the pair of flavour-tagging assignments of the OS and SS algorithms used to identify the \Bds-meson flavour at production, and $\vec{\eta}=\left(\eta_{\rm OS},\,\eta_{\rm SS}\right)$ is the pair of associated mistag probabilities defined in Sec.~\ref{sec:flavourTagging}. The variables $\xi_{\rm OS}$ and $\xi_{\rm SS}$ can assume the discrete values $+1$ when the candidate is tagged as \Bds, $-1$ when the candidate is tagged as \Bdsb, and zero for untagged candidates. The functions $\Omega_{\rm sig}(\vec{\xi},\vec{\eta})$ and $\bar{\Omega}_{\rm sig}(\vec{\xi},\vec{\eta})$ are the PDFs of the variables $\vec{\xi}$ and $\vec{\eta}$ for a \Bds or a \Bdsb meson, respectively. Their definitions are given in Sec.~\ref{sec:flavourTagging}.
The functions $H_{+}\left(t,\,\delta_t\right)$ and $H_{-}\left(t,\,\delta_t\right)$ are defined as
\begin{eqnarray}
  H_{+}\left(t,\,\delta_t\right) & = & \left[e^{-\Gamma_{d,s} t^{\prime}}\cosh{\left(\frac{\Delta\Gamma_{d,s}}{2}t^{\prime}\right)}\right]\otimes R\left(t-t^{\prime}|\delta_t\right)\, g_{\rm sig}\left(\delta_t\right)\,\varepsilon_{\rm sig}\left(t\right),\label{eq:flavourSpecificHfunctions} \\
  H_{-}\left(t,\,\delta_t\right) & = & \left[e^{-\Gamma_{d,s} t^{\prime}}\cos{\left(\Delta m_{d,s}t^{\prime}\right)}\right]\otimes R\left(t-t^{\prime}|\delta_t\right)\, g_{\rm sig}\left(\delta_t\right)\,\varepsilon_{\rm sig}\left(t\right),\nonumber
\end{eqnarray}
where $R\left(t-t^{\prime}|\delta_t\right)$ and $g_{\rm sig}\left(\delta_t\right)$ are the decay-time resolution model and the PDF of the per-event decay-time uncertainty $\delta_t$, respectively, discussed in Sec.~\ref{sec:timeResolution}, and $\varepsilon_{\rm sig}(t)$ is the time-dependent efficiency in reconstructing and selecting signal decays.

If the final state $f$ is a \CP eigenstate, as for the \BdTopipi and \BsToKK decays, the decay-time PDF is given by
\begin{equation}
	\begin{split}\label{eq:decayTimeB2HH}
	f_{\CP}\left(t,\,\delta_t,\vec{\xi},\vec{\eta}\right) = K_{\CP} & \left\{ \left[\left(1-A_{\rm P}\right)\Omega_{\rm sig}\left(\vec{\xi},\vec{\eta}\right)+\left(1+A_{\rm P}\right)\bar{\Omega}_{\rm sig}\left(\vec{\xi},\vec{\eta}\right)\right]I_{+}\left(t,\,\delta_t\right)+\right. \\
	& \left. \left[\left(1-A_{\rm P}\right)\Omega_{\rm sig}\left(\vec{\xi},\vec{\eta}\right)-\left(1+A_{\rm P}\right)\bar{\Omega}_{\rm sig}\left(\vec{\xi},\vec{\eta}\right)\right]I_{-}\left(t,\,\delta_t\right) \right\},
	\end{split}
\end{equation}
where $K_{\CP}$ is a normalisation factor and the functions $I_{+}\left(t\right)$ and $I_{-}\left(t\right)$ are
\begin{eqnarray}
			I_{+}\left(t,\,\delta_t\right) & = & \left\{e^{-\Gamma_{d,s} t^{\prime}}\left[\cosh{\left(\frac{\Delta\Gamma_{d,s}}{2}t^{\prime}\right)}+A_{f}^{\Delta\Gamma}\sinh{\left(\frac{\Delta\Gamma_{d,s}}{2}t^{\prime}\right)}\right]\right\}\otimes \nonumber \\
			                               &   & R\left(t-t^{\prime}|\delta_t\right)\, g_{\rm sig}\left(\delta_t\right)\, \varepsilon_{\rm sig}\left(t\right), \\
			I_{-}\left(t,\,\delta_t\right) & = & \left\{e^{-\Gamma_{d,s} t^{\prime}}\left[C_f\cos{\left(\Delta m_{d,s}t^{\prime}\right)}-S_f\sin{\left(\Delta m_{d,s}t^{\prime}\right)}\right]\right\}\otimes \nonumber \\
			                               &   & R\left(t-t^{\prime}|\delta_t\right)\, g_{\rm sig}\left(\delta_t\right)\, \varepsilon_{\rm sig}\left(t\right). \nonumber
\end{eqnarray}

It is instructive to see how the equations above would become in the absence of experimental effects. The final-state detection asymmetry $A_{\rm F}$ would have a zero value. In the limit of perfect flavour tagging, \ie absence of untagged candidates and mistag probabilities equal to zero with full agreement between OS and SS taggers, the function $\Omega_{\rm sig}(\vec{\xi},\vec{\eta})$ ($\bar{\Omega}_{\rm sig}(\vec{\xi},\vec{\eta})$) would become identically equal to 1 (0) if $\xi_{\rm OS,SS} = 1$, and to 0 (1) if $\xi_{\rm OS,SS} = -1$. The case of perfect determination of the decay time would be obtained by replacing the product of functions $R\left(t-t^{\prime}|\delta_t\right)\, g_{\rm sig}\left(\delta_t\right)$ with a product of Dirac delta functions, $\delta(t-t^{\prime})\, \delta(\delta_t)$. Finally, in the absence of a time dependence of the efficiency, the function $\varepsilon_{\rm sig}\left(t\right)$ would assume constant value.

The expressions for the decay-time PDFs of the cross-feed background components are
determined from Eqs.~\eqref{eq:decayTimeB2KPI} and~\eqref{eq:decayTimeB2HH}, assuming that the decay time calculated under the wrong mass hypothesis is equal to that calculated using the correct hypothesis. This assumption is verified using samples of simulated decays.

\begin{figure}[t]
  \begin{center}
    \includegraphics[width=0.45\textwidth]{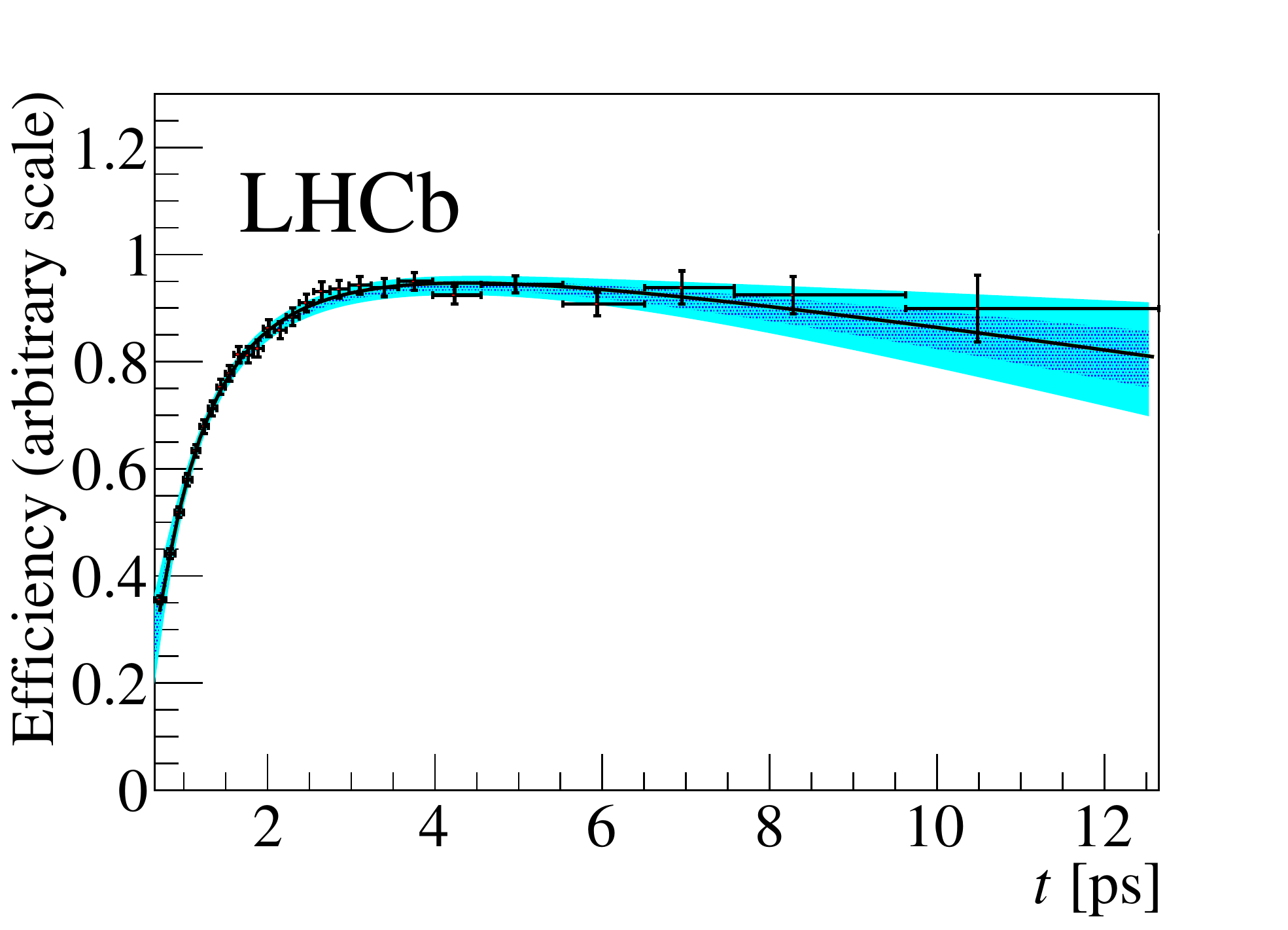}
    \includegraphics[width=0.45\textwidth]{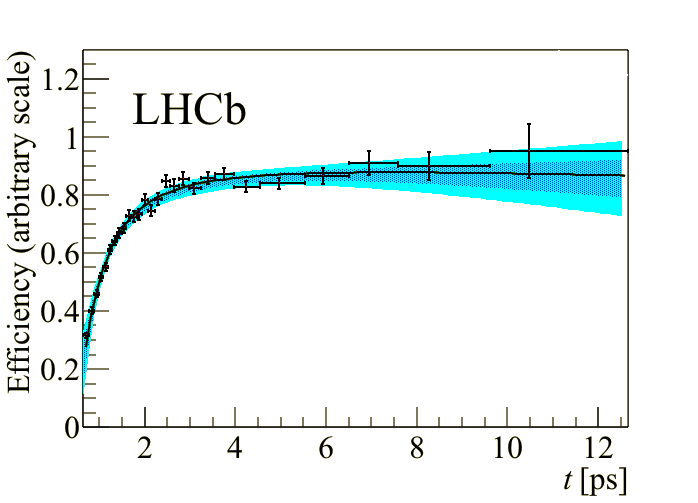}
    \includegraphics[width=0.45\textwidth]{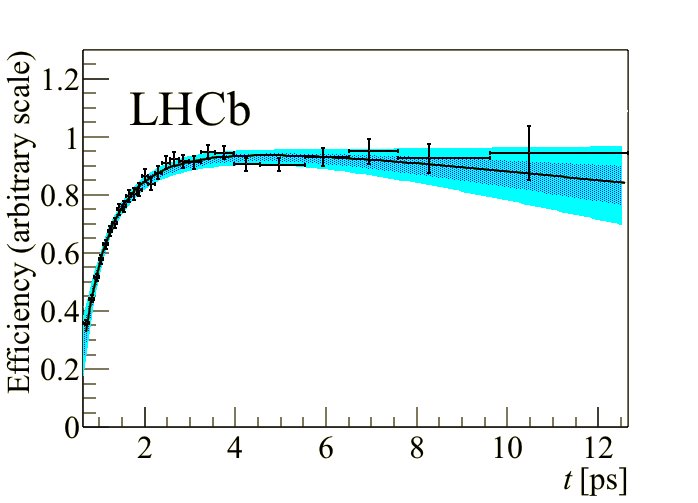}
    \includegraphics[width=0.45\textwidth]{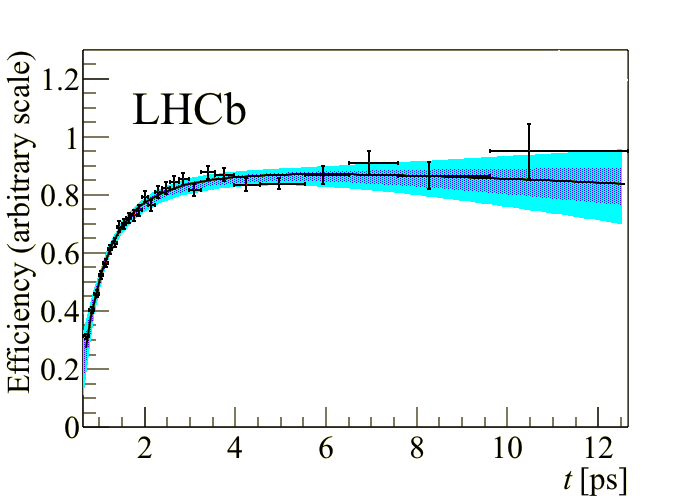}
 \end{center}
  \caption{\small Efficiencies as a function of decay time for (top left) \BdToKpi, (top right) \BsTopiK, (bottom left) \BdTopipi and (bottom right) \BsToKK decays. The black line is the result of the best fit of Eq.~\eqref{eq:accSigFunc} to the histograms, obtained as described in the text. The dark and bright areas correspond to the 68\% and 95\% confidence intervals, respectively.}
  \label{fig:accSigPlot}
\end{figure}

The efficiency $\varepsilon_{\rm sig}\left(t\right)$ is parameterised using the empirical function
\begin{equation} \label{eq:accSigFunc}
	\varepsilon_{\rm sig}\left(t\right) \propto \left[d_0-\mathrm{erf}\left(d_1 t^{d_2}\right)\right]\,\left(1-d_3 t\right),
\end{equation}
where $\mathrm{erf}$ denotes the error function and $d_i$ are parameters determined using the \BdToKpi decay, whose untagged time-dependent decay rate is a pure exponential with $\Gd=0.6588\pm0.0017\invps$~\cite{HFLAV16}. The yield of the \BdToKpi decay is determined in bins of decay time, by means of unbinned maximum likelihood fits to the \Kp\pim invariant-mass spectrum, using the model described in Sec.~\ref{sec:massModel}. The resulting histogram is then divided by a histogram built from an exponential function with decay constant equal to the central value of \Gd and arbitrary normalisation. By fitting the function in Eq.~\eqref{eq:accSigFunc} to the final histogram, the parameters $d_i$ are determined and fixed in the final fit to the data. The absolute scale of the efficiency function in Eq.~\eqref{eq:accSigFunc} is irrelevant in the likelihood maximisation since its value is absorbed into the global normalisation of the PDFs. For the other two-body decays under study, the same efficiency histogram is used, but with a small correction in order to take into account the differences between the various decay modes. The correction consists in multiplying the histogram by the ratio between the time-dependent efficiencies for the \BdToKpi and the other modes, as determined from simulated decays. The final histograms and corresponding time-dependent efficiencies for the \BdToKpi, \BsTopiK, \BdTopipi and \BsToKK decays are reported in Fig.~\ref{fig:accSigPlot}. 

The parameterisation of the decay-time distribution for combinatorial background in the \Kp\pim sample is studied by using the high-mass sideband from data, defined as $5.6 < m < 5.8\gevcc$. It is empirically found that the PDF can be written as
\begin{equation}\label{eq:combinatorialTime}
	\begin{split}
		f_{\rm comb}\left(t,\delta_t,\psi,\vec{\xi},\vec{\eta}\right) & = K_{\rm comb} \left( 1-\psi A_{\rm comb} \right ) \Omega_{\rm comb}(\vec{\xi},\vec{\eta})\, g_{\rm comb}(\delta_t)\, \times \\
		& \left[ f_{\rm comb}\, e^{-\Gamma_{\rm comb}t}+\left(1-f_{\rm comb}\right)e^{-\Gamma^{\prime}_{\rm comb}t} \right ]\varepsilon_{\rm comb}\!\left(t\right),
	\end{split}
\end{equation}
where $K_{\rm comb}$ is a normalisation factor; $\Omega_{\rm comb}(\vec{\xi},\vec{\eta})$ is the PDF of $\vec{\xi}$ and $\vec{\eta}$ for combinatorial-background candidates; $g_{\rm comb}(\delta_t)$ is the distribution of the per-event decay-time uncertainty $\delta_t$ for combinatorial background, discussed in Sec.~\ref{sec:timeResolution}; $A_{\rm comb}$ is the charge asymmetry of the combinatorial background; and $\Gamma_{\rm comb}$, $\Gamma^{\prime}_{\rm comb}$ and $f_{\rm comb}$ are free parameters to be determined by the fit. The function $\varepsilon_{\rm comb}\!\left(t\right)$ is an effective function, analogous to the time-dependent efficiency for signal decays. The parameterisation
\begin{equation}\label{eq:bkgCombAcc}
  \varepsilon_{\rm comb}\!\left(t\right) \propto 1-\mathrm{erf}\left(\frac{a_{\rm comb}-t}{a_{\rm comb} t}\right),
\end{equation}
where $a_{\rm comb}$ is a free parameter, provides a good description of the data. For the \pip\pim and $\Kp\!\Km$ samples, the same expression as in Eq.~\eqref{eq:combinatorialTime} is used, with $A_{\rm comb}$ set to zero.

The decay-time distribution of the three-body background component in the \Kp\pim sample is described using the same PDF as in Eq.~\eqref{eq:decayTimeB2KPI}, but with independent parameters entering the flavour-tagging PDF and an independent effective oscillation frequency. In addition, the time-dependent efficiency function in Eq.~\eqref{eq:flavourSpecificHfunctions} is parameterised as $\varepsilon_{\rm sig}(t)=\sum_{i=0}^6{c_{i}b_i(t)}$, following the procedure outlined in Ref.~\cite{CubicSpline}, where $b_i(t)$ are cubic spline functions and $c_i$ are coefficients left free to be adjusted during the final fit to data.

For the \pip\pim and $\Kp\!\Km$ samples, the decay-time distribution of three-body partially reconstructed backgrounds is parameterised using the PDF
\begin{equation}\label{eq:physTime}
  f_{\rm 3\mhyphen body}\left(t,\delta_t,\vec{\xi},\vec{\eta}\right) = K_{\rm \threebody} \, \Omega_{\rm 3\mhyphen body}(\vec{\xi},\vec{\eta})\, g_{\rm 3\mhyphen body}(\delta_t)\, e^{-\Gamma_{\rm 3\mhyphen body}t}\varepsilon_{\rm 3\mhyphen body}\left(t\right),
\end{equation}
where $K_{\rm \threebody}$ is a normalisation factor, and $\Omega_{\rm \threebody}(\vec{\xi},\vec{\eta})$ and $g_{\rm \threebody}(\delta_t)$ are the analogues of $\Omega_{\rm comb}(\vec{\xi},\vec{\eta})$ and $g_{\rm comb}(\delta_t)$ of Eq.~\eqref{eq:combinatorialTime}, respectively. The function $\varepsilon_{\rm \threebody}\left(t\right)$ is parameterised as in Eq.~\eqref{eq:bkgCombAcc}, with an independent parameter $a_{\rm \threebody}$, instead of $a_{\rm comb}$, left free to be adjusted by the fit.


\section{Flavour tagging}\label{sec:flavourTagging}

Flavour tagging is a fundamental ingredient to measure \CP asymmetries with \Bds-meson decays to \CP eigenstates. The sensitivity to the coefficients $C_{f}$ and $S_{f}$ governing the time-dependent \CP asymmetry defined in Eq.~\eqref{eq:acpTDDef} is directly related to the tagging power, defined as $\varepsilon_{\rm eff} = \sum_i{|\xi_i|\,(1-2\eta_i)^2}/N$, where $\xi_i$ and $\eta_i$ are the tagging decision and the associated mistag probability, respectively, for the $i$-th of the $N$ candidates.

Two classes of algorithms (OS and SS) are used to determine the initial flavour of the signal \Bds meson. The OS taggers~\cite{LHCb-PAPER-2011-027} exploit the fact that in \proton\proton collisions beauty quarks are almost exclusively produced in \bbbar pairs. Hence the flavour of the decaying signal \Bds meson can be determined by looking at the charge of the lepton, either muon or electron, originating from semileptonic decays, and of the kaon from the $\bquark\to\cquark\to\squark$ decay transition of the other \bquark hadron in the event. An additional OS tagger is based on the inclusive reconstruction of the opposite \bquark-hadron decay vertex and on the computation of a \pt-weighted average of the charges of all tracks associated to that vertex. 
For each OS tagger, the probability of misidentifying the flavour of the \Bds meson at production (mistag probability, $\eta$) is estimated by means of an artificial neural network, and is defined in the range $0 \leq \eta \leq 0.5$. When the response of more than one OS tagger is available per candidate, the different decisions and associated mistag probabilities are combined into a unique decision $\xi_{\rm OS}$ and a single $\eta_{\rm OS}$. The SS taggers are based on the identification of the particles produced in the hadronisation of the beauty quarks. In contrast to OS taggers, that to a very good approximation act equally on \Bd and \Bs mesons, SS taggers are specific to the nature of the \Bds meson under study. The additional \dquarkbar (\dquark) or \squarkbar (\squark) quarks produced in association with a \Bd (\Bdb) or a \Bs (\Bsb) meson, respectively, can form charged pions and protons, in the \dquark-quark case, or charged kaons, in the \squark-quark case. In this paper, so-called SS\pion and SS\proton taggers~\cite{LHCb-PAPER-2016-039} are used to determine the initial flavour of \Bd mesons, while the SS\kaon tagger~\cite{LHCb-PAPER-2015-056} is used for \Bs mesons.

The multivariate algorithms used to determine the values of $\eta_{\rm OS}$ and $\eta_{\rm SS}$ are trained using specific \B-meson decay channels and selections. The differences between the training samples and the selected signal \Bds mesons can lead to an imperfect determination of the mistag probability. Hence, a more accurate estimate, denoted as $\omega$ hereafter, is obtained by means of a calibration procedure that takes into account the specific kinematics of selected signal \Bds mesons. 
In the OS case, the relation between $\eta$ and $\omega$ is calibrated using \BdToKpi and \BsTopiK decays. In the SS\pion and SS\proton cases, only \BdToKpi decays are used. Once the calibration procedure is applied, the information provided by the two taggers is combined into a unique tagger, SSc, with decision $\xi_{\rm SSc}$ and mistag probability $\eta_{\rm SSc}$, as discussed in App.~\ref{sec:ssdCombination}. In the SS\kaon case, the small yield of the \BsTopiK decay is insufficient for a precise calibration. Hence, a large sample of \BsToDspi decays is used instead. The procedure is described in App.~\ref{sec:sskCalibration}.

Flavour-tagging information enters the PDF describing the decay-time distribution of the signals by means of the $\Omega_{\rm sig}(\vec{\xi},\,\vec{\eta})$ and $\bar{\Omega}_{\rm sig}(\vec{\xi},\,\vec{\eta})$ PDFs in Eqs.~\eqref{eq:decayTimeB2KPI} and~\eqref{eq:decayTimeB2HH}, and the same parameterisation is also adopted for the cross-feed backgrounds. Similar PDFs are used also for the combinatorial and three-body backgrounds. The full description of these PDFs is given in App.~\ref{sec:flavourTaggingAppendix}, together with the details and the results of the calibration procedure.


\section{Decay-time resolution}\label{sec:timeResolution}

\begin{figure}[t]
  \begin{center}
	\includegraphics[width=0.45\textwidth]{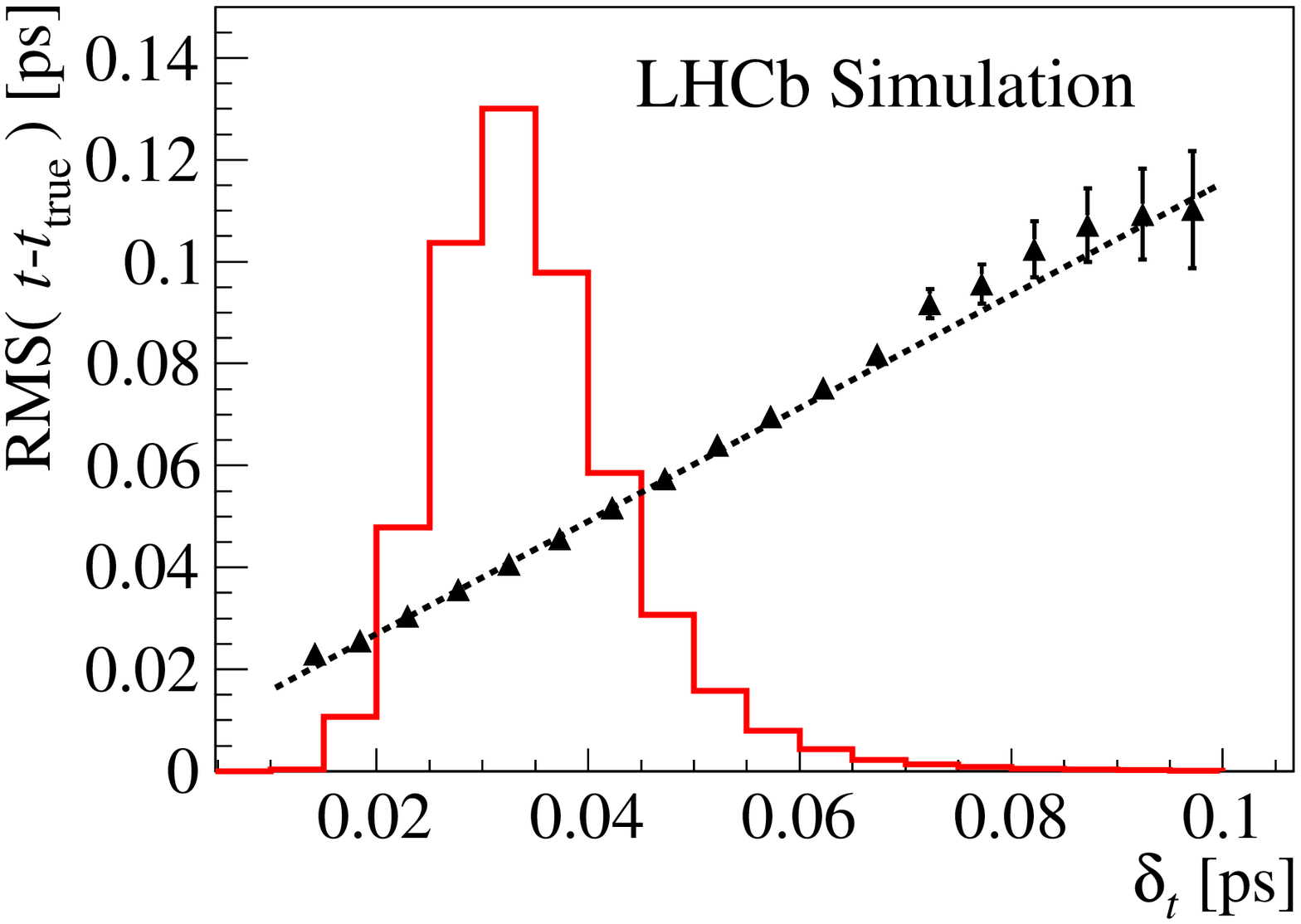}
  \includegraphics[width=0.45\textwidth]{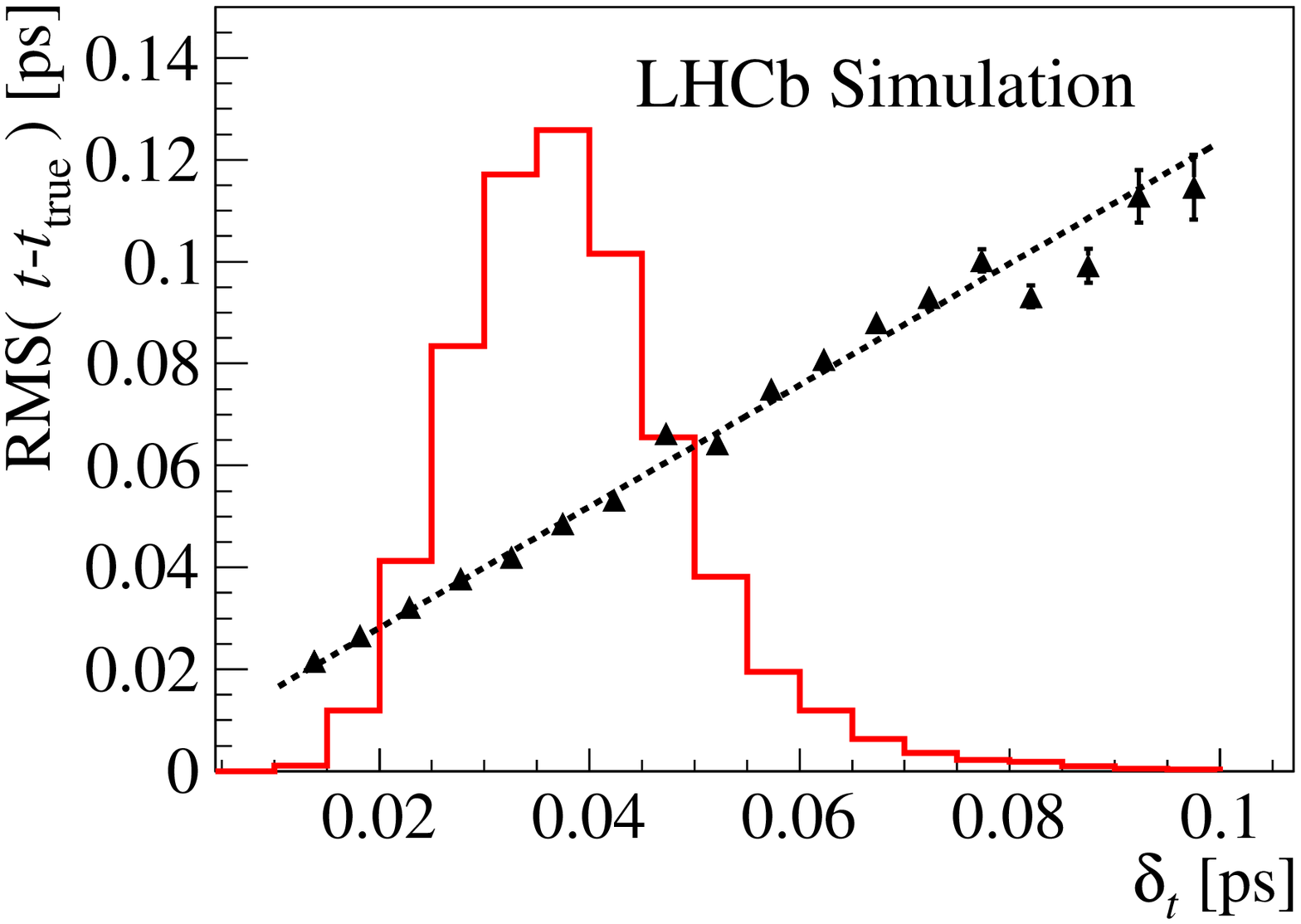}
 \end{center}
  \caption{\small The triangles represent the standard deviation of the difference between the reconstructed ($t$) and true decay ($t_{\rm true}$) time versus $\delta_t$ for simulated (left) \BsTopiK and (right) \decay{\Bs}{\Dsm\pip} decays. The dotted lines are the results of linear-function fits. The histograms represent the corresponding $\delta_t$ distributions with arbitrary normalisations.}
  \label{fig:resoDependence1PV}
\end{figure}

\begin{figure}[t]
  \begin{center}
	\includegraphics[width=0.45\textwidth]{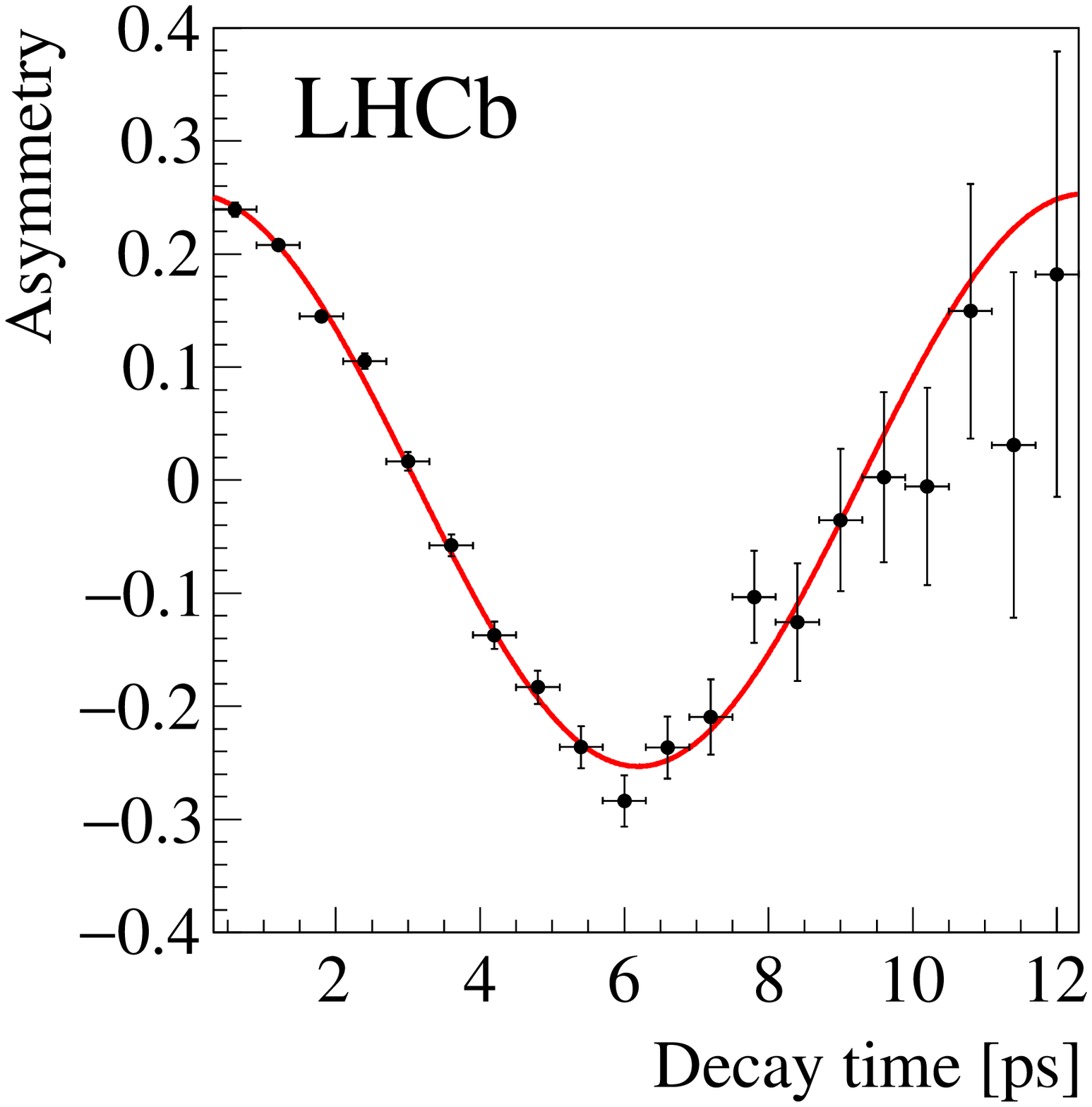}
	\includegraphics[width=0.45\textwidth]{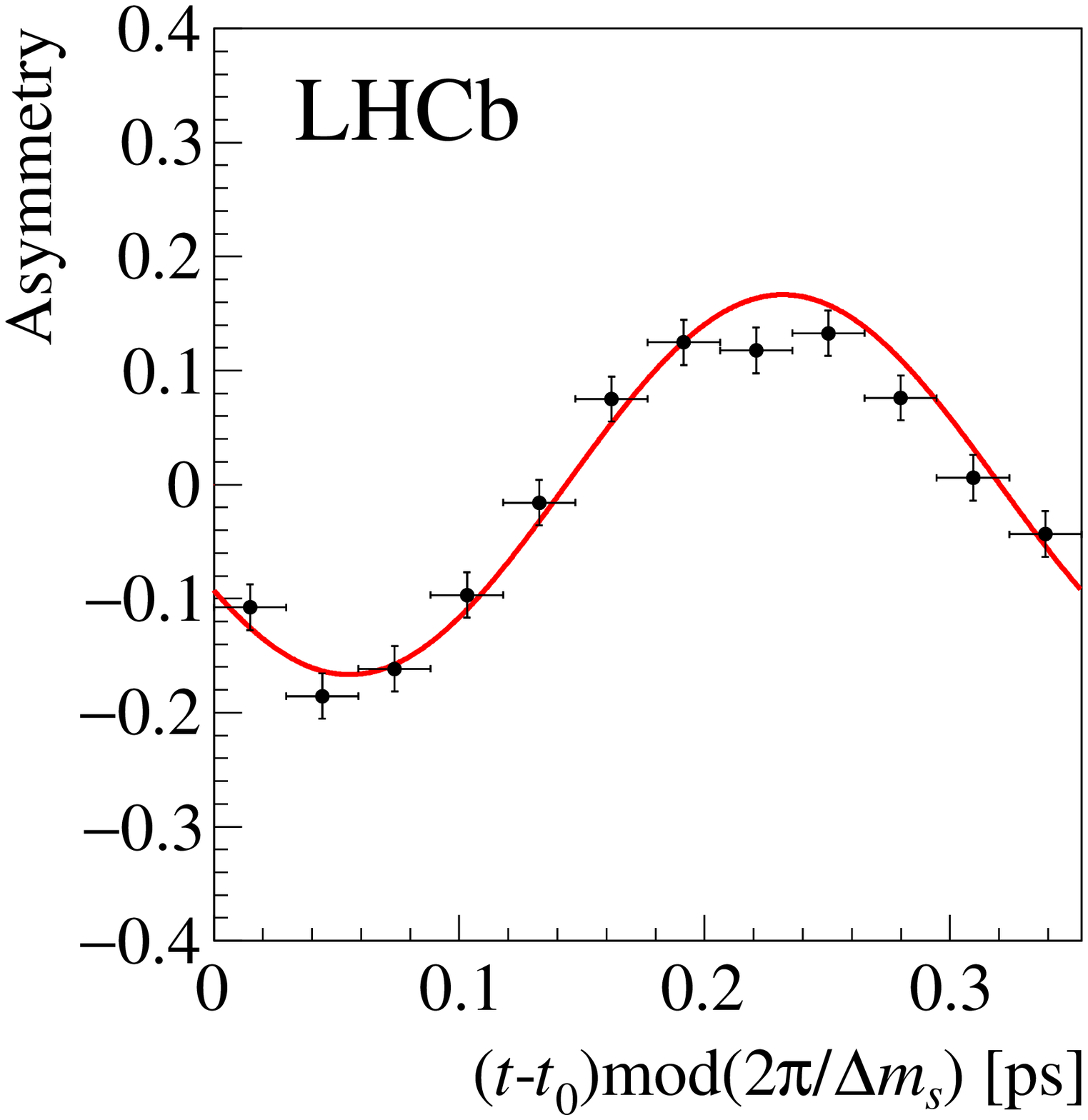}
 \end{center}
  \caption{\small Time-dependent asymmetries of (left) \decay{\Bd}{\Dm\pip} and (right) \decay{\Bs}{\Dsm\pip} decays obtained from data. The results of the best fits are superimposed. The time-dependent asymmetry of the \decay{\Bs}{\Dsm\pip} decays is folded into one mixing period $2\pi/\dms$ of the \Bs meson. The parameter $t_0 = 0.3\ps$ corresponds to the minimum value allowed by the selection.}
  \label{fig:resoFitB2DPi}
\end{figure}

The model to describe the decay-time resolution is obtained from the study of signal and \decay{\Bs}{\Dsm\pip} decays in simulation. It is found that the resolution function $R\left(t-t^{\prime}|\delta_t\right)$ is well described by the sum of two Gaussian functions with a shared mean fixed to zero and widths that depend on the decay-time uncertainty $\delta_t$, which varies on a candidate-by-candidate basis. The value of $\delta_t$ is determined for each \B candidate by combining the information of momentum, invariant mass, decay length and their corresponding uncertainties. The two widths are parameterised as
\begin{eqnarray}\label{eq:sigmaTimeDependency}
	\sigma_1(\delta_t) & = & q_0 + q_1\,(\delta_t-\hat{\delta}_t),\label{eq:sigmaTimeDependency1} \\
	\sigma_2(\delta_t) & = & r_{\sigma}\,\sigma_1(\delta_t),\nonumber
\end{eqnarray}
where $\hat{\delta_t} = 30\fs$ is approximately equal to the mean value of the $\delta_t$ distribution. It is also found that the parameters $q_0$, $q_1$, $r_{\sigma}$ and the relative fraction of the two Gaussian functions are very similar between signal and \decay{\Bs}{\Dsm\pip} decays. However, the simulation also shows the presence of a small component with long tails, that could be accommodated with a third Gaussian function with larger width. For simplicity the double Gaussian function is used in the baseline model, and a systematic uncertainty associated with this approximation is discussed in Sec.~\ref{sec:systematics}. Figure~\ref{fig:resoDependence1PV} shows the dependence on $\delta_t$ of the standard deviation of the difference between the reconstructed and true decay time for simulated \BsTopiK and \decay{\Bs}{\Dsm\pip} decays. This dependence is found to be well modelled by a straight line.
The parameter $r_{\sigma}$ and the relative contribution of the first Gaussian function are fixed to $3.0$ and $0.97$, respectively, as determined from full simulation. The values of the parameters $q_0$ and $q_1$ are determined from data by means of OS-tagged time-dependent fits to a sample of \decay{\Bs}{\Dsm\pip} decays, where the combined response of the OS taggers is calibrated using a sample of \decay{\Bd}{\Dm\pip} decays. Figure~\ref{fig:resoFitB2DPi} shows the time-dependent asymmetries of the \decay{\Bd}{\Dm\pip} and \decay{\Bs}{\Dsm\pip} decays, with the result of the fit superimposed.
The numerical results are $q_0 = 46.1 \pm 4.1\fs$ and $q_1 = 0.81 \pm 0.38$, with a correlation coefficient $\rho(q_0,q_1) = -0.32$. Residual small differences between signal and \decay{\Bs}{\Dsm\pip} decays, as seen in full simulation, are taken into account in the determination of the uncertainties on $q_0$ and $q_1$. If a simpler but less effective model based on a single Gaussian function with constant width were used, the value of such a width would have been approximately equal to 50\fs.

The distributions of $\delta_t$ for the signal components, $g_{\rm sig}\left(\delta_t\right)$, are modelled using background-subtracted histograms. For combinatorial and three-body backgrounds, they are described using histograms obtained by studying the high- and low-mass sidebands.


\section{Detection asymmetry between \boldmath{\Kp\pim} and \boldmath{\Km\pip} final states}\label{sec:asymmetrydet}

In this section the determination of the nuisance experimental detection asymmetry, needed to determine the \CP asymmetries \ACPBd~and \ACPBs, is described. This asymmetry arises because charge-conjugate final states are selected with different efficiencies. To excellent approximation, it can be expressed as the sum of two contributions
\begin{equation}\label{eq:rawAsymCorrection}
  A_{\rm F} = A^{\Km\pip}_{\rm D} + A^{\Km\pip}_{\rm PID},
\end{equation}
where $A^{\Km\pip}_{\rm D}$ is the asymmetry between the efficiencies of the \Km\pip and \Kp\pim final states without the application of the PID requirements and $A^{\Km\pip}_{\rm PID}$ is the asymmetry between the efficiencies of the PID requirements selecting the \Km\pip and \Kp\pim final states.

\subsection{Final-state detection asymmetry}\label{sec:recoAsym}

The final-state detection asymmetry is determined using $\Dp \to \Km \pip \pip$ and $\Dp \to \Kzb \pip$ control modes, with the neutral kaon decaying to \pip\pim, following the approach described in Ref.~\cite{LHCb-PAPER-2014-013}. Assuming negligible \CP violation in Cabibbo-favoured \D-meson decays, the asymmetries between the measured yields of \Dp and \Dm decays can be written as
\begin{eqnarray}\label{eq:rawAsymKpipi}
A_{\rm RAW}^{\Km\pip\pip} & = & A_{\rm P}^{\Dp}+A_{\rm D}^{\Km\pip}+A_{\rm D}^{\pip}, \label{eq:rawAsymKpipi1} \\
A_{\rm RAW}^{\Kzb\pion} & = & A_{\rm P}^{\Dp}+A_{\rm D}^{\pip}-A_{\rm D}^{\Kz},\label{eq:rawAsymKpipi2}
\end{eqnarray}
where $A_{\rm P}^{\Dp}$ is the asymmetry between the production cross-sections of \Dp and \Dm mesons, and $A^{\pip}_{\rm D}$ ($A^\Kz_{\rm D}$) is the asymmetry between the detection efficiencies of \pip (\Kz) and \pim (\Kzb) mesons. The difference between Eqs.~\eqref{eq:rawAsymKpipi1} and~\eqref{eq:rawAsymKpipi2} leads to
\begin{equation}\label{eq:adKPI}
A_{\rm D}^{\Km\pip} = A_{\rm RAW}^{\Km\pip\pip}-A_{\rm RAW}^{\Kzb\pip}-A_{\rm D}^{\Kz}.
\end{equation}
The asymmetry $A_{\rm D}^{\Kz}$ was determined to be $\left(0.054 \pm 0.014\right)\%$~\cite{LHCb-PAPER-2014-013}. 
The asymmetries $A_{\rm P}^{\Dp}$ and $A_{\rm D}^{\pion}$ could depend on the kinematics of the \Dp and \pip mesons. To achieve better cancellation of these nuisance asymmetries in Eq.~\eqref{eq:adKPI}, the momentum and \pt of the \Dp and \pip mesons from the $\Dp \to \Km \pip \pip$ sample are simultaneously weighted to match the corresponding distributions in the $\Dp \to \Kzb \pip$ sample. 
Because of the sizeable difference in the interaction cross-sections of positive and negative kaons with the detector material, $A_{\rm D}^{\Km\pip}$ is determined in bins of kaon momentum.
By taking into account the momentum distribution of the kaons from \BdToKpi and \BsTopiK decays, the values of $A_{\rm D}^{\Km\pip}$ for the two decay modes are found to be consistent, and the numerical result is
\begin{equation}
	A_{\rm D}^{\Km\pip}\left(\BdToKpi\right) = -A_{\rm D}^{\Km\pip}\left(\BsTopiK\right) = \left(-0.91 \pm 0.14\right)\%. \label{eq:adkpiBs2KPI}
\end{equation}
The different sign of the corrections for the \BdToKpi and \BsTopiK decays is a consequence of the opposite definition of the final states $f$ and $\bar{f}$ for the two modes.

\subsection{Asymmetry induced by PID requirements}\label{sec:pidAsym}

The PID asymmetry is determined using the calibration samples discussed in Sec.~\ref{sec:selection}. Using  \decay{\Dstarp}{\Dz(\Km\pip)\pip} decays, the asymmetry between the PID efficiencies of the \Kp\pim and \Km\pip final states is determined in bins of momentum, pseudorapidity and azimuthal angle of the two final-state particles. Several different binning schemes are used, and the average and standard deviation of the PID asymmetries determined in each scheme are used as central value and uncertainty for $A^{\Km\pip}_{\rm PID}$, respectively. The corrections for the two decays are found to be consistent, and the numerical result is
\begin{equation}
	A^{\Km\pip}_{\rm PID}\left(\BdToKpi\right) = -A^{\Km\pip}_{\rm PID}\left(\BsTopiK\right) = \left(-0.04 \pm 0.25\right)\%. \label{eq:apidB2KPI}
\end{equation}


\section{Fit results}\label{sec:fitResult}

\begin{figure}[tb]
  \begin{center}
    \includegraphics[width=0.45\textwidth]{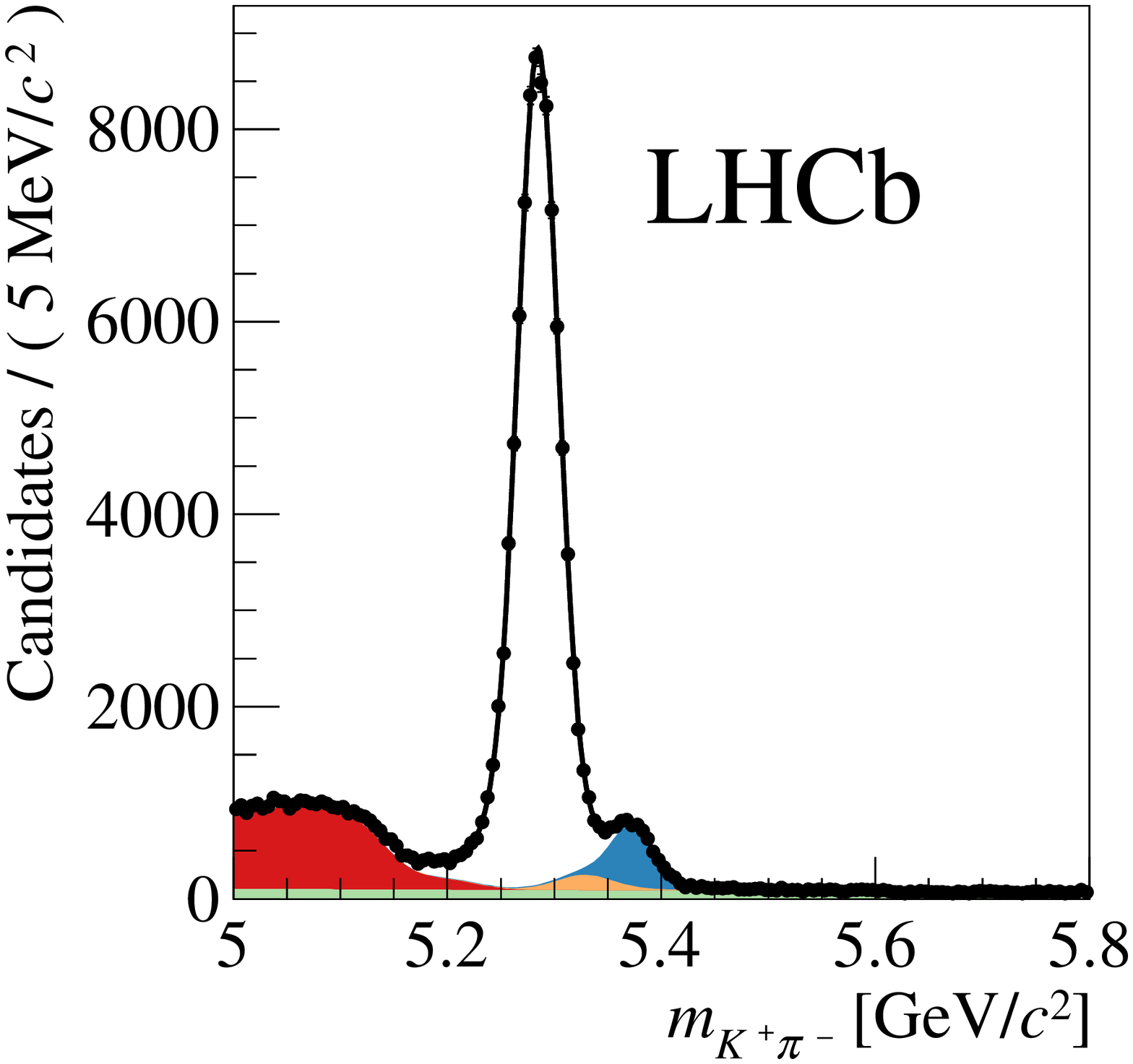}
    \includegraphics[width=0.45\textwidth]{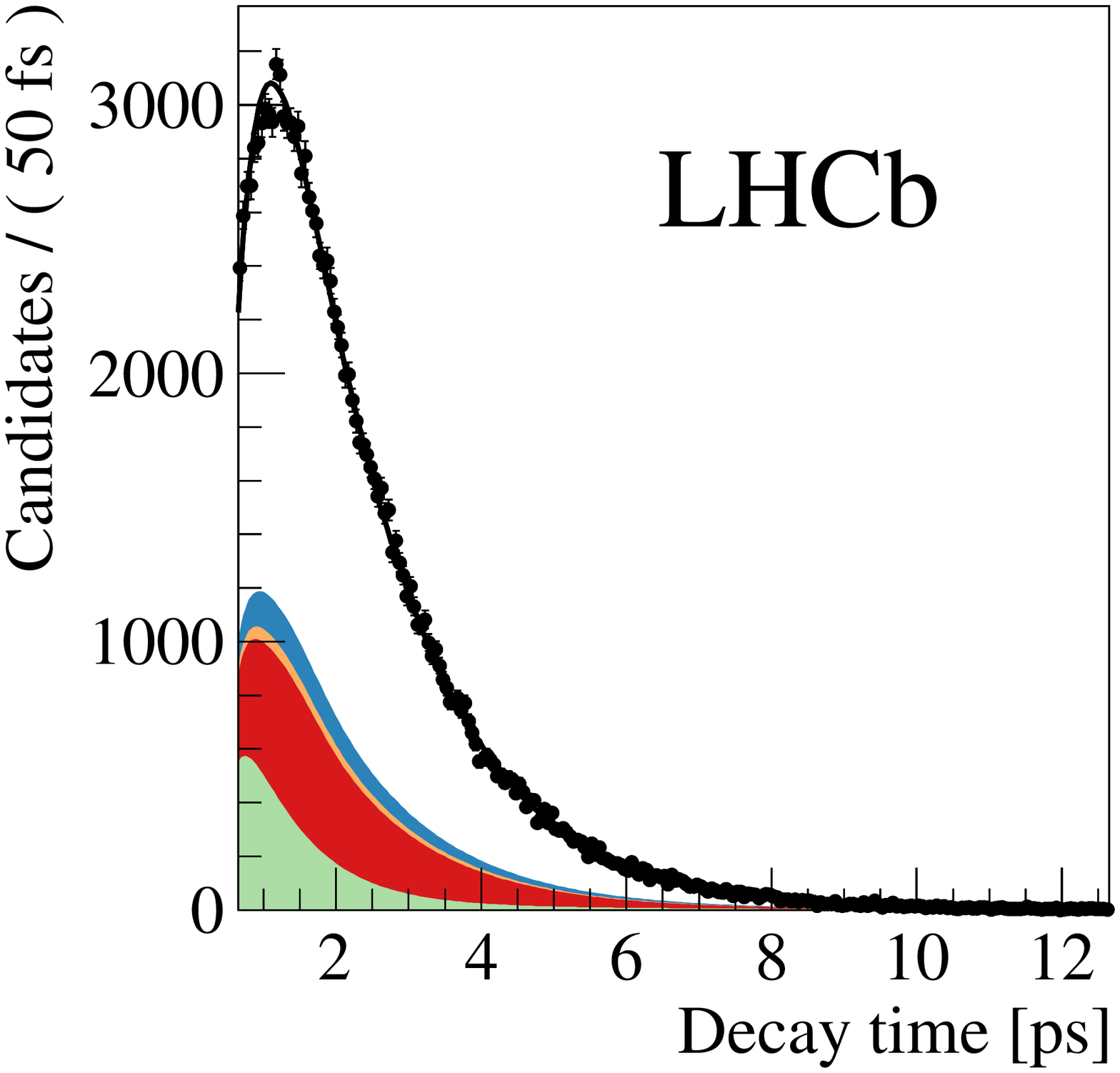}
    \includegraphics[width=0.45\textwidth]{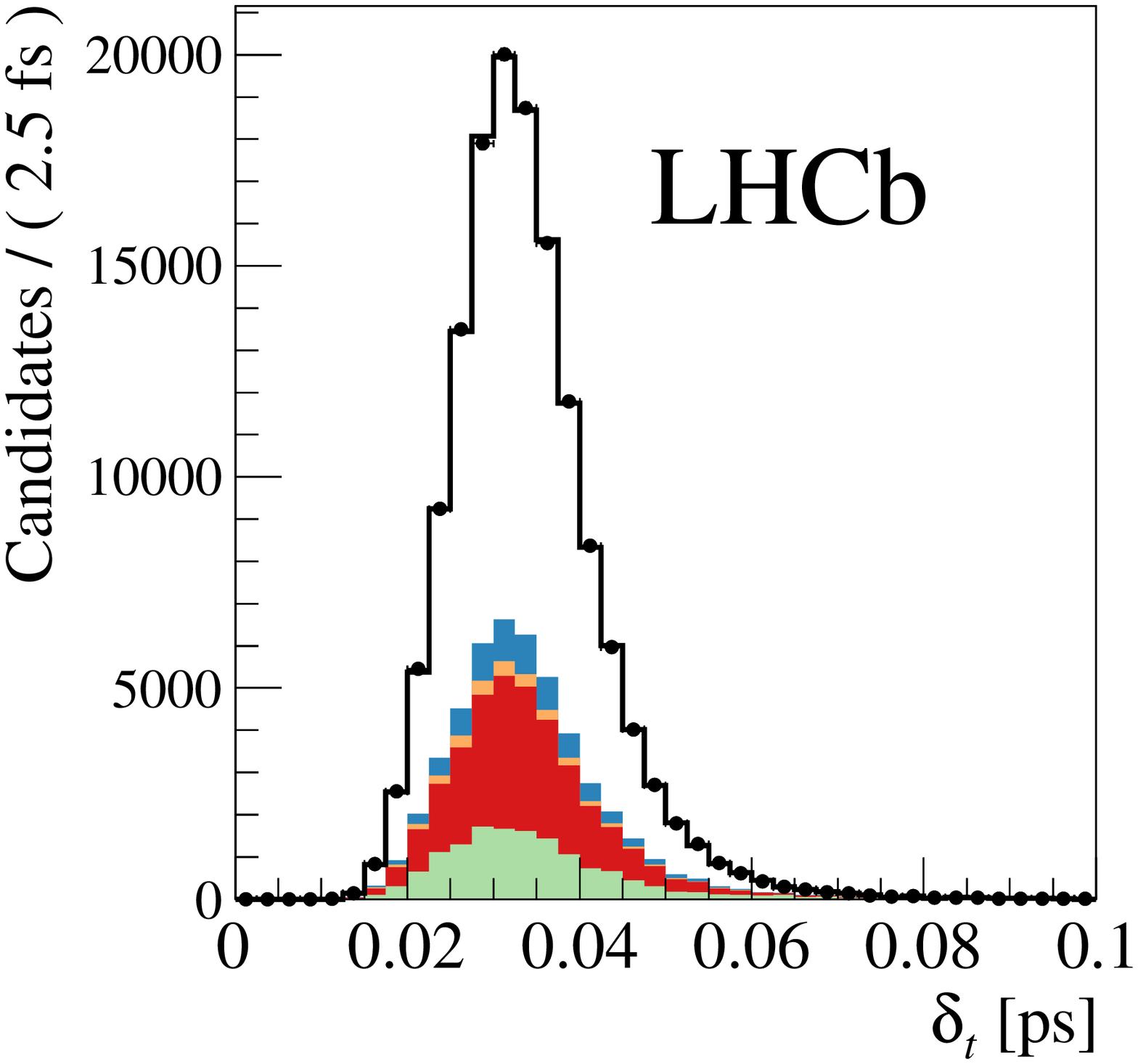}
    \includegraphics[width=0.45\textwidth]{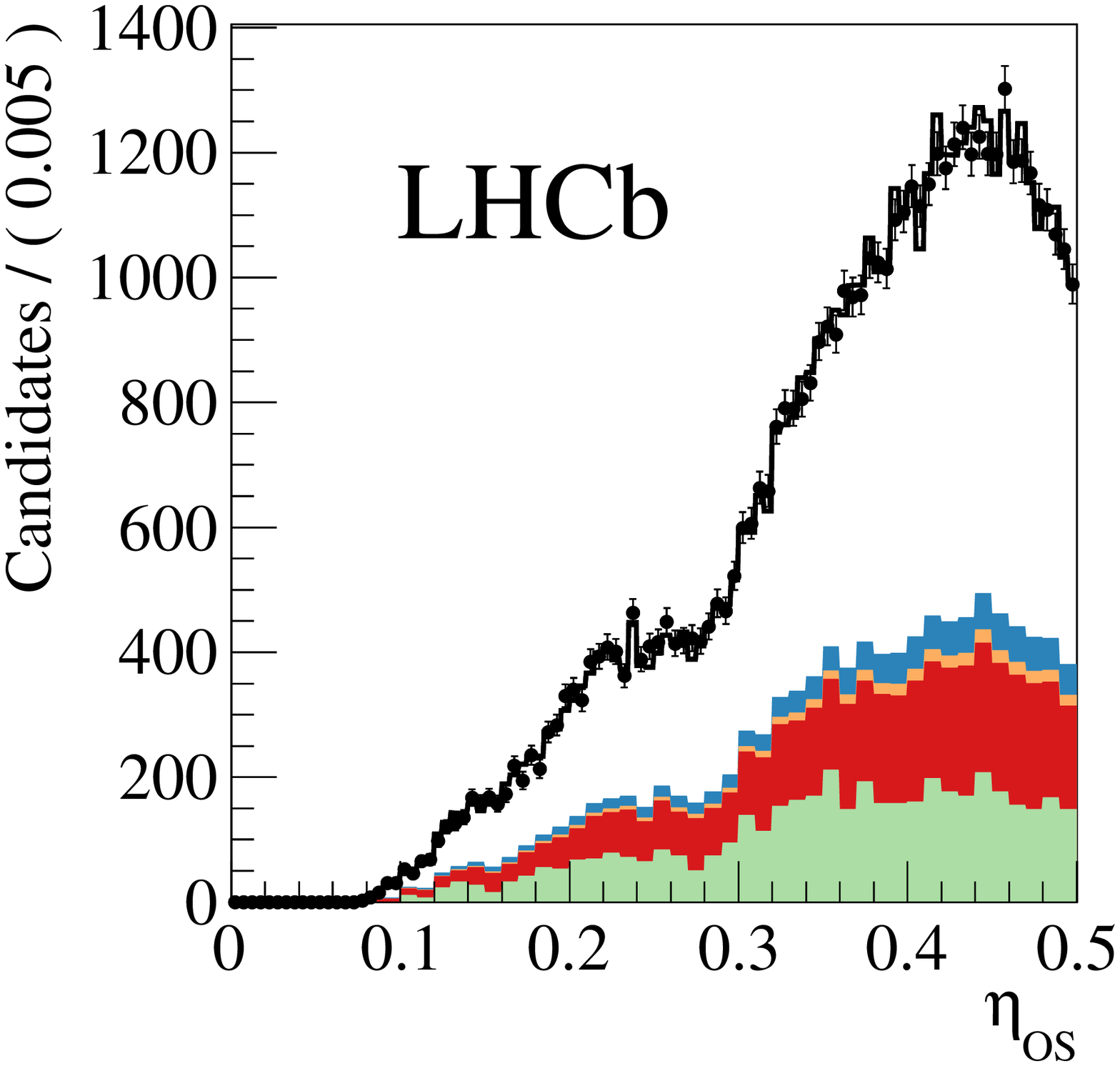}
    \includegraphics[width=0.45\textwidth]{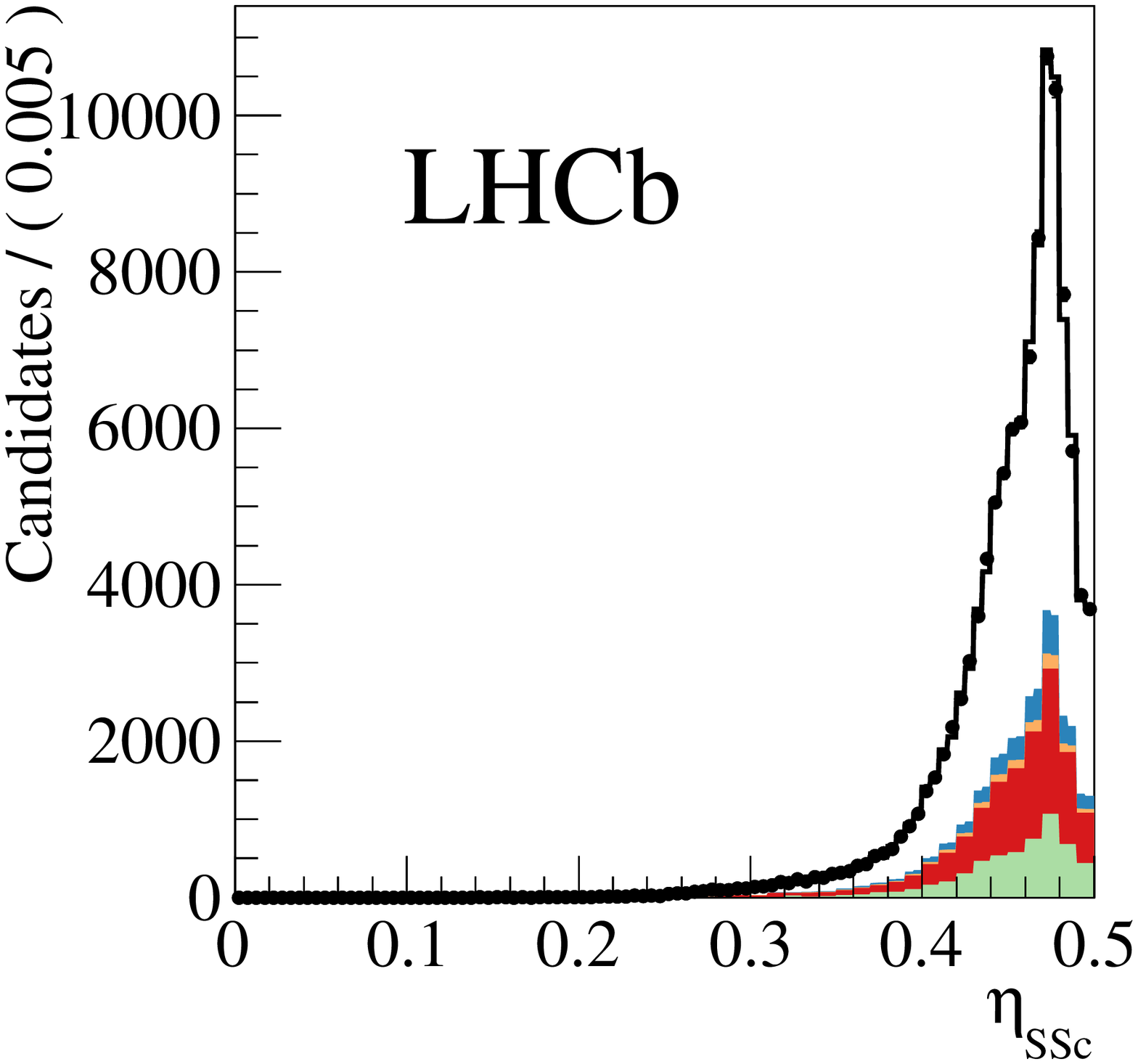}
    \includegraphics[width=0.45\textwidth]{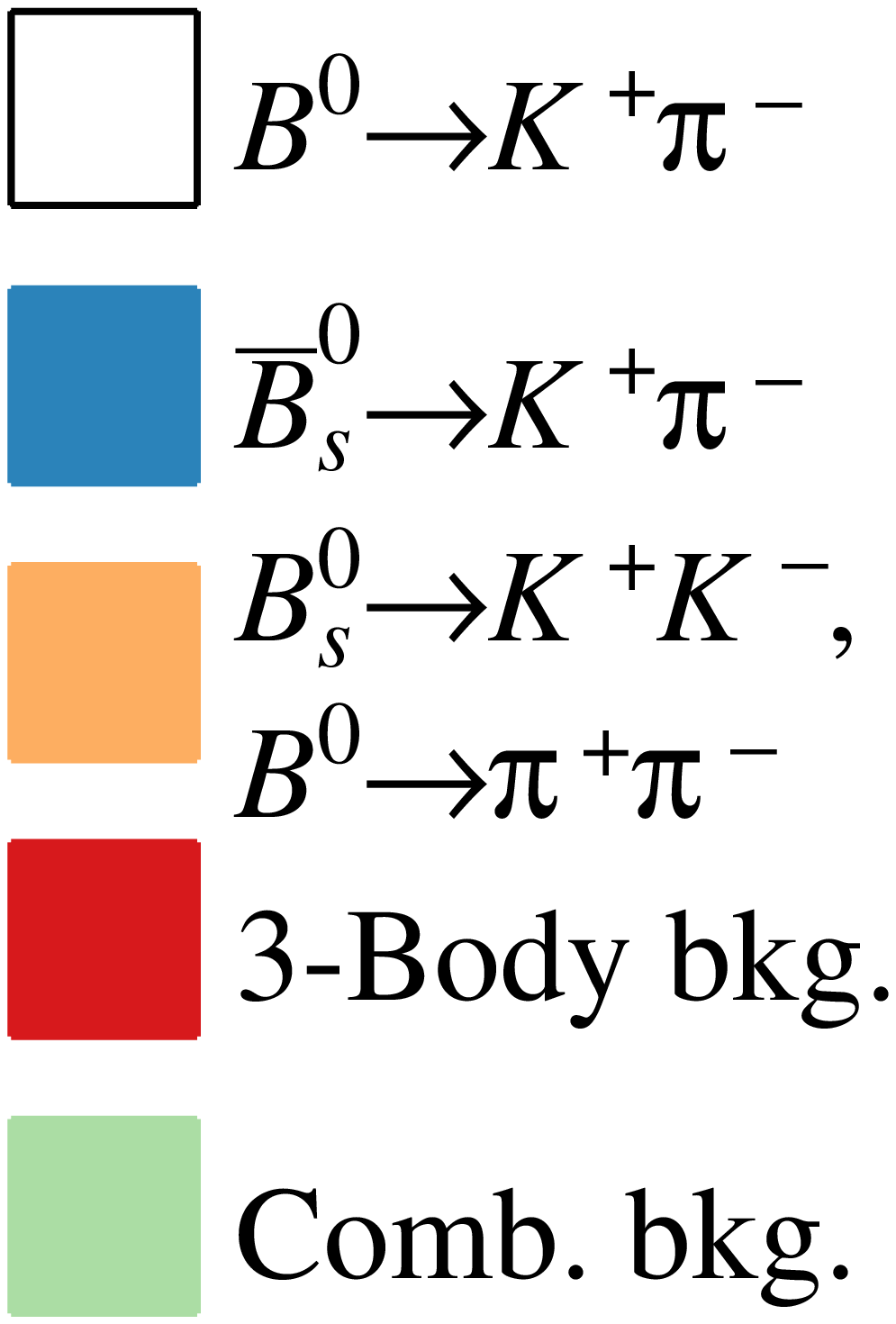}
 \end{center}
  \caption{\small Distributions of (top left) invariant mass, (top right) decay time, (middle left) decay-time uncertainty, (middle right) $\eta_{\rm OS}$, and (bottom) $\eta_{\rm SSc}$ for candidates in the \Kpm\pimp sample. The result of the simultaneous fit is overlaid. The individual components are also shown.}
  \label{fig:plotsKPI}
\end{figure}

\begin{figure}[tb]
  \begin{center}
    \includegraphics[width=0.45\textwidth]{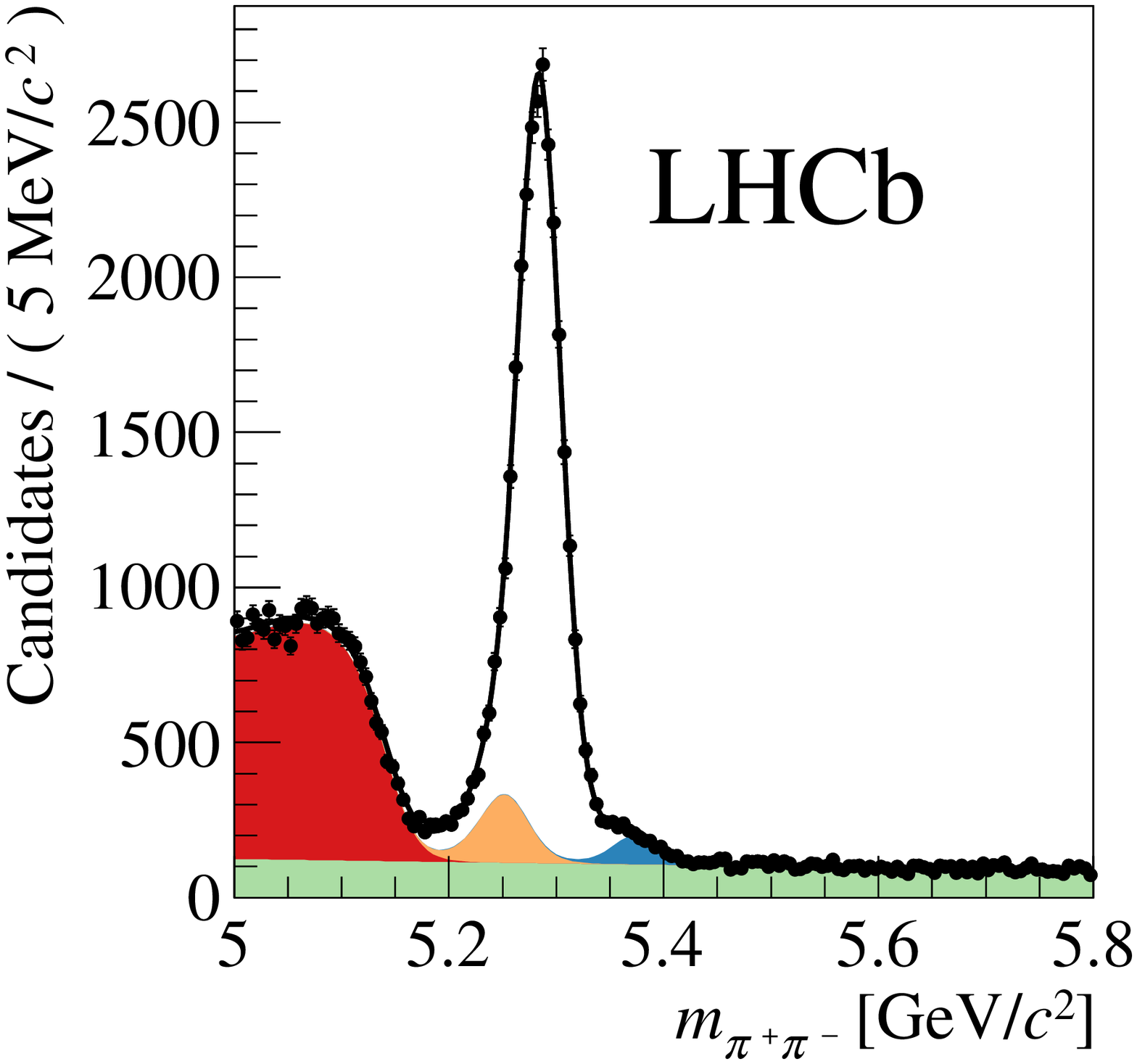}
    \includegraphics[width=0.45\textwidth]{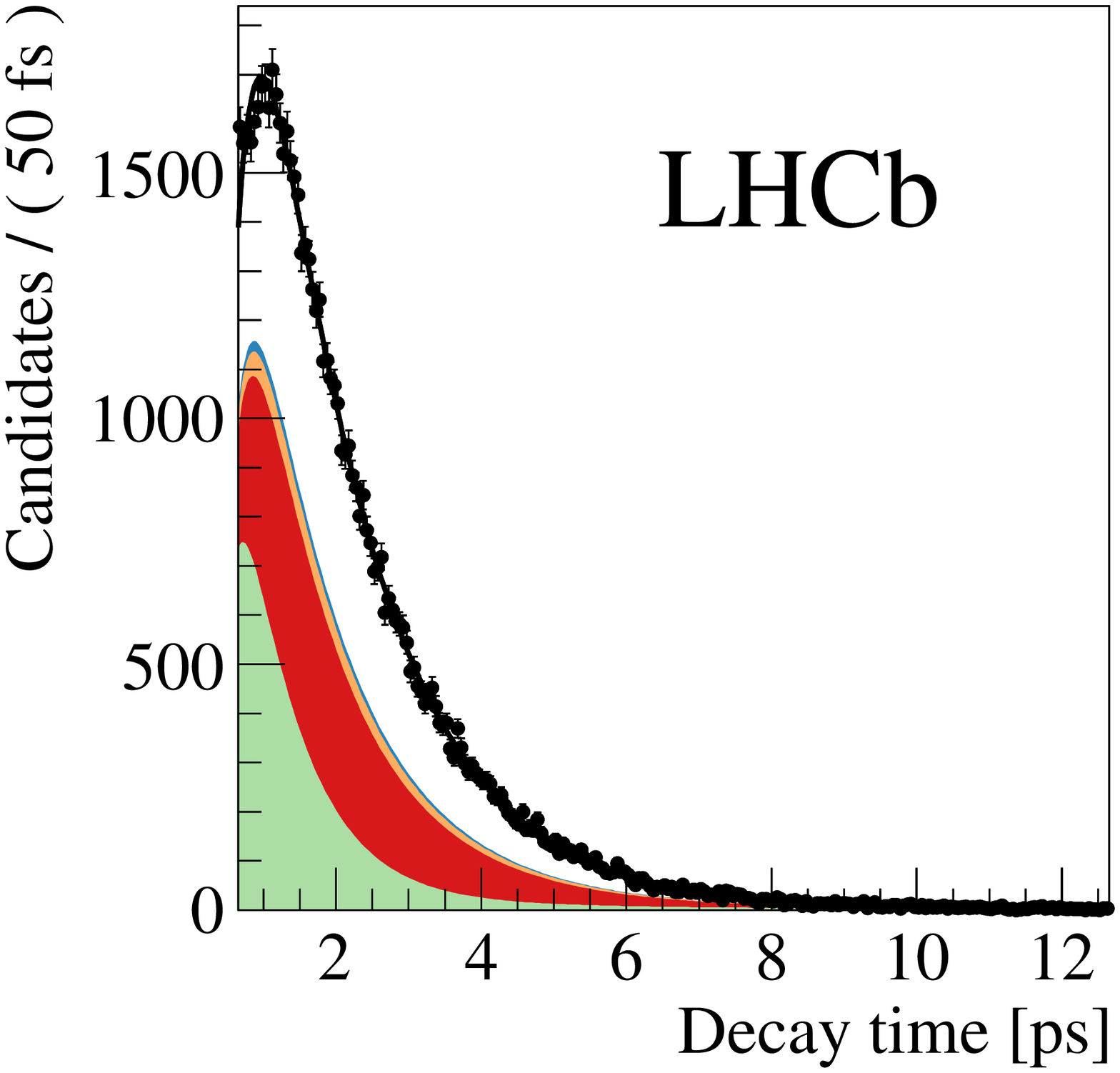}
    \includegraphics[width=0.45\textwidth]{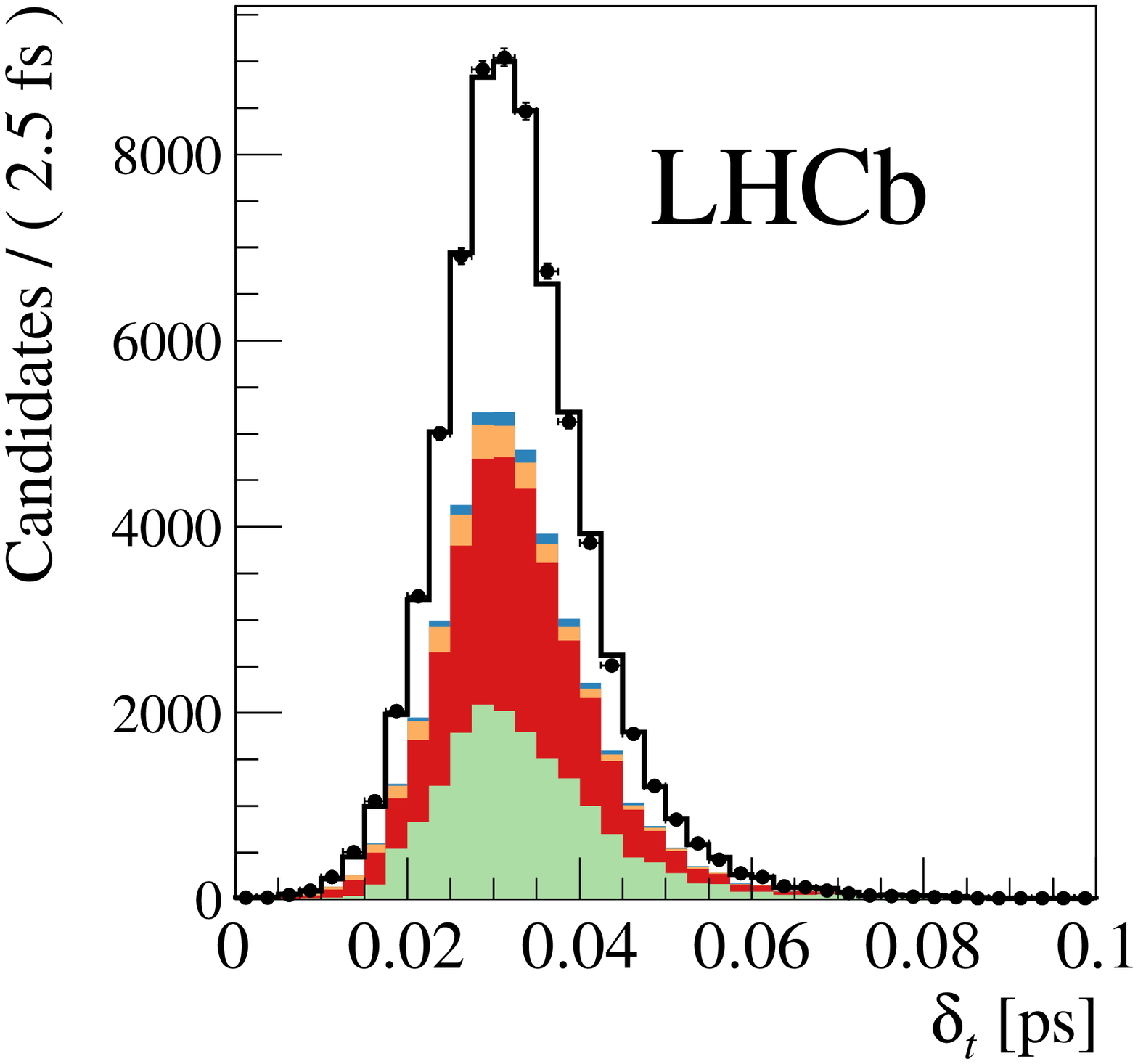}
    \includegraphics[width=0.45\textwidth]{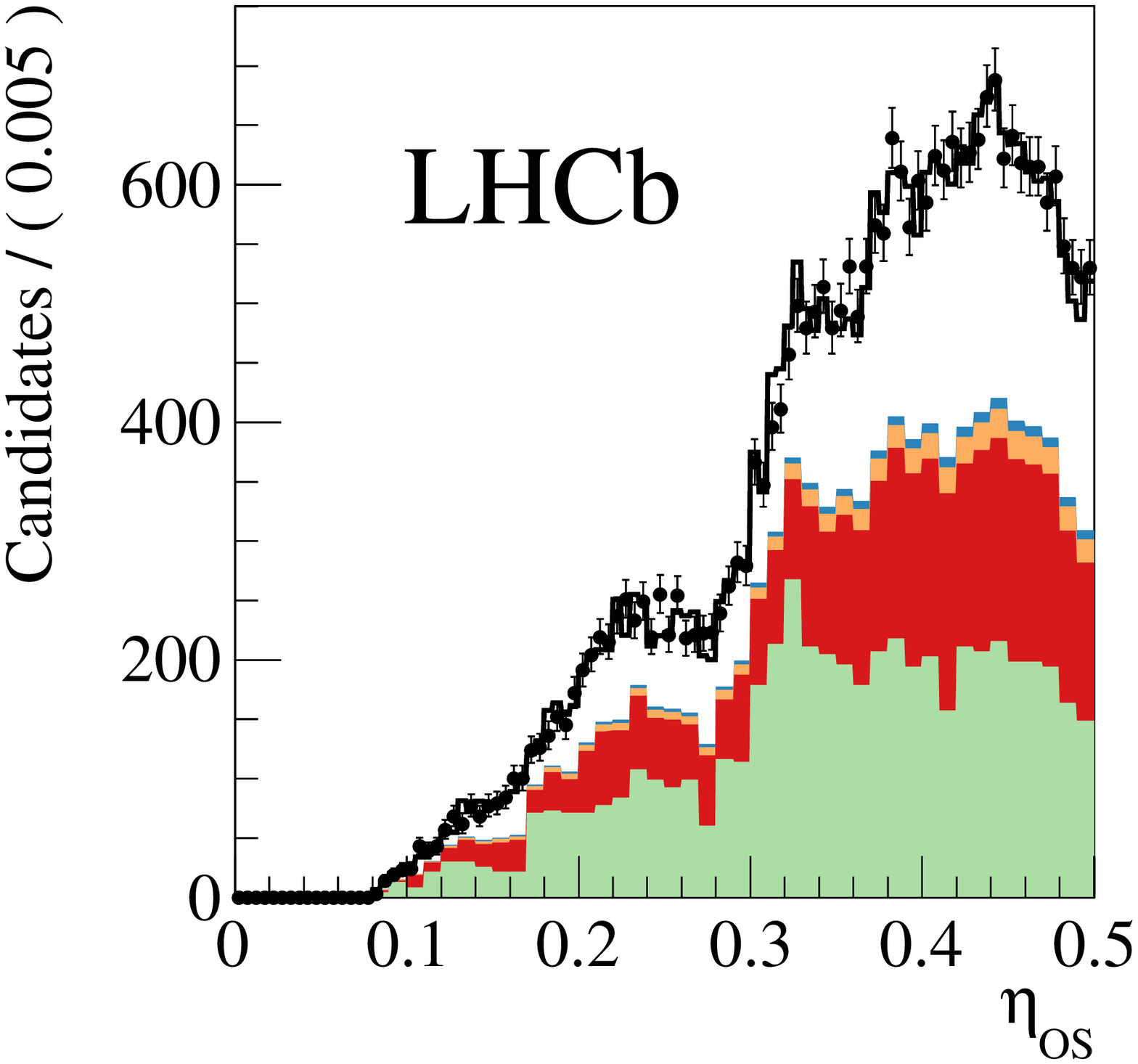}
    \includegraphics[width=0.45\textwidth]{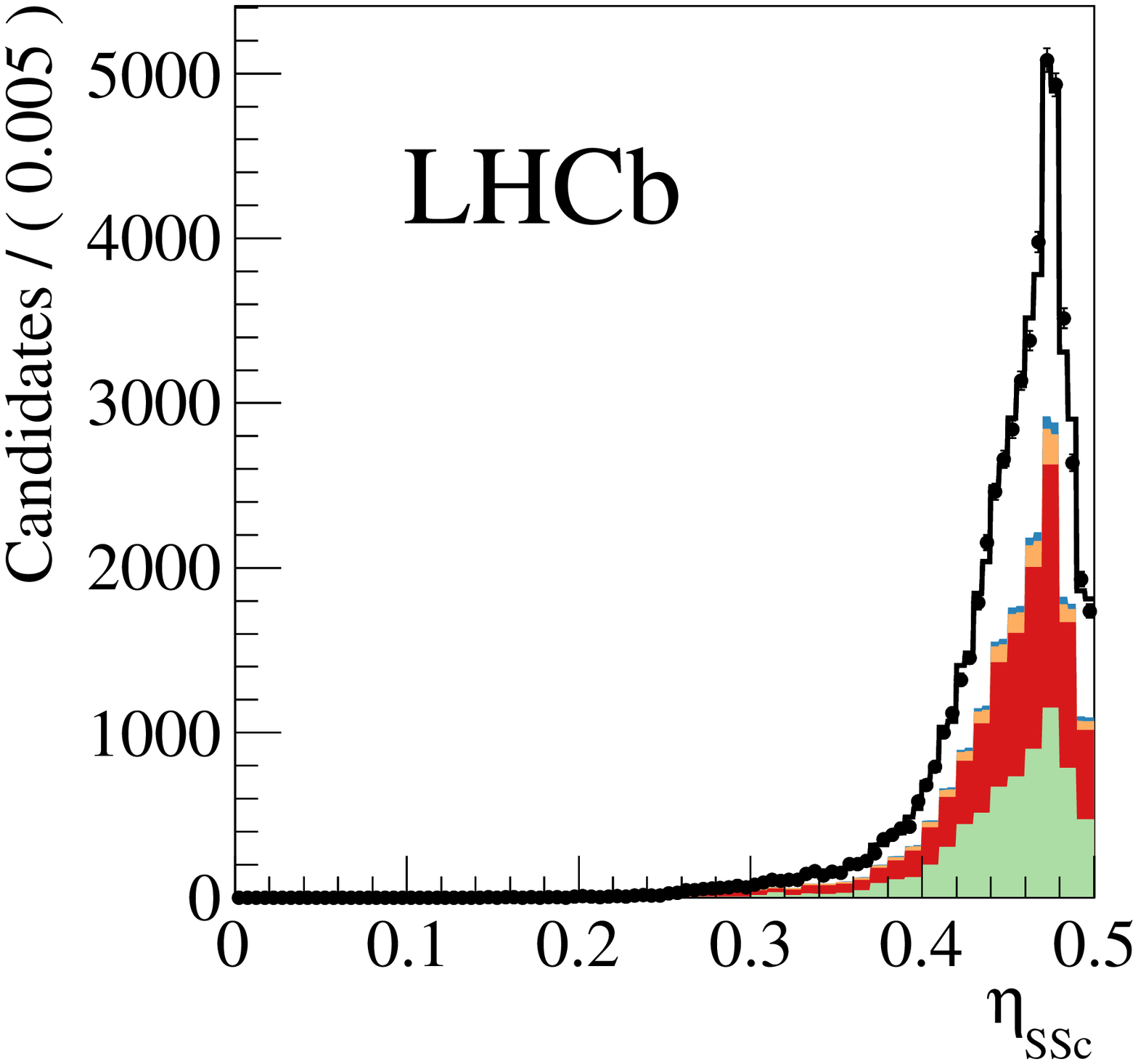}
    \includegraphics[width=0.45\textwidth]{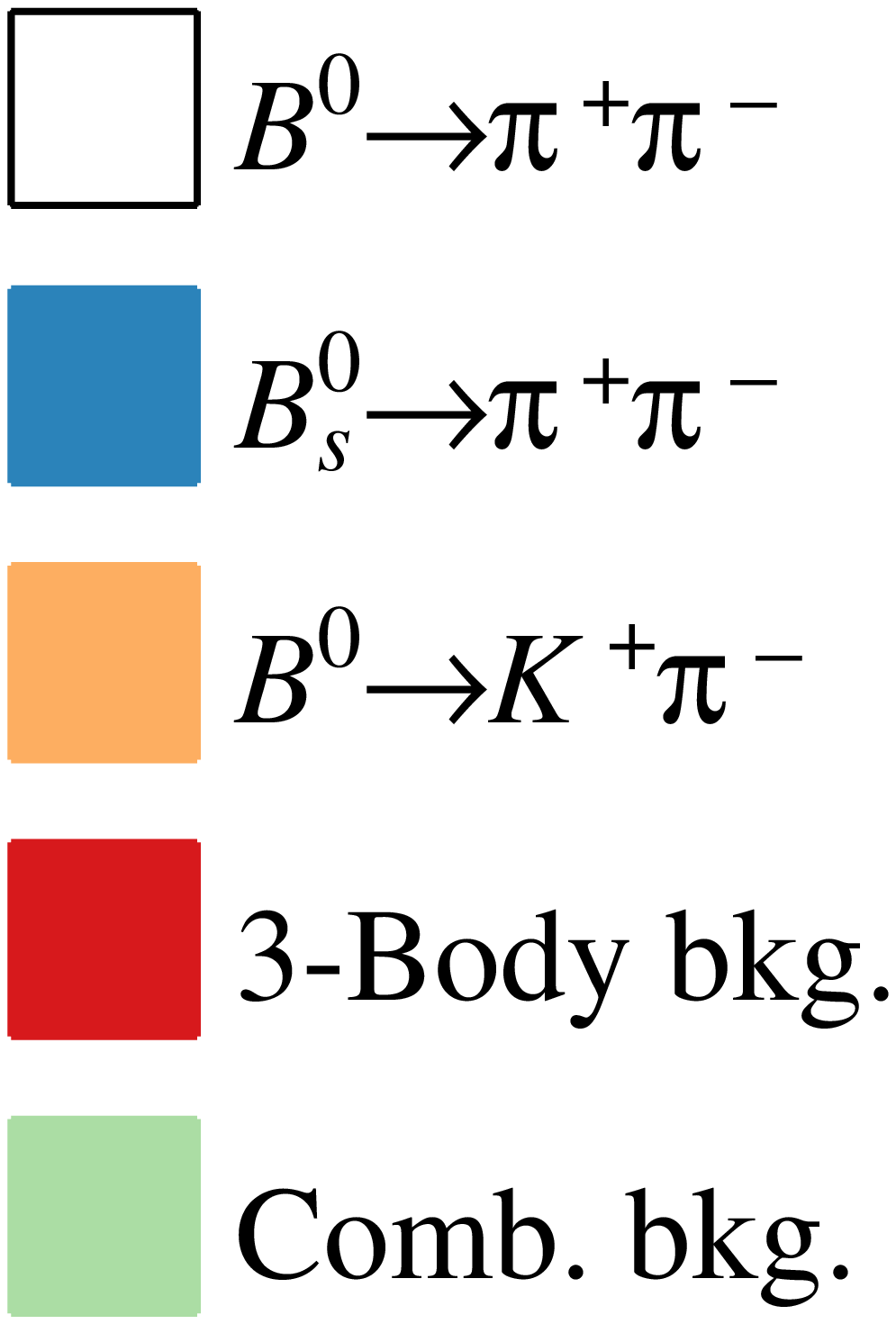}
 \end{center}
  \caption{\small Distributions of (top left) invariant mass, (top right) decay time, (middle left) decay-time uncertainty, (middle right) $\eta_{\rm OS}$, and (bottom) $\eta_{\rm SSc}$ for candidates in the \pip\pim sample. The result of the simultaneous fit is overlaid. The individual components are also shown.}
  \label{fig:plotsPIPI}
\end{figure}

\begin{figure}[tb]
  \begin{center}
    \includegraphics[width=0.45\textwidth]{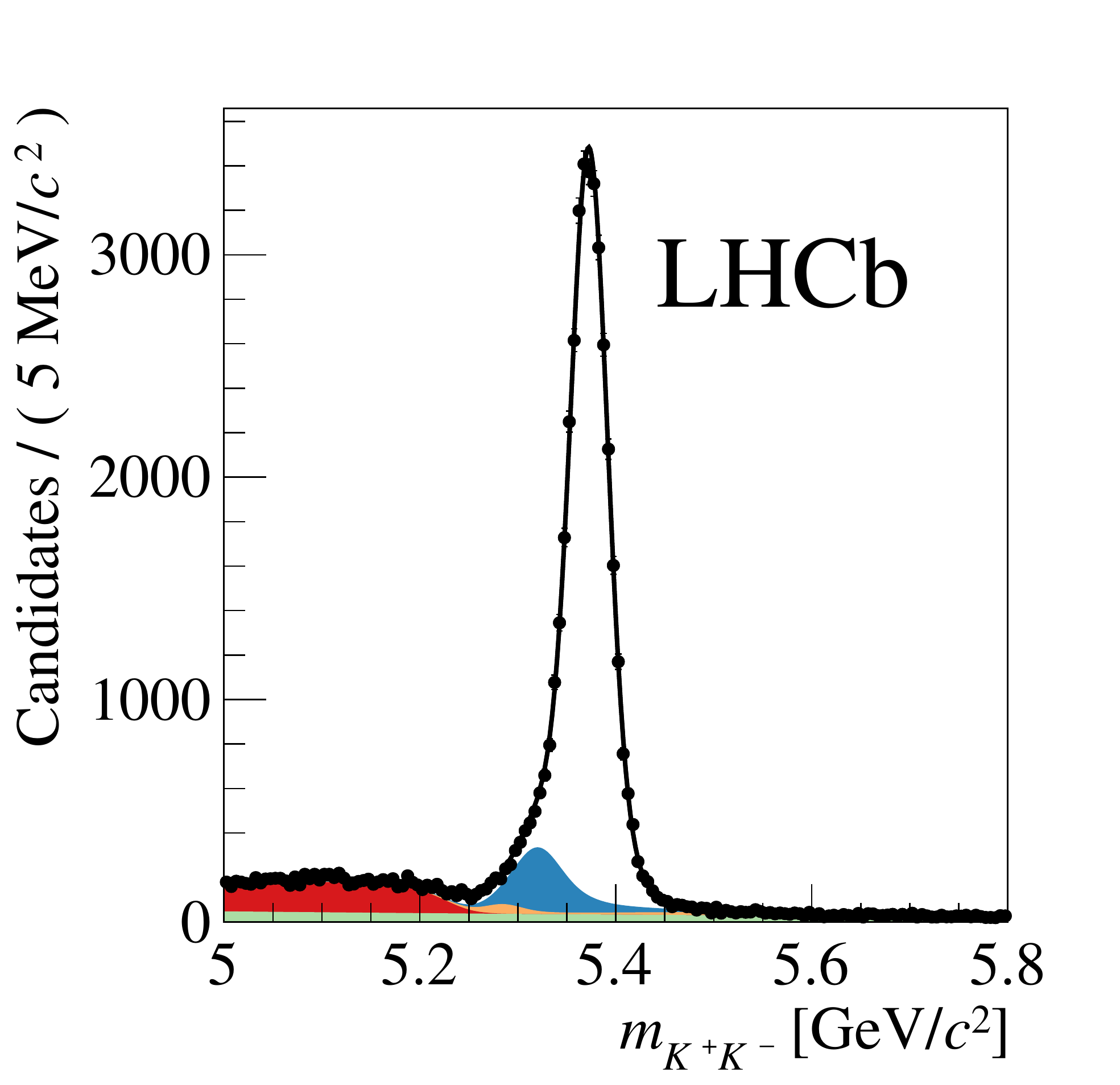}
    \includegraphics[width=0.45\textwidth]{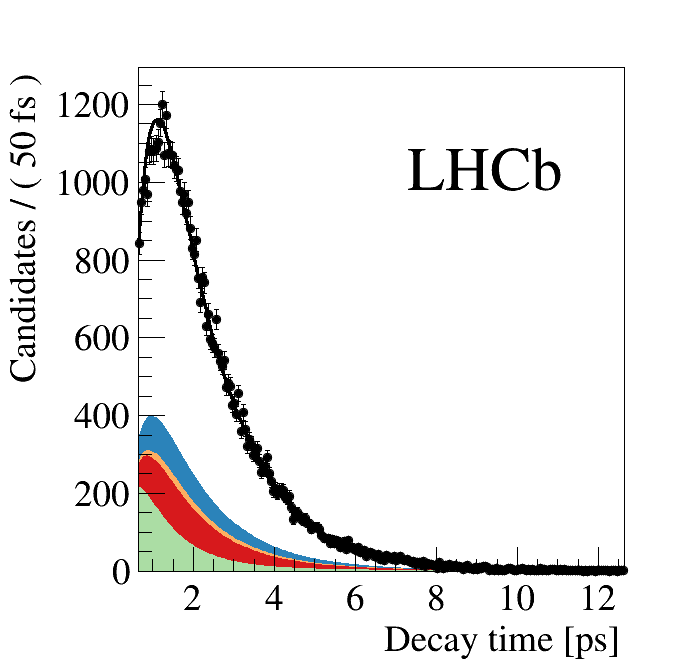}
    \includegraphics[width=0.45\textwidth]{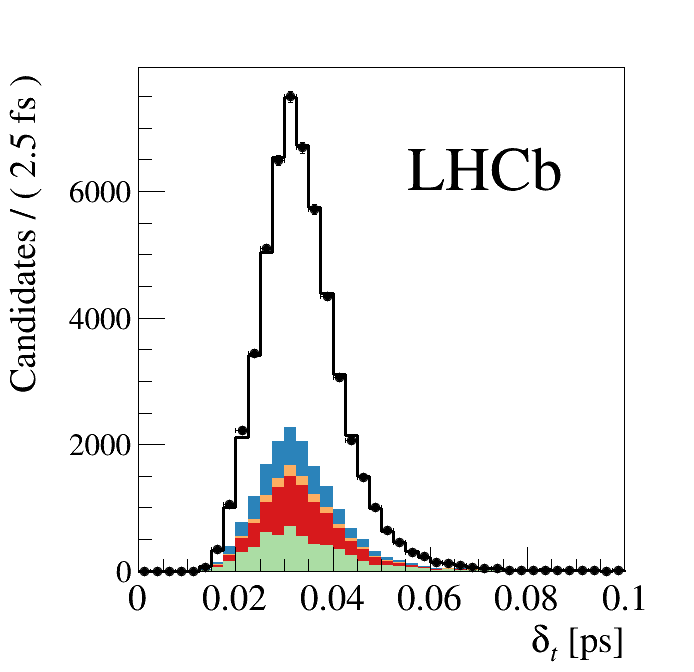}
    \includegraphics[width=0.45\textwidth]{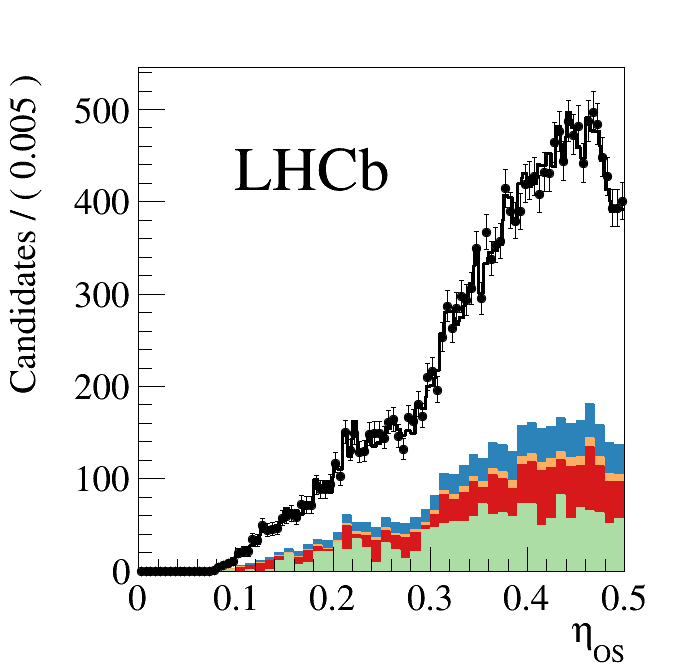}
    \includegraphics[width=0.45\textwidth]{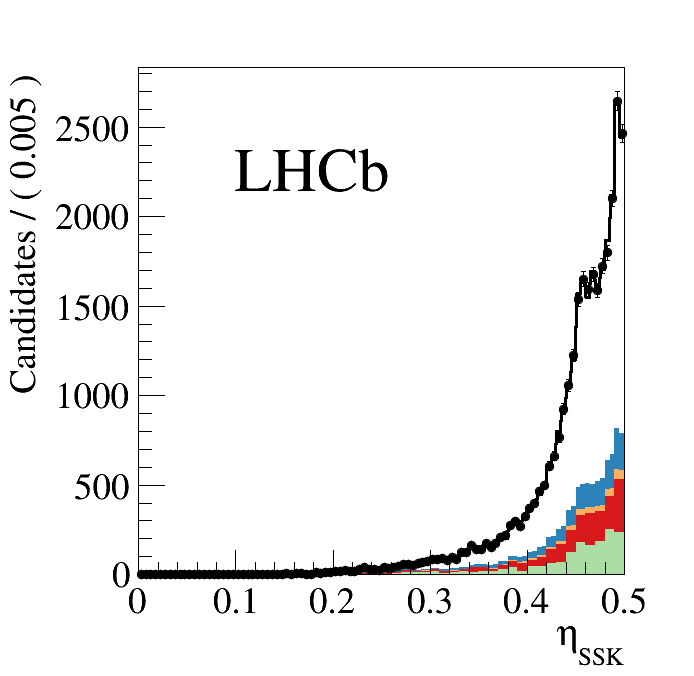}
    \includegraphics[width=0.45\textwidth]{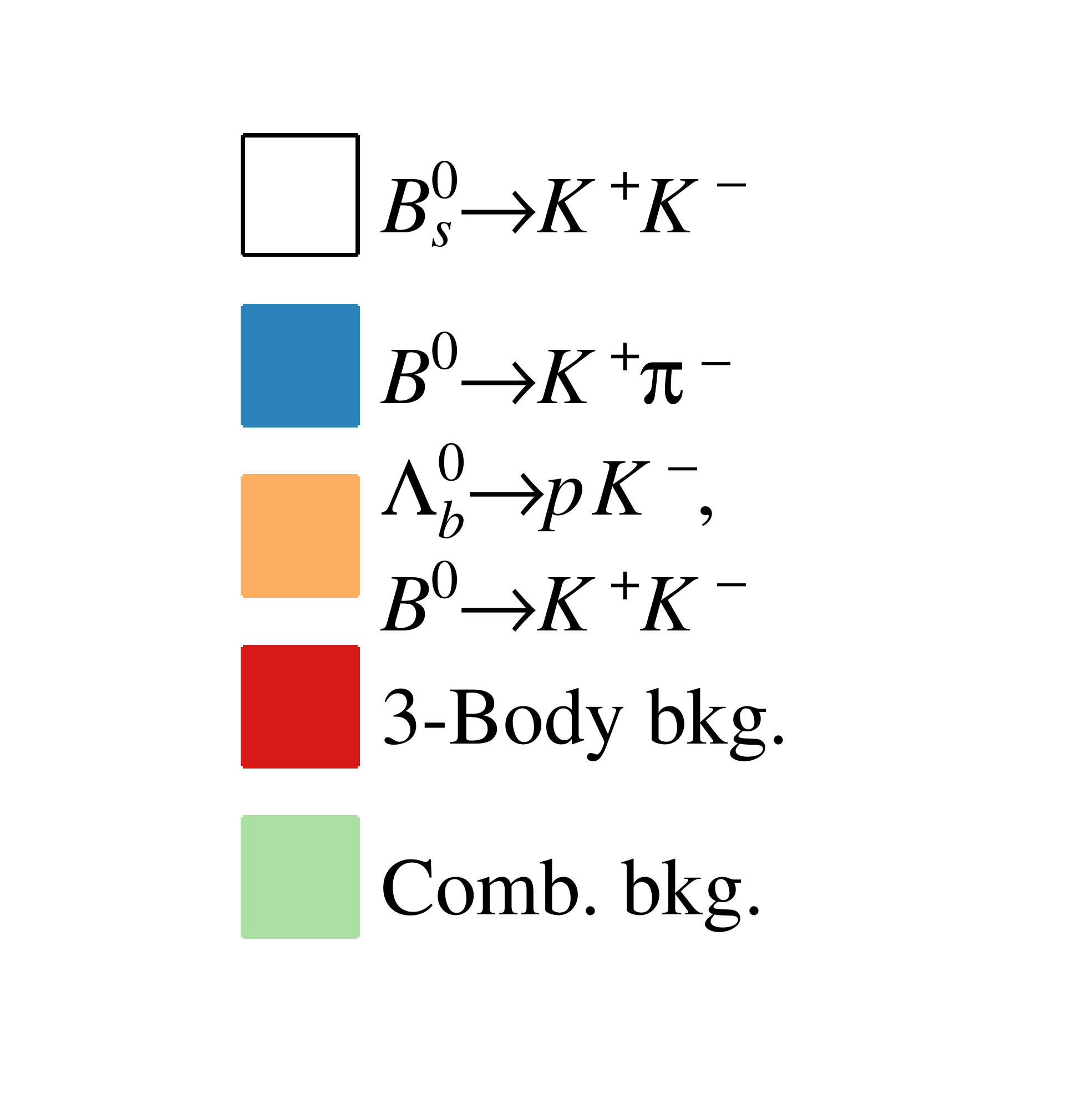}
 \end{center}
  \caption{\small Distributions of (top left) invariant mass, (top right) decay time, (middle left) decay-time uncertainty, (middle right) $\eta_{\rm OS}$, and (bottom) $\eta_{{\rm SS}\kaon}$ for candidates in the $\Kp\!\Km$ sample. The result of the simultaneous fit is overlaid. The individual components are also shown.}
  \label{fig:plotsKK}
\end{figure}

The simultaneous fit to the invariant mass, the decay time and its uncertainty, and the tagging decisions and their associated mistag probabilities for the \Kp\pim, \pip\pim and $\Kp\!\Km$ final states determines the coefficients \Cpipi, \Spipi, \CKK, \SKK, \ADGKK~and the \CP asymmetries \ACPBd~and \ACPBs. In the fits the parameters $\Delta m_{\dquark(\squark)}$, $\Gamma_{\dquark(\squark)}$, and $\Delta\Gamma_{\dquark(\squark)}$ are fixed to the central values reported in Table~\ref{tab:lifetimeParameters}. The signal yields are $N(\BdTopipi) = 28\hspace{0.5mm}650 \pm 230$, $N(\BsToKK) = 36\hspace{0.5mm}840 \pm 220$, $N(\BdToKpi) = 94\hspace{0.5mm}220 \pm 340$ and $N(\BsTopiK) = 7030 \pm 120$, where uncertainties are statistical only.
\begin{table}[t]
  \caption{ \small Values of the parameters \dmd, \dms, \Gd, \Gs and \DGs~\cite{HFLAV16}, fixed to their central values in the fit to the data. For \Gs and \DGs the correlation factor between the two quantities is also reported. The decay width difference \DGd is fixed to zero.}
  \begin{center}
    \begin{tabular}{l|c}
      Parameter        & Value \\
      \hline
      \dmd             & $0.5065 \pm 0.0019\invps$ \\
      \Gd              & $0.6579 \pm 0.0017\invps$ \\
      \DGd             & $\phantom{0.0000 }0\phantom{ 0.0000\invps}$ \\
      \dms             & $17.757 \pm 0.021\phantom{0}\invps$ \\
      \Gs              & $0.6654 \pm 0.0022\invps$ \\
      \DGs             & $\phantom{0}0.083 \pm 0.007\phantom{0}\invps$ \\
      $\rho(\Gs,\DGs)$ & $\phantom{0.00}-0.292\phantom{ 0.0000\invps}$ \\
    \end{tabular}
  \end{center}
  \label{tab:lifetimeParameters}
\end{table}
The one-dimensional distributions of the measured variables used in the fit, with the results of the fit overlaid, are shown in Figs.~\ref{fig:plotsKPI}, \ref{fig:plotsPIPI} and~\ref{fig:plotsKK}. 
\begin{table}[t]
  \caption{\small Tagging powers for the \BdTopipi and \BsToKK decays (last two rows), with a breakdown of the OS and SS contributions.}
  \begin{center}\begin{tabular}{l|c}
      Flavour tagger  & Tagging power (\%)\\
      \hline
      OS              & $2.94 \pm 0.17$ \\
      \hline
      SS\pion         & $0.81 \pm 0.13$ \\
      SS\proton       & $0.42 \pm 0.17$ \\
      SSc             & $1.17 \pm 0.11$ \\
      \hline
      SS\kaon         & $0.71 \pm 0.12$ \\
      \hline
      Total \BdTopipi & $4.08 \pm 0.20$ \\
      Total \BsToKK   & $3.65 \pm 0.21$ \\
    \end{tabular}
  \end{center}
  \label{tab:taggingPowerSummary}
\end{table}

\begin{figure}[tb]
  \begin{center}
	\includegraphics[width=0.38\textwidth]{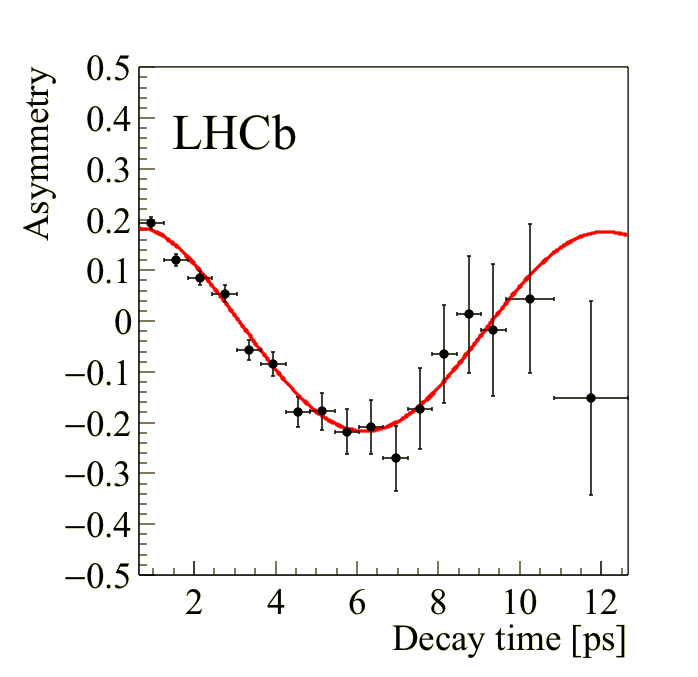}
	\includegraphics[width=0.38\textwidth]{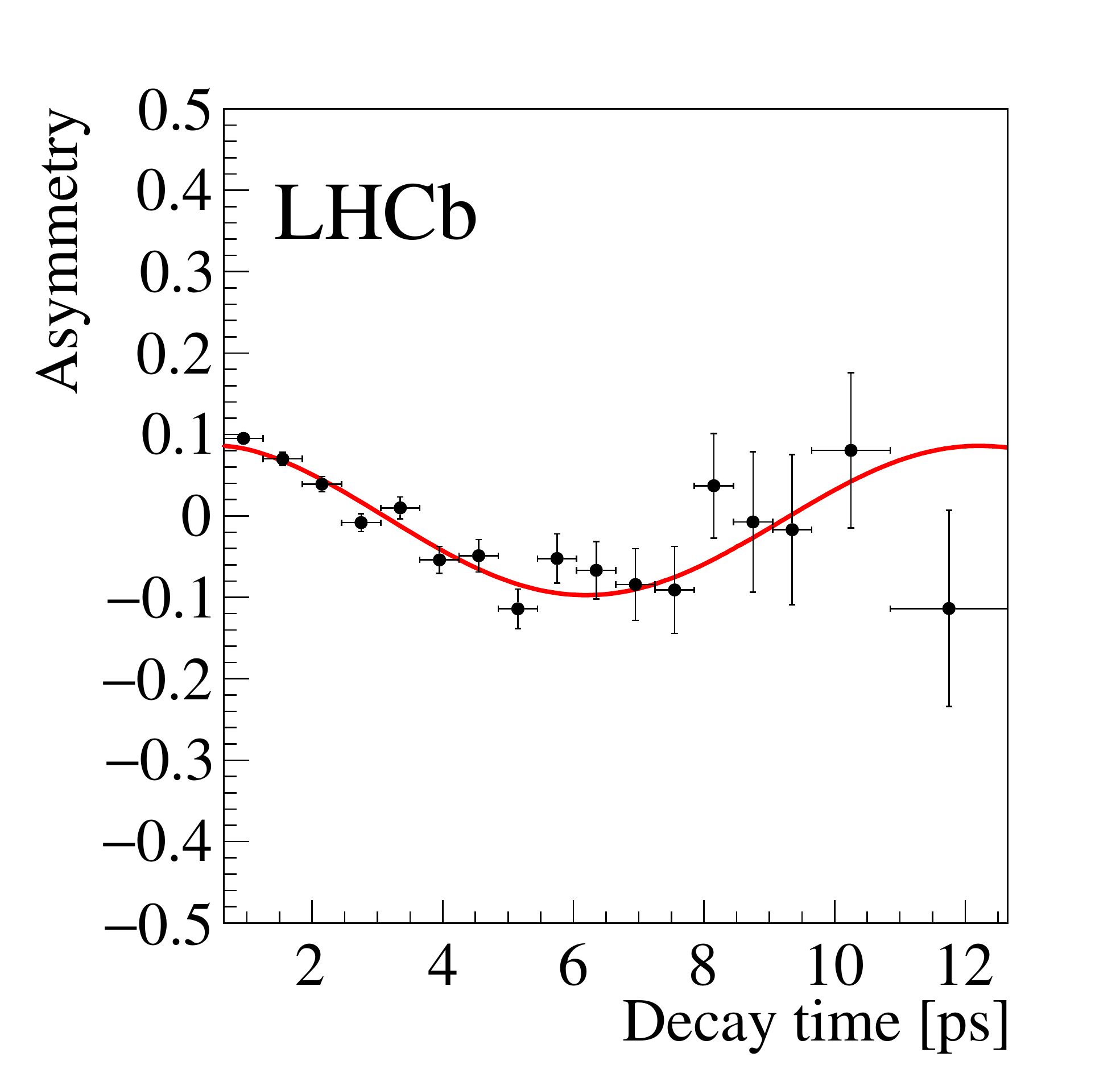}
 \end{center}
  \caption{\small Time-dependent asymmetries for \Kpm\pimp candidates with invariant-mass values in the interval $5.20 < m < 5.32\gevcc$: (left) using the OS-tagging decision and (right) the SS-tagging decision. The result of the simultaneous fit is overlaid.}
  \label{fig:rawAsymmetryKPI}
\end{figure}

\begin{figure}[tb]
  \begin{center}
	\includegraphics[width=0.38\textwidth]{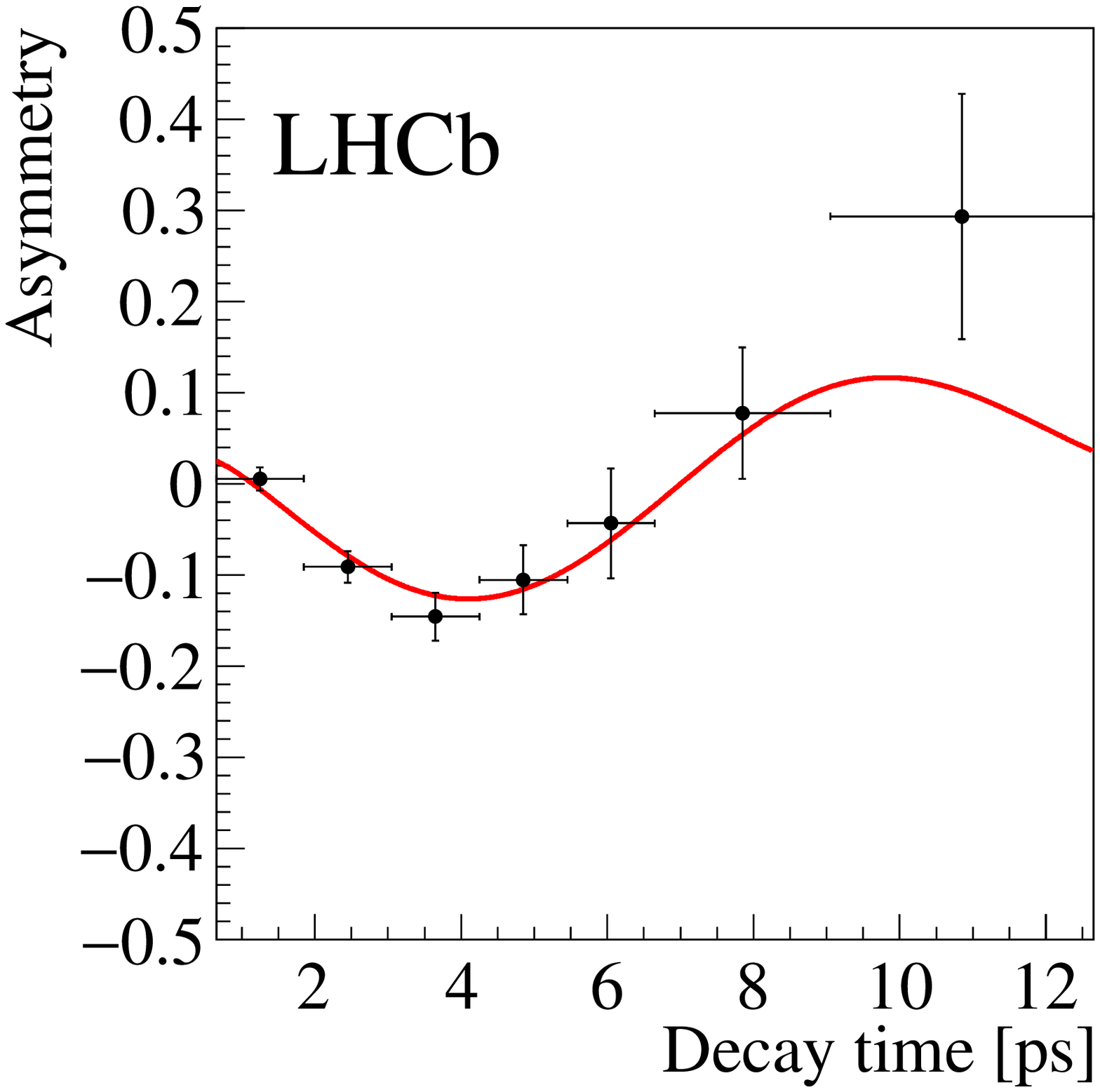}
	\includegraphics[width=0.38\textwidth]{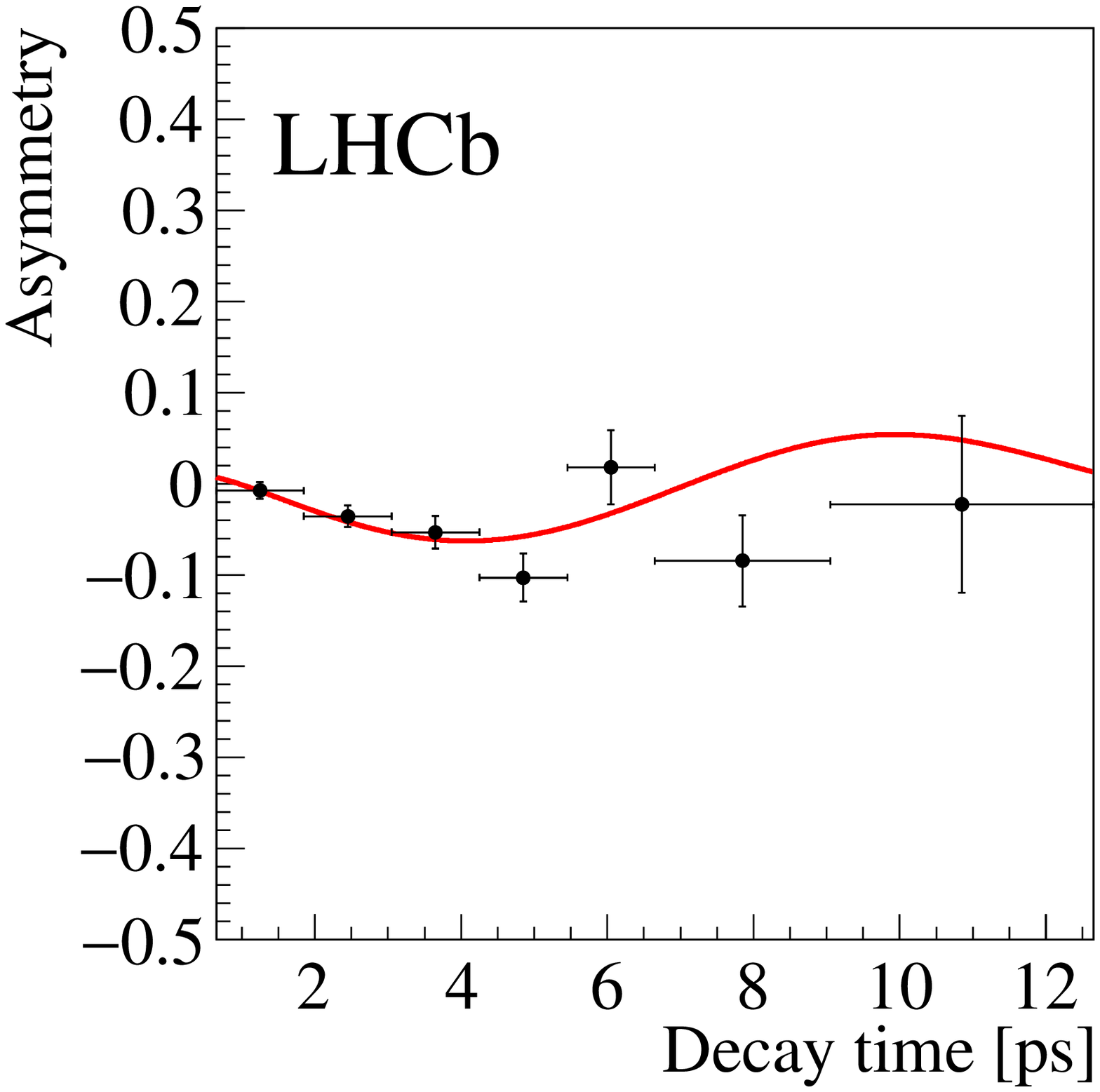}
	\includegraphics[width=0.38\textwidth]{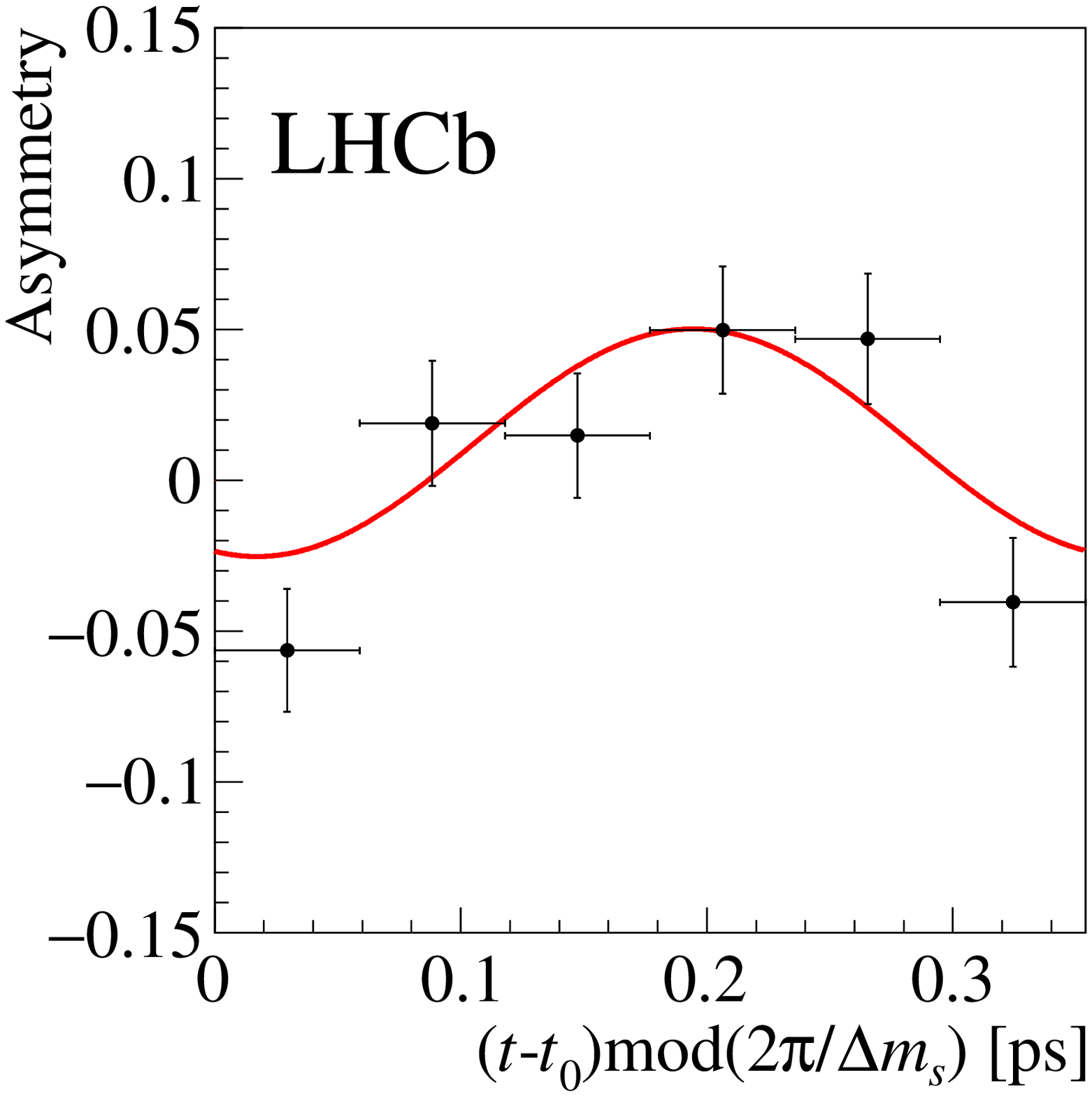}
	\includegraphics[width=0.38\textwidth]{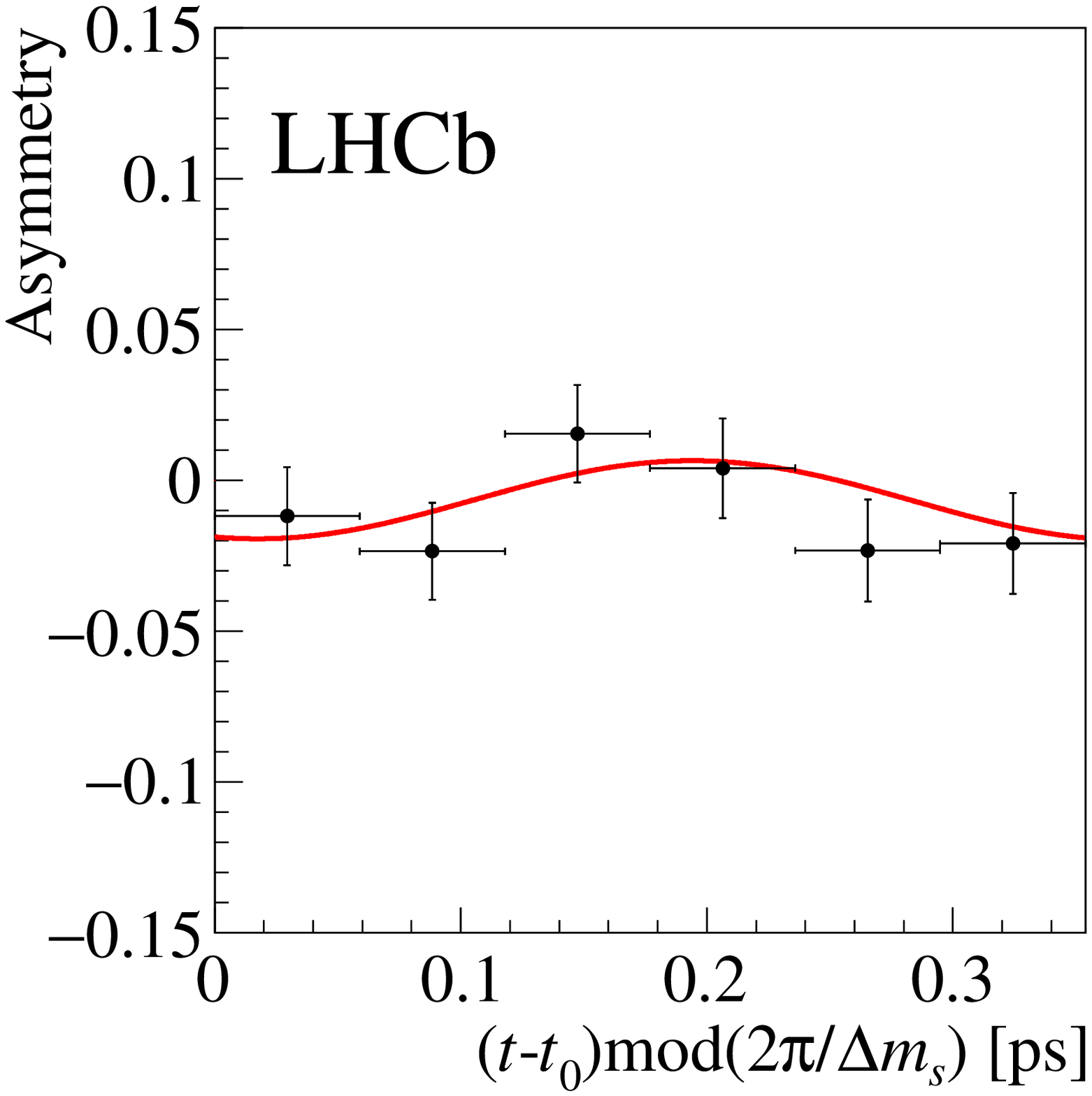}
 \end{center}
  \caption{\small Time-dependent asymmetries for (top) \pip\pim and (bottom) $\Kp\!\Km$ candidates with mass values in the intervals $5.20 < m < 5.35\gevcc$ and $5.30 < m < 5.44\gevcc$, respectively: (left) using the OS-tagging decision and (right) using either the SSc-tagging decision (for the \pip\pim candidates) or the SS\kaon-tagging decision (for the $\Kp\!\Km$ candidates). The result of the simultaneous fit is overlaid.}
  \label{fig:rawAsymCPEigenstate}
\end{figure}

The time-dependent asymmetries, obtained separately by using the OS or the SS tagging decisions, for candidates in the region $5.20 < m < 5.32\gevcc$ in the \Kp\pim spectrum, dominated by the \BdToKpi decay, are shown in Fig.~\ref{fig:rawAsymmetryKPI}. 
The calibration parameters of the OS and SSc taggers determined during the fit, mainly from \BdToKpi decays, are reported in Table~\ref{tab:taggingCalibrationResults} in App.~\ref{sec:flavourTaggingAppendix}. 
The production asymmetries for the \Bd and \Bs mesons are determined to be $(0.19 \pm 0.60)\%$ and $(2.4 \pm 2.1)\%$, respectively, where uncertainties are statistical only. They are consistent with the expectations from Ref.~\cite{LHCb-PAPER-2016-062}. The time-dependent asymmetries for \pip\pim candidates with mass values lying in the interval $5.20 < m < 5.35\gevcc$, and for $\Kp\!\Km$ candidates in the interval $5.30 < m < 5.45\gevcc$, both dominated by the corresponding signals, are shown in Fig.~\ref{fig:rawAsymCPEigenstate}, again separately for the OS and SS tagging decision. The tagging powers for the \BdTopipi and \BsToKK decays, together with a breakdown of the OS and SS contributions, are reported in Table~\ref{tab:taggingPowerSummary}. 
The results for the \CP-violating quantities are
\begin{eqnarray*}
  \Cpipi & = & -0.34\phantom{0} \pm 0.06,\\
  \Spipi & = & -0.63\phantom{0} \pm 0.05,\\
  \CKK   & = & \phantom{-}0.20\phantom{0} \pm 0.06,\\
  \SKK   & = & \phantom{-}0.18\phantom{0} \pm 0.06,\\
  \ADGKK & = & -0.79\phantom{0} \pm 0.07, \\
  \ACPBd & = & -0.084 \pm 0.004, \\
  \ACPBs & = & \phantom{-}0.213 \pm 0.015, 
\end{eqnarray*}
where the uncertainties are statistical only and the central values of \ACPBd~and \ACPBs~have been corrected for the \Kp\pim detection asymmetry. In this analysis the selection requirements and the flavour tagging performances for the various decay modes differ with respect to previous \lhcb publications~\cite{LHCb-PAPER-2013-018,LHCb-PAPER-2013-040}. For this reason, the statistical uncertainties are improved and do not follow a simple scaling rule with the integrated luminosity.


\section{Systematic uncertainties}\label{sec:systematics}

Two different strategies are adopted to determine systematic uncertainties on the \CP-violating parameters: to account for the knowledge of external inputs whose values are fixed in the fit, the fit to the data is repeated a large number of times, each time modifying the values of these parameters; when accounting for systematic uncertainties on the fitting model, several pseudoexperiments are performed according to the baseline model, and both the baseline model and modified models are used to fit the generated data. In either case the distribution of the difference between the baseline and alternative results for the \CP asymmetries is built, and the sum in quadrature of the mean and root-mean-square of the distribution is used to assign a systematic uncertainty. A detailed breakdown of the systematic uncertainties described in this Section is reported in Table~\ref{tab:systSummary}.

The alternative models used to determine systematic uncertainties associated with the choices of the invariant-mass shapes consist in turn of: substituting the invariant-mass resolution function used for signals and cross-feed backgrounds with a single Gaussian function; fixing the parameters governing the tails of the Johnson functions and their relative amount to the same values for all signals, namely to those of the \BdToKpi decay; and modelling the combinatorial-background model with a linear function.

To determine a systematic uncertainty associated with the knowledge of the efficiency as a function of the decay time, $\varepsilon_{\rm sig}(t)$, different sets of the parameters governing the efficiency functions are generated, according to their uncertainties and correlations. A systematic uncertainty associated with the choice of the decay-time model for the cross-feed backgrounds is evaluated by using an alternative model where the \CP asymmetry of the \BdToKpi component in the \pip\pim and $\Kp\!\Km$ final-state samples, and the $C_f$ and $S_f$ parameters of the \BdTopipi and \BsToKK components in the \Kp\pim final-state sample, are fixed to zero. A systematic uncertainty associated with the choice of the decay-time model for the combinatorial background is evaluated using a uniform decay-time efficiency function for this component in the alternative model. A systematic uncertainty associated with the model adopted for the three-body background is evaluated by performing the fits to pseudoexperiments, removing candidates with invariant-mass values lower than 5.2\gevcc, and removing the components describing this background from the model.

Systematic uncertainties associated with the calibration of the per-event decay-time resolution are due to the uncertainties on the parameters $q_0$ and $q_1$ and to the simulation-driven assumption that the resolution model is well described by a double Gaussian function. Different values for $q_0$ and $q_1$ are generated according to their uncertainties and correlations, and then are repeatedly used to fit the data. 
In addition, an alternative model for the decay-time resolution is used to assess a systematic uncertainty, including an additional contribution described by a third Gaussian function. The relative contributions of the three Gaussian functions and the ratios between their widths are determined from simulation, and the overall calibration of the new model is performed applying the same procedure outlined in Sec.~\ref{sec:timeResolution}. 
A systematic uncertainty associated with the uncertainties on the parameters reported in Table~\ref{tab:lifetimeParameters} is determined by repeating the simultaneous fit using different fixed values, generated according to their uncertainties and correlations. 

Systematic uncertainties associated with the calibration of the OS and SSc flavour-tagging responses are determined by replacing the linear relation between $\eta_{\rm OS(SS)}$ and $\omega_{\rm OS(SS)}$ of Eq.~\eqref{eq:taggingCalibrationParameters} with a second-order polynomial. A systematic uncertainty associated with the calibration of the SS\kaon flavour-tagging response is determined by varying the calibration parameters reported in Table~\ref{tab:calibrationSSkNNTotal} according to their uncertainties and correlations. Finally, the uncertainties on the PID and detection asymmetries reported in Eqs.~\eqref{eq:apidB2KPI} and \eqref{eq:adkpiBs2KPI} are accounted for as systematic uncertainties on \ACPBd~and \ACPBs.

The total systematic uncertainties are obtained as the quadratic sum of the individual contributions, and are smaller than the corresponding statistical uncertainties for all parameters but \ADGKK. The dominating systematic uncertainty for \ADGKK~is related to the knowledge of how the efficiency varies with the decay time. Since such a dependence is determined from data, using the \BdToKpi decay, the size of the associated uncertainty will be reduced with future data. 
\begin{landscape}
\begin{table}[t]
  \caption{\small Systematic uncertainties on the various \CP-violating parameters. When present, the dash indicates that the uncertainty is not applicable to the given case.}
  \begin{center}
    \begin{tabular}{l|ccccccc}
        Source of uncertainty           & $\Cpipi$ & \Spipi   & \CKK     & \SKK     & \ADGKK   & \ACPBd     & \ACPBs   \\
        \hline
		Time-dependent efficiency       & $0.0011$      & $0.0004$      & $0.0020$      & $0.0017$      & $0.0778$      & $0.0004$      & $0.0002$   \\
		Time-resolution calibration     & $0.0014$      & $0.0013$      & $0.0108$      & $0.0119$      & $0.0051$      & $0.0001$      & $0.0001$   \\
		Time-resolution model           & $0.0001$      & $0.0005$      & $0.0002$      & $0.0002$      & $0.0003$      & negligible    & negligible \\
		Input parameters                & $0.0025$      & $0.0024$      & $0.0092$      & $0.0107$      & $0.0480$      & negligible    & $0.0001$   \\
		OS-tagging calibration          & $0.0018$      & $0.0021$      & $0.0018$      & $0.0019$      & $0.0001$      & negligible    & negligible \\
		SS\kaon-tagging calibration     & {\textemdash} & {\textemdash} & $0.0061$      & $0.0086$      & $0.0004$      & {\textemdash} & {\textemdash} \\
		SSc-tagging calibration         & $0.0015$      & $0.0017$      & {\textemdash} & {\textemdash} & {\textemdash} & negligible    & negligible \\
		Cross-feed time model           & $0.0075$      & $0.0059$      & $0.0022$      & $0.0024$      & $0.0003$      & $0.0001$      & $0.0001$   \\ 
		Three-body bkg.                 & $0.0070$      & $0.0056$      & $0.0044$      & $0.0043$      & $0.0304$      & $0.0008$      & $0.0043$   \\
		Comb.-bkg. time model           & $0.0016$      & $0.0016$      & $0.0004$      & $0.0002$      & $0.0019$      & $0.0001$      & $0.0005$   \\

		Signal mass model (reso.)       & $0.0027$      & $0.0025$      & $0.0015$      & $0.0015$      & $0.0023$      & $0.0001$      & $0.0041$   \\
		Signal mass model (tails)       & $0.0007$      & $0.0008$      & $0.0013$      & $0.0013$      & $0.0016$      & negligible    & $0.0003$   \\
		Comb.-bkg. mass model           & $0.0001$      & $0.0003$      & $0.0002$      & $0.0002$      & $0.0016$      & negligible    & $0.0001$   \\

		PID asymmetry                   & {\textemdash} & {\textemdash} & {\textemdash} & {\textemdash} & {\textemdash} & $0.0025$      & $0.0025$   \\
		Detection asymmetry             & {\textemdash} & {\textemdash} & {\textemdash} & {\textemdash} & {\textemdash} & $0.0014$      & $0.0014$   \\
		\hline
		Total                           & $0.0115$      & $0.0095$      & $0.0165$      & $0.0191$      & $0.0966$      & $0.0030$      & $0.0066$   \\
    \end{tabular}
  \end{center}
  \label{tab:systSummary}
\end{table}
\end{landscape}

\section{Conclusions}\label{sec:conclusions}

Measurements are presented of time-dependent \CP violation in \BdTopipi and \BsToKK decays, and of the \CP asymmetries in \BdToKpi and \BsTopiK decays, based on a data sample of \proton\proton collisions corresponding to an integrated luminosity of 3.0\invfb collected with the \lhcb detector at centre-of-mass energies of 7 and 8\tev. The results are
\begin{eqnarray*}\label{eq:finalResults}
  \Cpipi & = & -0.34\phantom{0} \pm 0.06\phantom{0} \pm 0.01,\\
  \Spipi & = & -0.63\phantom{0} \pm 0.05\phantom{0} \pm 0.01,\\
  \CKK   & = & \phantom{-}0.20\phantom{0} \pm 0.06\phantom{0} \pm 0.02,\\
  \SKK   & = & \phantom{-}0.18\phantom{0} \pm 0.06\phantom{0} \pm 0.02,\\
  \ADGKK & = & -0.79\phantom{0} \pm 0.07\phantom{0} \pm 0.10, \\
  \ACPBd & = & -0.084 \pm 0.004 \pm 0.003, \\
  \ACPBs & = & \phantom{-}0.213 \pm 0.015 \pm 0.007, 
\end{eqnarray*}
where the first uncertainties are statistical and the second systematic. They supersede with much improved precision those of Refs.~\cite{LHCb-PAPER-2013-040,LHCb-PAPER-2013-018}. The corresponding statistical correlation matrix is reported in Table~\ref{tab:correlationMatrixFinal}. Taking into account the sizes of statistical and systematic uncertainties, correlations due to the latter can be neglected.
\begin{table}[bt]
  \caption{\small Statistical correlations among the \CP-violating parameters.}
  \begin{center}
    \begin{tabular}{l|ccccccc}
            & \Cpipi                       & \Spipi                        & \CKK                         & \SKK                         & \ADGKK                       & \ACPBd                & \ACPBs\rule[-1.3ex]{0pt}{0pt}            \\
     \hline
     \Cpipi & $\phantom{-0.}1\phantom{00}$ & $\phantom{-}0.448$            & $-0.006$ & $-0.009$          & $\phantom{-}0.000$           & $-0.009$                     & $\phantom{-}0.003$\rule{0pt}{2.6ex} \\
     \Spipi &                              & $\phantom{-0.}1\phantom{00}$  & $-0.040$ & $-0.006$          & $\phantom{-}0.000$           & $\phantom{-}0.008$           & $\phantom{-}0.000$ \\
     \CKK   &                              &                               & $\phantom{-0.}1\phantom{00}$ & $-0.014$                     & $\phantom{-}0.025$           & $\phantom{-}0.006$           & $\phantom{-}0.001$ \\
     \SKK   &                              &                               &                              & $\phantom{-0.}1\phantom{00}$ & $\phantom{-}0.028$           & $-0.003$                     & $\phantom{-}0.000$ \\
     \ADGKK &                              &                               &                              &                              & $\phantom{-0.}1\phantom{00}$ & $\phantom{-}0.001$           & $\phantom{-}0.000$ \\
     \ACPBd &                              &                               &                              &                              &                              & $\phantom{-0.}1\phantom{00}$ & $\phantom{-}0.043$ \\
     \ACPBs &                              &                               &                              &                              &                              &                            & $\phantom{-0.}1\phantom{00}$ 
    \end{tabular}
  \end{center}
  \label{tab:correlationMatrixFinal}
\end{table}
The measurements of \Cpipi, \Spipi, \ACPBd~and \ACPBs~are the most precise from a single experiment to date, and are in good agreement with previous determinations~\cite{Lees:2012mma,Adachi:2013mae,Duh:2012ie,Aaltonen:2014vra}. Those of \CKK~and \SKK~are in good agreement with the previous \lhcb result~\cite{LHCb-PAPER-2013-040}. By summing in quadrature the statistical and systematic uncertainties and neglecting the small correlations between \CKK, \SKK~and \ADGKK, the significance for $(\CKK,\,\SKK,\,\ADGKK)$ to differ from $(0,\,0,\,-1)$ is determined by means of a $\chi^2$ test statistic to be $4.0$ standard deviations. This result constitutes the strongest evidence for time-dependent \CP violation in the \Bs-meson sector to date. As a cross-check, the distribution of the variable $Q$, defined by $Q^2=\left(\CKK\right)^2+\left(\SKK\right)^2+\left(\ADGKK\right)^2$, is studied by generating, according to the multivariate Gaussian function defined by their uncertainties and correlations, a large sample of values for the variables \CKK, \SKK~and \ADGKK. The distribution of $Q$ is found to be Gaussian, with mean $0.83$ and width $0.12$.

The measurements of \ACPBd~and \ACPBs~allow a test of the validity of the SM, as suggested in Ref.~\cite{LIPKIN2005126}, by checking the equality
\begin{equation}\label{eq:lipkinTest}
\Delta = \frac{\ACPBd}{\ACPBs}+\frac{\mathcal{B}\left(\BsTopiK\right)}{\mathcal{B}\left(\BdToKpi\right)}\frac{\tau_d}{\tau_s} = 0,
\end{equation}
where $\mathcal{B}\left(\BdToKpi\right)$ and $\mathcal{B}\left(\BsTopiK\right)$ are \CP-averaged branching fractions, and $\tau_\dquark$ and $\tau_\squark$ are the \Bz and \Bs mean lifetimes, respectively. Using the world averages for $f_\squark/f_\dquark\times \mathcal{B}\left(\BsTopiK\right)/\mathcal{B}\left(\BdToKpi\right)$ and $\tau_\squark/\tau_\dquark$~\cite{HFLAV16} and the measurement of the relative hadronisation fraction between \Bs and \Bz mesons $f_\squark/f_\dquark = 0.259 \pm 0.015$~\cite{fsfd}, the value $\Delta = -0.11 \pm 0.04 \pm 0.03$ is obtained, where the first uncertainty is from the measurements of the \CP asymmetries and the second is from the input values of the branching fractions, the lifetimes and the hadronisation fractions. No evidence for a deviation from zero of $\Delta$ is observed with the present experimental precision.

These new measurements will enable improved constraints to be set on the CKM \CP-violating phases, using processes whose amplitudes receive significant contributions from loop diagrams both in the mixing and decay of \Bds mesons~\cite{Fleischer:2010ib,Ciuchini:2012gd,LHCb-PAPER-2014-045}. Comparisons with tree-level determinations of the same phases will provide tests of the SM and constrain possible new-physics contributions.


\section*{Acknowledgements}
%
%
\noindent We express our gratitude to our colleagues in the CERN
accelerator departments for the excellent performance of the LHC. We
thank the technical and administrative staff at the LHCb
institutes. We acknowledge support from CERN and from the national
agencies: CAPES, CNPq, FAPERJ and FINEP (Brazil); MOST and NSFC
(China); CNRS/IN2P3 (France); BMBF, DFG and MPG (Germany); INFN
(Italy); NWO (The Netherlands); MNiSW and NCN (Poland); MEN/IFA
(Romania); MinES and FASO (Russia); MinECo (Spain); SNSF and SER
(Switzerland); NASU (Ukraine); STFC (United Kingdom); NSF (USA).  We
acknowledge the computing resources that are provided by CERN, IN2P3
(France), KIT and DESY (Germany), INFN (Italy), SURF (The
Netherlands), PIC (Spain), GridPP (United Kingdom), RRCKI and Yandex
LLC (Russia), CSCS (Switzerland), IFIN-HH (Romania), CBPF (Brazil),
PL-GRID (Poland) and OSC (USA). We are indebted to the communities
behind the multiple open-source software packages on which we depend.
Individual groups or members have received support from AvH Foundation
(Germany), EPLANET, Marie Sk\l{}odowska-Curie Actions and ERC
(European Union), ANR, Labex P2IO and OCEVU, and R\'{e}gion
Auvergne-Rh\^{o}ne-Alpes (France), Key Research Program of Frontier
Sciences of CAS, CAS PIFI, and the Thousand Talents Program (China),
RFBR, RSF and Yandex LLC (Russia), GVA, XuntaGal and GENCAT (Spain),
Herchel Smith Fund, the Royal Society, the English-Speaking Union and
the Leverhulme Trust (United Kingdom).

\clearpage
{\noindent\normalfont\bfseries\Large Appendix}
\appendix
\section{Flavour-tagging details}\label{sec:flavourTaggingAppendix}

\subsection{Formalism}

The functions $\Omega_{\rm sig}(\vec{\xi},\,\vec{\eta})$ and $\bar{\Omega}_{\rm sig}(\vec{\xi},\,\vec{\eta})$ in Eqs.~\eqref{eq:decayTimeB2KPI} and~\eqref{eq:decayTimeB2HH} are 
\begin{eqnarray}\label{eq:omegaOS}
    \Omega_{\rm sig}(\vec{\xi},\,\vec{\eta}) & = & \Omega^{\rm OS}_{\rm sig}(\xi_{\rm OS},\,\eta_{\rm OS})\,\Omega^{\rm SS}_{\rm sig}(\xi_{\rm SS},\,\eta_{\rm SS}), \\
    \bar{\Omega}_{\rm sig}(\vec{\xi},\,\vec{\eta}) & = & \bar{\Omega}^{\rm OS}_{\rm sig}(\xi_{\rm OS},\,\eta_{\rm OS})\,\bar{\Omega}^{\rm SS}_{\rm sig}(\xi_{\rm SS},\,\eta_{\rm SS}), \nonumber
\end{eqnarray} 
where $\Omega^{\rm tag}_{\rm sig}(\xi_{\rm tag},\,\eta_{\rm tag})$ and $\bar{\Omega}^{\rm tag}_{\rm sig}(\xi_{\rm tag},\,\eta_{\rm tag})$ (with ${\rm tag} \in \{{\rm OS},\,{\rm SS}\}$) are
\begin{equation}\label{eq:tagginProbs}
    \begin{split}
        \Omega^{\rm tag}_{\rm sig}(\xi_{\rm tag},\,\eta_{\rm tag}) = & \delta_{\xi_{\rm tag},\,1}\,\varepsilon^{\rm tag}_{\rm sig}\,\left[1-\omega_{\rm tag}(\eta_{\rm tag})\right]\,h^{\rm tag}_{\rm sig}(\eta_{\rm tag})\,+\\
                                                 & \delta_{\xi_{\rm tag},\,-1}\,\varepsilon^{\rm tag}_{\rm sig}\,\omega_{\rm tag}(\eta_{\rm tag})\,h^{\rm tag}_{\rm sig}(\eta_{\rm tag})\,+\\
                                                 & \delta_{\xi_{\rm tag},\,0}\,(1-\varepsilon^{\rm tag}_{\rm sig})\,U(\eta_{\rm tag}), \\
        \bar{\Omega}^{\rm tag}_{\rm sig}(\xi_{\rm tag},\,\eta_{\rm tag}) = & \delta_{\xi_{\rm tag},\,-1}\,\bar{\varepsilon}^{\rm tag}_{\rm sig}\,\left[1-\bar{\omega}_{\rm tag}(\eta_{\rm tag})\right]\,h^{\rm tag}_{\rm sig}(\eta_{\rm tag})\,+\\
                                                 & \delta_{\xi_{\rm tag},\,1}\,\bar{\varepsilon}^{\rm tag}_{\rm sig}\,\bar{\omega}_{\rm tag}(\eta_{\rm tag})\,h^{\rm tag}_{\rm sig}(\eta_{\rm tag})\,+\\
                                                 & \delta_{\xi_{\rm tag},\,0}\,(1-\bar{\varepsilon}^{\rm tag}_{\rm sig})\,U(\eta_{\rm tag}).
    \end{split}
\end{equation}
The symbol $\delta_{\xi_{\rm tag},\,i}$ stands for the Kronecker delta function, $\varepsilon^{\rm tag}_{\rm sig}$ ($\bar{\varepsilon}^{\rm tag}_{\rm sig}$) is the probability that the flavour of a \Bds (\Bdsb) meson is tagged, $\omega_{\rm tag}(\eta_{\rm tag})$ ($\bar{\omega}_{\rm tag}(\eta_{\rm tag})$) is the calibrated mistag probability as a function of $\eta_{\rm tag}$ for a \Bds (\Bdsb) meson, $h^{\rm tag}_{\rm sig}(\eta_{\rm tag})$ is the PDF describing the distribution of $\eta_{\rm tag}$ for tagged events, and $U(\eta_{\rm tag})$ is a uniform distribution of $\eta_{\rm tag}$. It is empirically found that, to a good approximation, $\eta_{\rm tag}$ and $\omega_{\rm tag}$ are related by a linear function, \ie
\begin{eqnarray}\label{eq:taggingCalibrationParameters}
    \omega_{\rm tag}(\eta_{\rm tag}) & = & p^{\rm tag}_0\,+\,p^{\rm tag}_1\,(\eta_{\rm tag}\,-\,\hat{\eta}_{\rm tag}), \\
    \bar{\omega}_{\rm tag}(\eta_{\rm tag}) & = & \bar{p}^{\rm tag}_0\,+\,\bar{p}^{\rm tag}_1\,(\eta_{\rm tag}\,-\,\hat{\eta}_{\rm tag}),\nonumber
\end{eqnarray}
where $\hat{\eta}_{\rm tag}$ is a fixed value, chosen to be equal to the mean value of the $\eta_{\rm tag}$ distribution to minimise the correlation among the parameters. To reduce the correlation among $\varepsilon^{\rm tag}_{\rm sig}$ and $\bar{\varepsilon}^{\rm tag}_{\rm sig}$, and $p^{\rm tag}_0$, $\bar{p}^{\rm tag}_0$, $p^{\rm tag}_1$, and $\bar{p}^{\rm tag}_1$, these variables are conveniently parameterised as
\begin{eqnarray}
    \varepsilon^{\rm tag}_{\rm sig} & = & \hat{\varepsilon}^{\rm tag}_{\rm sig}(1+\Delta\varepsilon^{\rm tag}_{\rm sig}), \nonumber \\
    \bar{\varepsilon}^{\rm tag}_{\rm sig} & = & \hat{\varepsilon}^{\rm tag}_{\rm sig}(1-\Delta\varepsilon^{\rm tag}_{\rm sig}), \nonumber \\
    p^{\rm tag}_0 & = & \hat{p}^{\rm tag}_0(1+\Delta p^{\rm tag}_0), \\
    \bar{p}^{\rm tag}_0 & = & \hat{p}^{\rm tag}_0(1-\Delta p^{\rm tag}_0), \nonumber \\
    p^{\rm tag}_1 & = & \hat{p}^{\rm tag}_1(1+\Delta p^{\rm tag}_1), \nonumber \\
    \bar{p}^{\rm tag}_1 & = & \hat{p}^{\rm tag}_1(1-\Delta p^{\rm tag}_1),\nonumber
\end{eqnarray}
where $\hat{p}^{\rm tag}_{0,1}$ and $\Delta p^{\rm tag}_{0,1}$ are the average and the asymmetry between $p^{\rm tag}_{0,1}$ and $\bar{p}^{\rm tag}_{0,1}$, and $\hat{\varepsilon}^{\rm tag}_{\rm sig}$ and $\Delta\varepsilon^{\rm tag}_{\rm sig}$ are the average and the asymmetry between $\varepsilon^{\rm tag}_{\rm sig}$ and $\bar{\varepsilon}^{\rm tag}_{\rm sig}$. The PDF $h^{\rm OS}_{\rm sig}(\eta)$ is modelled using background-subtracted histograms of signal candidates. The description of $h^{\rm SS}_{\rm sig}(\eta)$ for the SS taggers is presented in Secs.~\ref{sec:ssdCombination} and~\ref{sec:sskCalibration}, respectively.

The PDF of $\xi_{\rm tag}$ and $\eta_{\rm tag}$ for the combinatorial background is empirically parameterised as
\begin{equation}\label{eq:omegaBkg}
    \begin{split}
        \Omega^{\rm tag}_{\rm comb}(\xi_{\rm tag},\,\eta_{\rm tag}) = & \delta_{\xi_{\rm tag},\,1}\varepsilon^{\rm tag}_{\rm comb}\,h^{\rm tag}_{\rm comb}(\eta_{\rm tag})\,+\,\delta_{\xi_{\rm tag},\,-1}\bar{\varepsilon}^{\rm tag}_{\rm comb}\,h^{\rm tag}_{\rm comb}(\eta_{\rm tag})\,+ \\
        & \delta_{\xi_{\rm tag},\,0}\,(1-\varepsilon^{\rm tag}_{\rm comb}-\bar{\varepsilon}^{\rm tag}_{\rm comb})\,U(\eta_{\rm tag}),
    \end{split}
\end{equation}
where $\varepsilon^{\rm tag}_{\rm comb}$ and $\bar{\varepsilon}^{\rm tag}_{\rm comb}$ are the efficiencies to tag a combinatorial-background candidate as \Bds or \Bdsb, respectively, $h^{\rm tag}_{\rm comb}(\eta_{\rm tag})$ is the PDF of $\eta_{\rm tag}$. As done for the signal model, the tagging efficiencies are parameterised as
\begin{eqnarray}
    \varepsilon^{\rm tag}_{\rm comb} & = & \frac{\hat{\varepsilon}^{\rm tag}_{\rm comb}}{2}(1+\Delta\varepsilon^{\rm tag}_{\rm comb}), \\
    \bar{\varepsilon}^{\rm tag}_{\rm comb} & = & \frac{\hat{\varepsilon}^{\rm tag}_{\rm comb}}{2}(1-\Delta\varepsilon^{\rm tag}_{\rm comb}),\nonumber
\end{eqnarray}
such that the fits determine the total efficiency to tag a combinatorial-background candidate as \Bds or \Bdsb ($\hat{\varepsilon}^{\rm tag}_{\rm comb}$), and the asymmetry between the two efficiencies ($\Delta\varepsilon^{\rm tag}_{\rm comb}$). The PDF $h^{\rm tag}_{\rm comb}(\eta_{\rm tag})$ is determined as a histogram from the high-mass sideband where only combinatorial background is present. The combined PDF of $\xi_{\rm OS}$, $\xi_{\rm SS}$, $\eta_{\rm OS}$ and $\eta_{\rm SS}$, analogously to the signal case, is given by
\begin{equation}\label{eq:omegaBkgComb}
  \Omega_{\rm comb}(\vec{\xi},\,\vec{\eta}) = \Omega^{\rm OS}_{\rm comb}(\xi_{\rm OS},\,\eta_{\rm OS})\cdot \Omega^{\rm SS}_{\rm comb}(\xi_{\rm SS},\,\eta_{\rm SS}).
\end{equation}

The PDF of $\xi_{\rm tag}$ and $\eta_{\rm tag}$ for three-body backgrounds in the \pip\pim and \Kp\!\Km spectra is empirically parameterised as
\begin{equation}\label{eq:omegaPhys}
    \begin{split}
        \Omega^{\rm tag}_{\threebody}(\xi_{\rm tag},\,\eta_{\rm tag}) = & \delta_{\xi_{\rm tag},\,1}\varepsilon^{\rm tag}_{\threebody}\,h^{\rm tag}_{\threebody}(\eta_{\rm tag})\,+\,\delta_{\xi_{\rm tag},\,-1}\bar{\varepsilon}^{\rm tag}_{\threebody}\,h^{\rm tag}_{\threebody}(\eta_{\rm tag})\,+ \\
        & \delta_{\xi_{\rm tag},\,0}\,(1-\varepsilon^{\rm tag}_{\threebody}-\bar{\varepsilon}^{\rm tag}_{\threebody})\,U(\eta_{\rm tag}),
    \end{split}
\end{equation}
where $\varepsilon^{\rm tag}_{\threebody}$ and $\bar{\varepsilon}^{\rm tag}_{\threebody}$ are the efficiencies to tag a background candidate as \Bds or \Bdsb, respectively, and $h^{\rm tag}_{\threebody}(\eta_{\rm tag})$ is the PDF of $\eta_{\rm tag}$. Also in this case the tagging efficiencies are parameterised as a function of the total efficiency ($\hat{\varepsilon}^{\rm tag}_{\threebody}$) and asymmetry ($\Delta\varepsilon^{\rm tag}_{\threebody}$)
\begin{eqnarray}
    \varepsilon^{\rm tag}_{\threebody} & = & \frac{\hat{\varepsilon}^{\rm tag}_{\threebody}}{2}(1+\Delta\varepsilon^{\rm tag}_{\threebody}), \\
    \bar{\varepsilon}^{\rm tag}_{\threebody} & = & \frac{\hat{\varepsilon}^{\rm tag}_{\threebody}}{2}(1-\Delta\varepsilon^{\rm tag}_{\threebody}).\nonumber
\end{eqnarray}
The PDF $h_{\threebody}^{\rm tag}(\eta_{\rm tag})$ is determined as a histogram from the low-mass sideband, where the residual contamination of combinatorial-background candidates is subtracted. As mentioned in Sec.~\ref{sec:timeModel}, for the \Kp\pim final-state sample the three-body background is parameterised in the same way as for the \BdToKpi decay, but with independent parameters for the flavour-tagging calibration.

The PDFs in Eqs.~\eqref{eq:omegaOS}, \eqref{eq:omegaBkgComb} and~\eqref{eq:omegaPhys} are valid if $\eta_{\rm OS}$ and $\eta_{\rm SS}$ are uncorrelated variables. This assumption is verified by means of background-subtracted~\cite{Pivk:2004ty} signals, and of candidates from the high- and low-mass sidebands for the combinatorial and three-body backgrounds, respectively. 

\subsection{Combination of the SS\boldmath{\pion} and SS\boldmath{\proton} taggers}\label{sec:ssdCombination}

\begin{figure}[t]
    \begin{center}
        \includegraphics[width=0.49\textwidth]{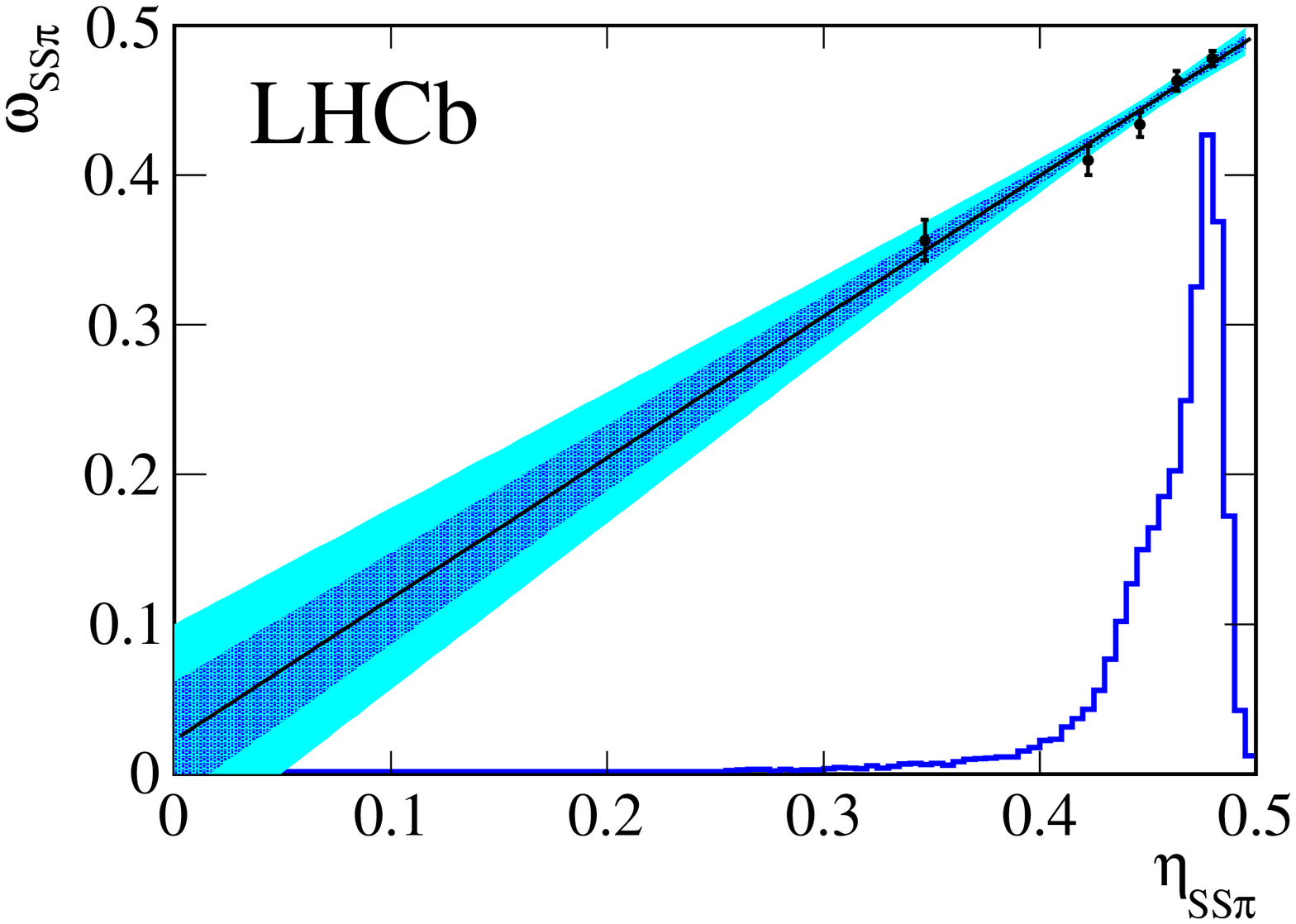}
        \includegraphics[width=0.49\textwidth]{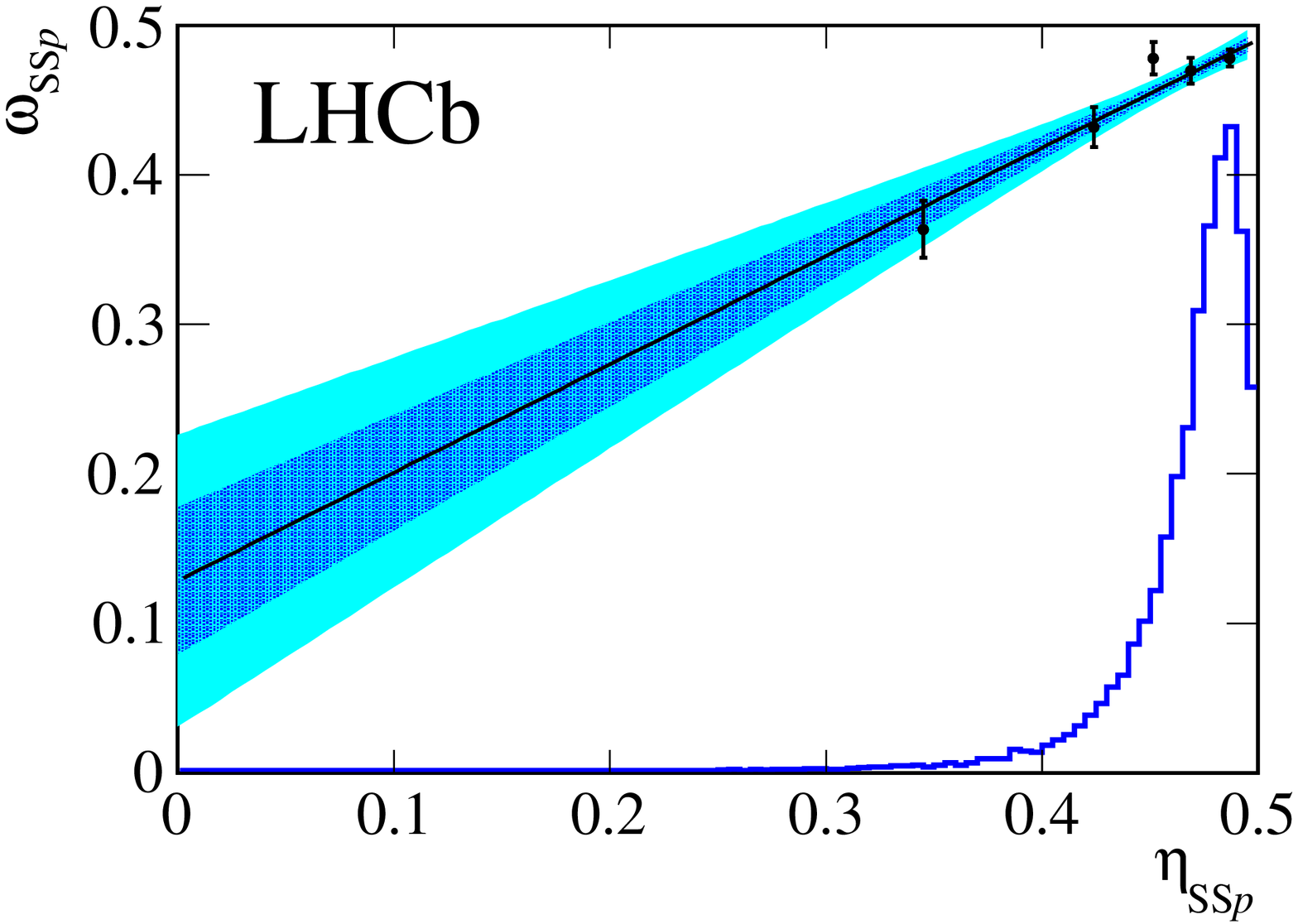}
        \includegraphics[width=0.49\textwidth]{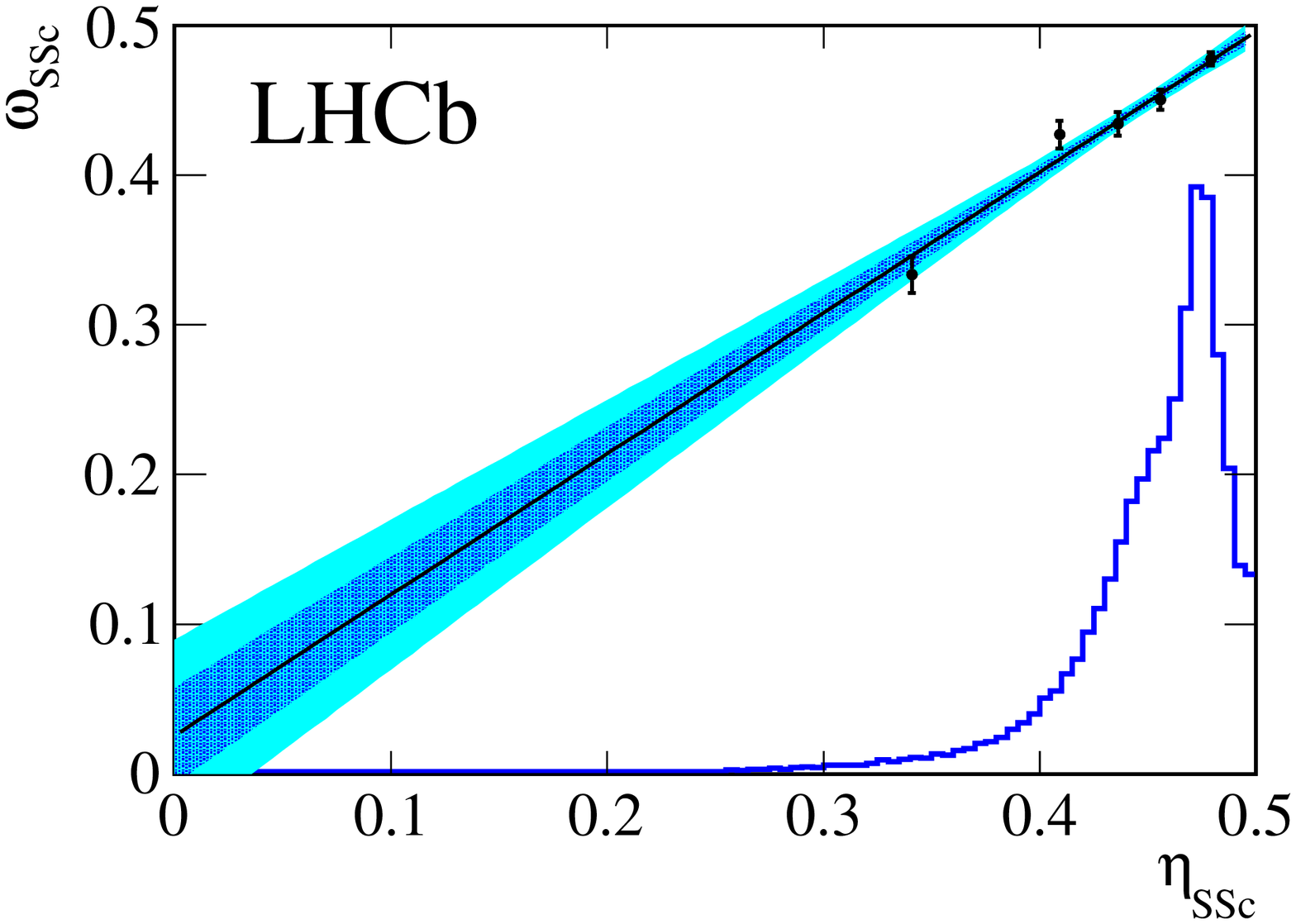}
        \includegraphics[width=0.49\textwidth]{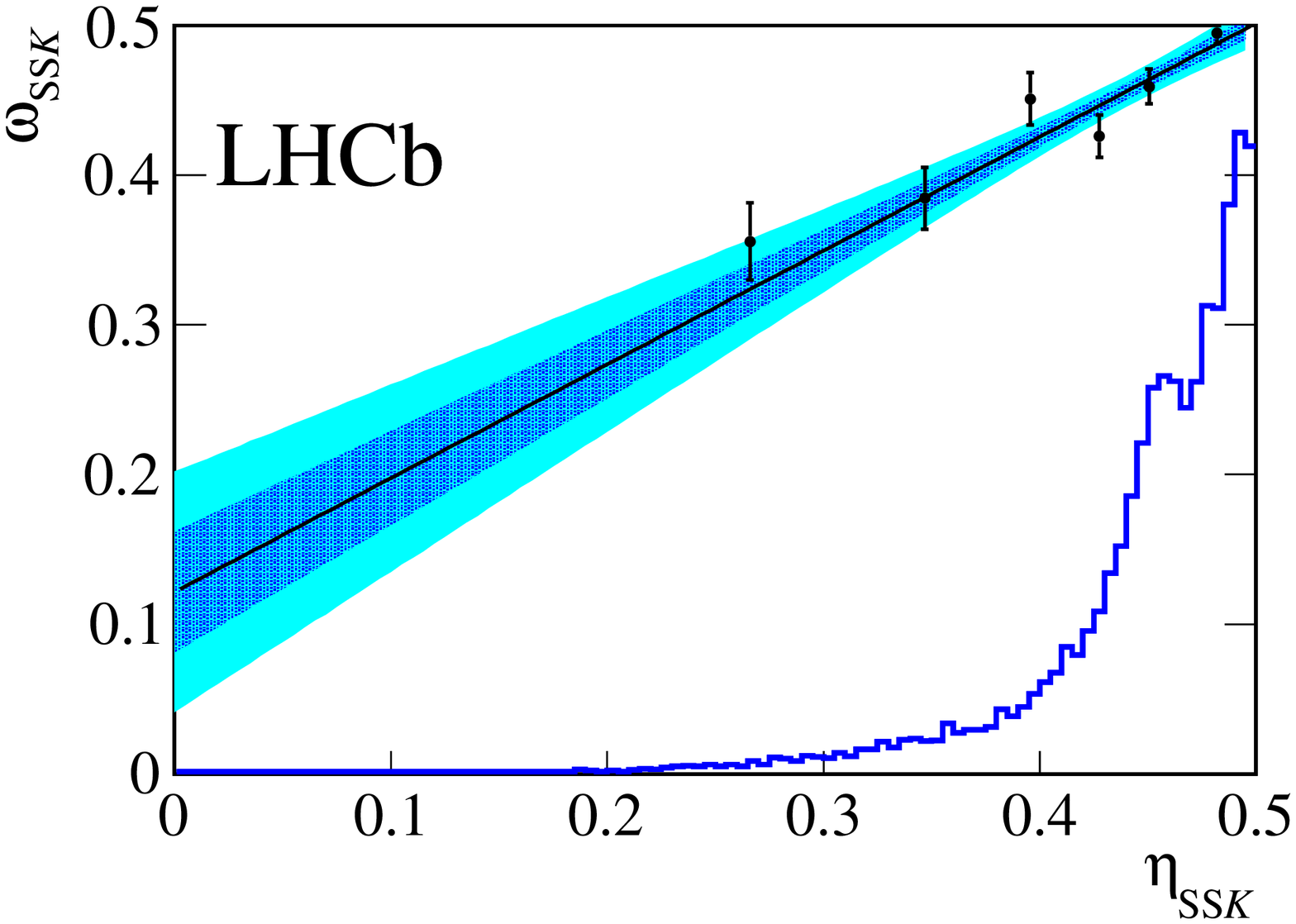}
    \end{center}
    \caption{\small Relation between $\omega_{\rm tag}$ on $\eta_{\rm tag}$ for (top left) SS\pion, (top right) SS\proton, (bottom left) SSc and (bottom right) SS\kaon taggers. The black dots represent the average value of $\omega_{\rm tag}$ in bins of $\eta_{\rm tag}$, as described in the text. The black straight line represents the linear relation between $\omega_{\rm tag}$ and $\eta_{\rm tag}$ obtained from the calibration procedure. The darker and brighter areas are the corresponding 68\% and 95\% confidence intervals, respectively. The distributions of $\eta_{\rm tag}$ are also reported as histograms with arbitrary normalisations.}
    \label{fig:fig_SS_cal}
\end{figure}

The SS\pion and SS\proton taggers are calibrated separately using background-subtracted \BdToKpi decays. By using the PDF in Eq.~\eqref{eq:decayTimeB2KPI} to perform a fit to the tagged decay-time distribution of these candidates, the parameters governing the relations in Eqs.~\eqref{eq:omegaOS} are determined separately for the two taggers. The calibration parameters determined from the fit are used to combine the two taggers into a unique one (SSc) with decision $\xi_{\rm SSc}$ and mistag probability $\eta_{\rm SSc}$. To validate the assumption of a linear relation between $\eta_{\rm tag}$ and $\omega_{\rm tag}$, the sample is split into bins of $\eta_{{\rm SS}\pion{\rm (SS}\proton{\rm )}}$, such that each subsample has approximately the same tagging power. The average mistag fraction in each bin is determined by means of a tagged time-dependent fit to the various subsamples. This check is performed separately for the SS\pion, SS\proton and SSc. The results of the calibration procedure and of the cross-check using the fits in bins of $\eta_{\rm SS\pion}$, $\eta_{\rm SS\proton}$ and~$\eta_{\rm SSc}$ are shown in Fig.~\ref{fig:fig_SS_cal}. The final calibration for $\eta_{\rm SSc}$ is performed during the final fit, and the values of the calibration parameters are reported later in Table~\ref{tab:taggingCalibrationResults}.

The PDFs $h^{\rm SS}_{\rm sig}(\eta_{\rm SSc})$ describing the $\eta_{\rm SSc}$ distributions for the signal \Bd mesons are determined using background-subtracted histograms of \BdToDpi decays. It is empirically found that the distribution of $\eta_{\rm SSc}$ has a sizeable dependence on the \Bd-meson \pt. Hence the \BdToDpi sample is weighted in order to equalise the \pt distribution to that of the signal.
\begin{table}[t]
  \caption{\small Values for the calibration parameters of the flavour tagging obtained from the fits. The values of $\hat{\eta}_{\rm OS}$ and $\hat{\eta}_{\rm SS}$ are fixed in the fit to $0.37$ and $0.44$, respectively.}
  \begin{center}\begin{tabular}{l|c}
      Parameter                           & Value \\
      \hline
      $\hat{p}^{\rm OS}_0$                & $\phantom{-}0.385 \pm 0.004$\rule{0pt}{2.6ex}\\
      $\Delta p^{\rm OS}_0$               & $\phantom{-}0.016 \pm 0.006$ \\
      $\hat{p}^{\rm OS}_1$                & $\phantom{-}1.02\phantom{0} \pm 0.04\phantom{0}$ \\
      $\Delta p^{\rm OS}_1$               & $\phantom{-}0.029 \pm 0.024$\rule[-1.3ex]{0pt}{0pt}\\
    \hline
      $\hat{p}^{\rm SSc}_0$               & $\phantom{-}0.438 \pm 0.003$\rule{0pt}{2.6ex} \\
      $\Delta p^{\rm SSc}_0$              & $\phantom{-}0.002 \pm 0.004$\\
      $\hat{p}^{\rm SSc}_1$               & $\phantom{-}0.96\phantom{0} \pm 0.07\phantom{0}$\\
      $\Delta p^{\rm SSc}_1$              & $-0.03\phantom{0} \pm 0.04\phantom{0}$\\
    \end{tabular}
  \end{center}
  \label{tab:taggingCalibrationResults}
\end{table}

\subsection{Calibration of the SS\boldmath{\kaon} tagger}\label{sec:sskCalibration}

To calibrate the response of the SS\kaon tagger, the natural control mode would be the \BsTopiK decay. However, the signal yield of this decay is approximately 8\% of that of the \BdToKpi decay, and 20\% of that of the \BsToKK decay. Hence the calibration parameters of the SS\kaon tagger would be affected by large uncertainties, limiting the precision on \CKK~and \SKK. Therefore, the calibration is performed with a large sample of \BsToDspi decays. Analogously to the SS\pion and SS\proton cases, the SS\kaon-calibration parameters are determined using an unbinned maximum likelihood fit to the tagged decay-time distribution of the \BsToDspi decay. The PDF used to fit the decay-time rate is the same as that for the SS\pion and SS\proton taggers. The fit is performed using the flavour-tagging information on a per-event basis, determining the calibration parameters directly. To check the linearity of the relation between $\eta_{{\rm SS}\kaon}$ and $\omega_{{\rm SS}\kaon}$, the sample is again divided in bins of $\eta_{{\rm SS}\kaon}$ and the average $\omega_{{\rm SS}\kaon}$ is determined in each bin (see Fig.~\ref{fig:fig_SS_cal}).

The SS\kaon tagger uses kaons coming from the hadronisation of the beauty quark to determine the flavour of the \Bs meson. As the kaon kinematics are correlated to those of the \Bs meson, the performance of the SS\kaon tagger also depends on the latter. To take into account the differences between the \Bs-meson kinematics and other relevant distributions in \BsToDspi and \BsToKK decays, due to the different topologies and selection requirements, a weighting procedure is applied to the \BsToDspi sample. It is empirically found that the distributions of the following variables need to be equalised: the transverse momentum, the pseudorapidity and the azimuthal angle of the \Bs meson, and the number of PVs and tracks in the events. The results of the fit to the weighted sample are reported in Table~\ref{tab:calibrationSSkNNTotal}.

The PDF $h^{{\rm SS}\kaon}_{\rm sig}(\eta_{{\rm SS}\kaon})$ for \BsToKK decays is determined using a background-subtracted histogram of the same weighted sample of \BsToDspi decays used for the calibration.

\begin{table}[t]
  \caption{ \small Calibration parameters for the SS\kaon tagger.}
  \begin{center}
    \begin{tabular}{l|c}
        Parameter                 & Value \\
        \hline
        $\hat{p}^{\rm SS\kaon}_0$               & $\phantom{-}0.456 \pm 0.005$ \\
        $\Delta p^{\rm SS\kaon}_0$              & $-0.011 \pm 0.005$ \\
        $\hat{p}^{\rm SS\kaon}_1$               & $\phantom{-}0.76\phantom{0} \pm 0.09\phantom{0}$ \\
        $\Delta p^{\rm SS\kaon}_1$              & $\phantom{-}0.03\phantom{0} \pm 0.05\phantom{0}$ \\
    \end{tabular}
  \end{center}
  \label{tab:calibrationSSkNNTotal}
\end{table}


\addcontentsline{toc}{section}{References}
\setboolean{inbibliography}{true}
\bibliographystyle{LHCb}
\bibliography{main,LHCb-PAPER,LHCb-CONF,LHCb-DP,LHCb-TDR}

\newpage


 
\newpage
\centerline{\large\bf LHCb collaboration}
\begin{flushleft}
\small
R.~Aaij$^{43}$,
B.~Adeva$^{39}$,
M.~Adinolfi$^{48}$,
Z.~Ajaltouni$^{5}$,
S.~Akar$^{59}$,
P.~Albicocco$^{18}$,
J.~Albrecht$^{10}$,
F.~Alessio$^{40}$,
M.~Alexander$^{53}$,
A.~Alfonso~Albero$^{38}$
% ...

{\footnotesize \it
% ...
$ ^{5}$Clermont Universit{\'e}, Universit{\'e} Blaise Pascal, CNRS/IN2P3, LPC, Clermont-Ferrand, France\\
% ...
$ ^{10}$Fakult{\"a}t Physik, Technische Universit{\"a}t Dortmund, Dortmund, Germany\\
% ...
$ ^{18}$INFN Laboratori Nazionali di Frascati, Frascati, Italy\\
% ...
$ ^{38}$ICCUB, Universitat de Barcelona, Barcelona, Spain\\
$ ^{39}$Instituto Galego de F{\'\i}sica de Altas Enerx{\'\i}as (IGFAE), Universidade de Santiago de Compostela, Santiago de Compostela, Spain\\
$ ^{40}$European Organization for Nuclear Research (CERN), Geneva, Switzerland\\
% ...
$ ^{43}$Nikhef National Institute for Subatomic Physics, Amsterdam, The Netherlands\\
% ...
$ ^{48}$H.H. Wills Physics Laboratory, University of Bristol, Bristol, United Kingdom\\
% ...
$ ^{53}$School of Physics and Astronomy, University of Glasgow, Glasgow, United Kingdom\\
% ...
$ ^{59}$University of Cincinnati, Cincinnati, OH, United States\\
% ...
}
\end{flushleft}

\end{document}